\documentclass[a4paper,12pt]{article}
\pdfoutput=1
\usepackage{url}
\usepackage{epsfig}
\usepackage{amsmath}
\usepackage{amssymb}
\usepackage{amsfonts}
\usepackage{bbm}
\usepackage{fleqn}
\usepackage[footnotesize]{caption}
\usepackage{graphicx}
\usepackage{mathrsfs}
\usepackage[center,footnotesize,hang]{subfigure}

\def\be{\begin{equation}}
\def\ee{\end{equation}}
\def\beq{\begin{equation}}
\def\eeq{\end{equation}}
\def\bea{\begin{eqnarray}}
\def\eea{\end{eqnarray}}

\def\<{\left\langle}
\def\>{\right\rangle}

\def\lsim{\stackrel{<}{{}_\sim}}
\def\gsim{\stackrel{>}{{}_\sim}}

\def\be{\begin{equation}}
\def\ee{\end{equation}}
\def\beq{\begin{equation}}
\def\eeq{\end{equation}}
\def\bea{\begin{eqnarray}}
\def\eea{\end{eqnarray}}

\newcommand{\newc}{\newcommand}
\newc{\ol}{\overline}
\newc{\wt}{\widetilde}
\newc{\bs}{\boldsymbol}
\newc{\ma}{\mathcal}
\newc{\vl}{\langle}
\newc{\vr}{\rangle}
\def\vev#1{\langle #1 \rangle}
\newc{\sg}{S}
\newc{\ug}{U}
\newc{\tg}{T}

\allowdisplaybreaks[2]
\begin{document}
\bibliographystyle{OurBibTeX}

\title{\hfill ~\\[-30mm]
                  \textbf{A to Z of Flavour with Pati-Salam
                                  }        }
\date{}
\author{\\[-5mm]
Stephen F. King\footnote{E-mail: {\tt king@soton.ac.uk}}\\ \\
  \emph{\small School of Physics and Astronomy, University of Southampton,}\\
  \emph{\small Southampton, SO17 1BJ, United Kingdom}\\[4mm]}

\maketitle

\begin{abstract}
\noindent
{We propose an elegant theory of flavour based on $A_4\times Z_5$ family symmetry with 
Pati-Salam unification which provides an excellent description of quark and lepton masses,
mixing and CP violation. The $A_4$ symmetry
unifies the left-handed families and its vacuum alignment 
determines the columns of Yukawa matrices.
 The $Z_5$ symmetry distinguishes the right-handed
families and its breaking controls 
CP violation in both the quark and lepton sectors.
The Pati-Salam symmetry relates the quark and lepton Yukawa matrices,
with $Y^u=Y^{\nu}$ and $Y^d\sim Y^e$. Using the 
see-saw mechanism with very hierarchical right-handed neutrinos
and CSD4 vacuum alignment, the model predicts the entire PMNS mixing matrix
and gives a Cabibbo angle $\theta_C\approx 1/4$. In particular,
for a discrete choice of $Z_5$ phases, it predicts maximal atmospheric mixing,
$\theta^l_{23}=45^\circ\pm 0.5^\circ$ and leptonic CP violating phase
$\delta^l=260^\circ \pm 5^\circ$.
The reactor angle prediction is
$ \theta^l_{13}=9^\circ\pm 0.5^\circ$, while the solar angle is
$34^\circ \gsim \theta^l_{12}\gsim 31^\circ$, for a lightest neutrino mass in the range 
$0 \lsim m_1 \lsim 0.5$ meV, corresponding to a normal
neutrino mass hierarchy and a very small rate for neutrinoless double beta decay.} 
 \end{abstract}
\thispagestyle{empty}
\vfill
\newpage
\setcounter{page}{1}

\section{Introduction}

The problem of understanding the quark and lepton masses, mixing angles and CP violating phases
remains one of the most fascinating puzzles in particle physics. 
Following the discovery of a Standard Model (SM)-like Higgs boson at the LHC \cite{Aad:2012tfa},
it seems highly plausible that quark masses, mixing angles and CP phase originate from
Yukawa couplings to a Higgs field. 
However the SM offers absolutely no insight into the origin or nature of these Yukawa couplings,
motivating approaches beyond the SM \cite{Raby:1995uv}.

In the quark sector, the Yukawa couplings are organised into 
$3\times 3$ quark Yukawa matrices $Y^u$ and $Y^d$,
which must be responsible for 
the quark mass hierarchies and 
small quark mixing angles, together with the CP phase. Similarly, 
the charged lepton Yukawa matrix $Y^e$ must lead to a mass hierarchy similar to 
that of the down-type quarks. The origin of small quark mixing and CP violation
and the strong mass hierarchies of the quarks and charged leptons, with an especially strong hierarchy in the up-type quark sector, is simply unexplained within the SM. The nine charged fermion masses,
three quark mixing angles, including the largest Cabibbo angle $\theta_C\approx 13^\circ$,
and the CP phase are all determined from experiment. 
From a more fundamental point of view, the three Yukawa matrices $Y^u$, $Y^d$ and $Y^e$ contain 54 undetermined Yukawa couplings leading to 13 physical observables with a 
calculable scale dependence \cite{Antusch:2013jca}.

Following the discovery of atmospheric neutrino oscillations by Super-Kamiokande 
in 1998 and solar neutrino oscillations by SNO in 2002 \cite{Strumia:2006db},
Daya Bay has recently accurately measured a non-zero reactor angle \cite{An:2013zwz} which rules out tri-bimaximal (TB)
\cite{Harrison:2002er} mixing. However, recent global fits 
\cite{Forero:2014bxa,Capozzi:2013csa,GonzalezGarcia:2012sz,update}
are consistent with tri-bimaximal-Cabibbo (TBC) 
\cite{King:2012vj} mixing, based on the TB atmospheric angle $\theta^l_{23}\approx 45^\circ$,
the TB solar angle $\theta^l_{12}\approx 35^\circ$ and a reactor angle 
$\theta^l_{13}\approx \theta_C/\sqrt{2}\approx 9^\circ$.
The extra parameters of the lepton sector include three neutrino masses, 
three lepton mixing angles and up to
three CP phases, although no leptonic CP violation has yet been observed and 
the lightest neutrino mass has not been measured.
The 9 additional neutrino observables, 
together with the 13 physical observables in the charged fermion sector,
requires 22 unexplained parameters in the flavour sector of the SM.
This provides a powerful motivation to search for 
theories of flavour (TOF) based on discrete family symmetry 
which contain fewer parameters \cite{King:2013eh}.

The origin of neutrino mass is presently unknown and certainly requires some extension of the SM, even if only by the addition of right-handed (RH) neutrinos which are singlets under the SM gauge group. Since such RH neutrinos may have large Majorana masses, in excess of the 
electroweak breaking scale, such a minimal extension naturally leads to the idea of a see-saw mechanism
\cite{Minkowski:1977sc},
resulting from a neutrino Yukawa matrix $Y^{\nu}$, 
together with a complex symmetric Majorana matrix $M_R$ of heavy right-handed neutrinos,
leading to a light effective Majorana neutrino mass matrix $m^{\nu}\sim v^2Y^{\nu}M_R^{-1}{Y^{\nu}}^T$,
where $v$ is the Higgs vacuum expectation value (VEV).
However the see-saw mechanism does not explain large lepton mixing angles,
with the smallest being the reactor angle $\theta^l_{13}\approx 9^\circ$,
nor does it address any of the flavour puzzles in the charged fermion sector.

The origin of large lepton mixing may be accounted for within the see-saw mechanism with the aid
of sequential dominance (SD) \cite{King:1998jw}. For example,
with an approximately diagonal $M_R$, the lightest 
right-handed neutrino $\nu_R^{\rm atm}$ may give the dominant contribution to 
the atmospheric neutrino mass $m_3$, the second lightest right-handed neutrino $\nu_R^{\rm sol}$ to the solar
neutrino mass $m_2$ and the heaviest, almost decoupled, 
right-handed neutrino $\nu_R^{\rm dec}$ may be responsible for the
lightest neutrino mass $m_1$.
The immediate prediction of SD is a normal neutrino mass hierarchy, 
$m_3>m_2\gg m_1$,
which will be tested 
in the near future. However SD also provides a simple way to account for maximal atmospheric
mixing and tri-maximal solar mixing by adding constraints to the first two columns of the neutrino Yukawa matrix $Y^{\nu}$, with the third column assumed to be approximately decoupled from the see-saw mechanism. 
In the diagonal $Y^e$ basis, 
if the dominant first column of $Y^{\nu}$
is proportional to $(0,1,1)^T$ then this implies a maximal atmospheric angle $\tan \theta^l_{23}\approx 1$
 \cite{King:2002nf}.
This could be achieved with a non-Abelian family symmetry such as $A_4$ \cite{Ma:2001dn}, if the first column is
generated by a triplet flavon field with a vacuum alignment proportional to $(0,1,1)^T$.
In such models, it has been shown that the vacuum alignment completely breaks the $A_4$ symmetry,
and such models are therefore referred to as ``indirect'' models \cite{King:2009ap}. Such ``indirect'' models are highly predictive and do not require such large discrete groups as the ``direct'' models where the Klein symmetry of the neutrino mass matrix is identified as a subgroup of the family symmetry
\cite{King:2013vna,Ishimori:2014jwa,Holthausen:2013vba}.

Constrained sequential dominance (CSD) \cite{King:2005bj}
involves the dominant right-handed neutrino $\nu_R^{\rm atm}$ mainly responsible for the atmospheric neutrino mass 
having couplings to $(\nu_e, \nu_{\mu}, \nu_{\tau})$ proportional to $(0,1,1)$, as above, while the subdominant 
right-handed neutrino $\nu_R^{\rm sol}$ giving the solar neutrino mass has various 
couplings to $(\nu_e, \nu_{\mu}, \nu_{\tau})$ as follows:
\begin{itemize}
\item CSD1: $(1,1,-1)$ leading to TB mixing with zero reactor angle $\theta^l_{13}\approx 0^\circ$ 
\cite{King:2005bj}.
\item CSD2: $(1,2,0)$ giving $\theta^l_{13}\approx 6^\circ$ 
\cite{Antusch:2011ic}.
\item CSD3: $(1,3,1)$ with a relative phase $\pm \pi/3$ 
giving $\theta^l_{13}\approx 8.5^\circ$ \cite{King:2013iva}.
\item CSD4: $(1,4,2)$, with a
relative phase $\pm 2\pi/5$ giving $\theta^l_{13}\approx 9^\circ$ \cite{King:2013iva,King:2013xba}.
\end{itemize}
``Indirect'' models of leptons have been constructed based on $A_4$ using both CSD3 
\cite{King:2013iva} and CSD4 \cite{King:2013xba} since these are the most promising from the 
point of view of the reactor angle. From the point of view of extending to the quark sector,
CSD4 seems to be the most promising since in unified models with $Y^u=Y^{\nu}$,
the second column is proportional to $(1,4,2)^T$. This simultaneously 
provides a prediction for both lepton mixing and the Cabibbo angle
$\theta_C\approx 1/4$ in the diagonal $Y^d\sim Y^e$ basis \cite{King:2013hoa}.

\begin{figure}[t]
\centering
\includegraphics[width=0.50\textwidth]{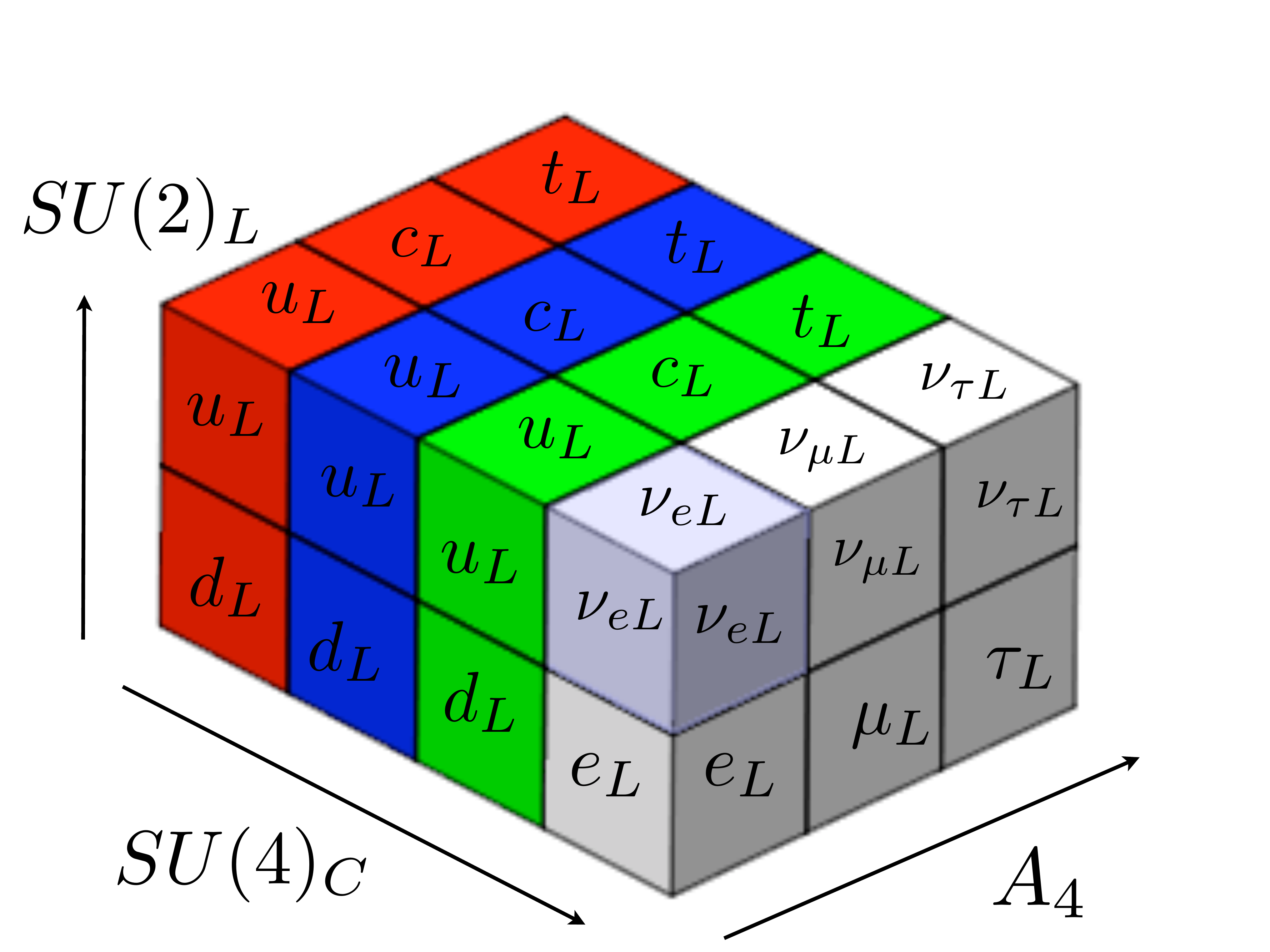}
\includegraphics[width=0.48\textwidth]{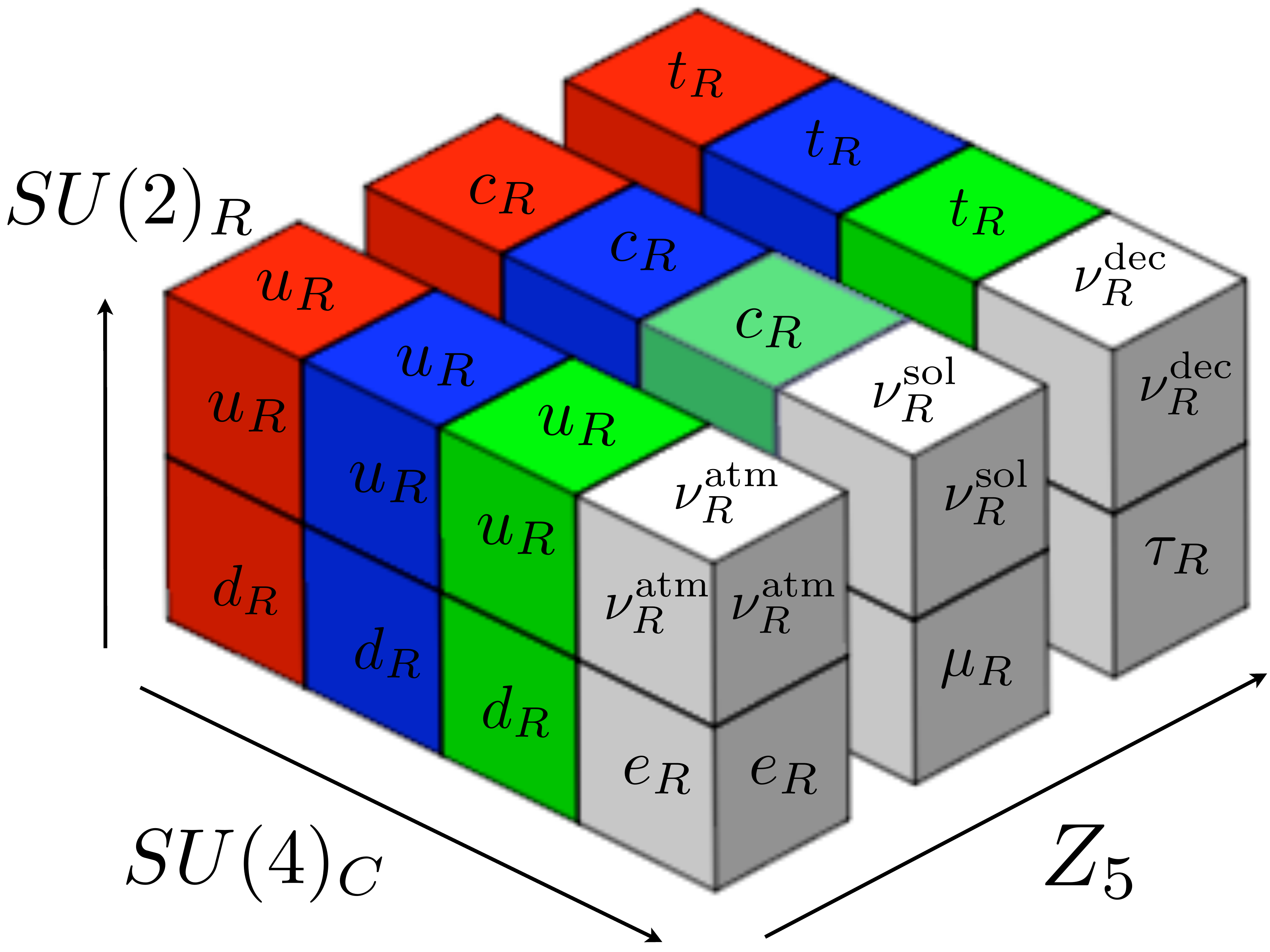}
\vspace*{-4mm}
    \caption{$A$ to $Z$ of flavour with Pati-Salam, where $A\equiv A_4$ and $Z\equiv Z_5$. 
         The left-handed families form a triplet of $A_4$ and are
         doublets of $SU(2)_L$.
The right-handed families are distinguished by $Z_5$
and are doublets of $SU(2)_R$.
The $SU(4)_C$ unifies the quarks and leptons with leptons as the fourth colour,
   depicted here as white.
  } \label{421fig}
\vspace*{-2mm}
\end{figure}

The model in \cite{King:2013hoa} was based on $A_4$ family symmetry
with $Z_3^4\times Z_5^5$ and
quark-lepton unification via the Pati-Salam (PS)
\cite{Pati:1974yy} gauge subgroup $SU(4)_{PS}\times SU(2)_L\times U(1)_R$ and the CSD4 alignment $(1,4,2)$. The small quark mixing angles arose from higher order (HO) corrections
appearing in $Y^u$ and $Y^{\nu}$, providing a theoretical error or noise
which blurred the PMNS predictions. Here we discuss an alternative $A_4$ model which has three advantages over the
previous model. Firstly it is more unified, being 
based on the full PS gauge group $SU(4)_{PS}\times SU(2)_L\times SU(2)_R$ \cite{Pati:1974yy}.
Secondly it introduces only a single $Z_5$ symmetry, replacing the rather 
cumbersome $Z_3^4\times Z_5^5$ symmetry. Thirdly, it accounts for small quark mixing angles already 
at the leading order (LO), with all Higher Order (HO) corrections being rather small, leading to more
precise predictions for the PMNS parameters, such as maximal atmospheric mixing. 
Unlike other $A_4\times {\rm PS}$ models (see e.g. \cite{King:2006np}), the present model does
not involve any Abelian $U(1)$ family symmetry. Instead the left-handed PS fermions are unified
into a triplet of $A_4$ while the right-handed PS fermions are distinguished by $Z_5$,
as in Fig.~\ref{421fig}.

In the present paper, then, we propose a rather elegant TOF 
based on the PS gauge group combined with a discrete
$A_4\times Z_5$ family symmetry.
PS unification relates quark and lepton Yukawa matrices and in particular 
predicts equal up-type quark and neutrino Yukawa matrices
$Y^u=Y^{\nu}$, leading to Dirac neutrino masses being equal to up, charm and top masses.
The see-saw mechanism then implies very hierarchical right-handed neutrinos.
The $A_4$ family symmetry
determines the structure of Yukawa matrices via the CSD4 vacuum alignment 
\cite{King:2013iva,King:2013xba},
with the three columns of $Y^u=Y^{\nu}$ being proportional to 
$(0,1,1)^T$, $(1,4,2)^T$ and $(0,0,1)^T$, respectively,
where each column has an overall phase
determined by $Z_5$ breaking,
which controls CP violation in both the quark and lepton sectors.
The down-type quark and charged lepton Yukawa matrices  
are both approximately equal and diagonal $Y^d\sim Y^e$, but contain small off-diagonal elements 
responsible for the small quark mixing angles $\theta^q_{13}$ and $\theta^q_{23}$.
The model predicts the Cabibbo angle $\theta_C\approx 1/4$, up to such small angle corrections.
The main limitation of the model is that it describes the 
fermion masses and small quark mixing angles by 16 free parameters.
The main success of the model is that, since there are 6 fewer parameters than the 22 flavour observables,
it predicts the entire PMNS lepton mixing matrix including the three lepton mixing angles and 
the three leptonic CP phases. The model may be tested quite soon via its prediction of maximal atmospheric mixing with a normal neutrino mass hierarchy.

The layout of the remainder of the paper is as follows.
In Section~\ref{overview}, we give a brief overview of the essential features of the model.
In Section~\ref{model}, we present the full model and show how the messenger sector can lead to 
effective operators, then discuss how these operators lead to Yukawa and Majorana mass matrices.
In Section~\ref{quarks}, we derive the quark masses and mixing, including CP violation, arising 
from the quark Yukawa matrices, first analytically, then numerically.
In Section~\ref{leptons}, we implement the see-saw mechanism, then 
consider the resulting neutrino masses and lepton mixing,
with modified Georgi-Jarlskog relations, before performing a full numerical analysis
of neutrino masses and lepton mixing, including CP violation.
In Section~\ref{higher}, we consider higher order corrections to the results and show that they are
small due to the particular messenger sector.
Finally Section~\ref{conclusions} concludes the paper.
$A_4$ group theory is discussed in Appendix~\ref{A4} and 
the origin of the light Higgs doublets $H_u$ and $H_d$ in Appendix~\ref{Higgs}.

\section{Overview of the model}
\label{overview}
\subsection{Symmetries of the model}
The model is based on the Pati-Salam gauge group \cite{Pati:1974yy}, 
with $A_4\times Z_5$ family symmetry,
\beq
SU(4)_{C} \times SU(2)_L \times SU(2)_R\times A_4 \times Z_5.
\label{422A4Z5}
\eeq
The quarks and leptons are unified in the PS representations as follows,
\bea
F_i& = & (4,2,1)_i = \left(\begin{array}{cccc}u&u&u&\nu\\
d&d&d&e\end{array}\right)_i \rightarrow  (Q_i,L_i), \nonumber \\
F^c_i & = & (\bar{4},1,2)_i = 
\left(\begin{array}{cccc}u^c&u^c&u^c&\nu^c\\
d^c&d^c&d^c&e^c\end{array}\right)_i \rightarrow (u^c_i,d^c_i,\nu^c_i,e^c_i ),
\label{ql}
\eea
\noindent where the SM multiplets $Q_i,L_i,u^c_i,d^c_i,\nu^c_i,e^c_i$ 
resulting from PS breaking are also shown and the 
subscript $i\ (=1,2,3)$ denotes the family index.
The left-handed quarks and leptons form an $A_4$ triplet $F$, 
while the three (CP conjugated) right-handed fields $F^c_i$ are $A_4$ singlets, distinguished by $Z_5$ charges $\alpha, \alpha^3,1$, for $i=1,2,3$, respectively.
Clearly the Pati-Salam model cannot be embedded into an $SO(10)$ Grand Unified Theory (GUT)
since different components 
of the 16-dimensional representation of $SO(10)$ would have to transform differently under $A_4\times Z_5$,
which is impossible. On the other hand, the PS gauge group and $A_4$ could emerge directly from
string theory (see e.g. \cite{Karozas:2014aha}).

\subsection{Pati-Salam breaking}
The Pati-Salam gauge group is broken at the GUT scale to the SM,
\beq
SU(4)_{C}\times SU(2)_L \times SU(2)_R\rightarrow SU(3)_C\times SU(2)_L \times U(1)_Y,
\eeq
by PS Higgs, 
$H^c$  and $\overline{H^c}$,
\bea
{H^c} & = & (\bar{4},1,2)= (u^c_H,d^c_H,\nu^c_H,e^c_H ), \nonumber \\
 {\overline{H^c}} & = & (4,1,2) = (\bar{u}^c_H,\bar{d}^c_H,\bar{\nu}^c_H,\bar{e}^c_H ).
\eea
These acquire vacuum expectation values (VEVs) in the ``right-handed neutrino'' directions, with equal VEVs
close to the GUT scale $2\times 10^{16}$ GeV,
\beq
 \vev{H^c}= \vev{\nu^c_H}=\vev{\overline{H^c}}=\vev{\bar{\nu}^c_H}\sim 2\times 10^{16} \ {\rm GeV},
\label{PS}
\eeq
so as to maintain supersymmetric gauge coupling unification.
Since the PS Higgs fields do not carry any $A_4\times Z_5$ charges,
the potential responsible for supersymmetric PS breaking considered in \cite{King:1997ia}
is assumed to be responsible for PS breaking here.

\subsection{CP violation }
Our starting point is to assume that the high energy theory, above the PS breaking scale,
conserves CP \cite{Antusch:2013wn}.
We shall further assume that CP is spontaneously broken by the complex VEVs of scalar fields which spontaneously break $A_4$ and $Z_5$. The scalars include $A_4$ triplets $\phi \sim 3$, $A_4$ singlets $\xi \sim 1$, and other one dimensional $A_4$ representations such as 
$\Sigma_u \sim 1'$ and 
$\Sigma_d \sim 1''$. In addition all of the above fields carry $Z_5$ charges denoted as 
the powers $\alpha^n$, where $\alpha= e^{2\pi i/5}$ and $n$ is an integer.
For example $\xi \sim \alpha^4$ under $Z_5$.
The group theory of $A_4$ is reviewed in Appendix~\ref{A4},
while $Z_5$ corresponds to $\alpha^5=1$.

Under a CP transformation, the $A_4$ singlet fields transform into their complex conjugates
\cite{Ding:2013bpa},
\beq
\xi \rightarrow \xi^*, \ \ \Sigma_u \rightarrow  \Sigma_u^*, \ \ \Sigma_d \rightarrow  \Sigma_d^*,
\eeq
where the complex conjugate fields transform in the complex conjugate representations under
$A_4\times Z_5$. For example if $\xi \sim \alpha^4$, under $Z_5$, then $\xi^* \sim \alpha$.
Similarly if $\Sigma_u \sim 1'$, $\Sigma_d \sim 1''$, under $A_4$, then
$\Sigma_u^* \sim 1''$, $\Sigma_d^* \sim 1'$. On the other hand, 
in the Ma-Rajarsakaran \cite{Ma:2001dn} 
basis of Appendix~\ref{A4}, for $A_4$ triplets $\phi \sim (\phi_1, \phi_2, \phi_3)$,
a consistent definition of CP symmetry requires the second
and third triplet components to swap under CP \cite{Ding:2013bpa},
\beq
\phi \rightarrow (\phi_1^*, \phi_3^*, \phi_2^*).
\eeq
CP violation has also been considered in a variety of other 
discrete groups \cite{deMedeirosVarzielas:2011zw}.
With the above definition of CP, all coupling constants $g$ and explicit masses $m$ are real due to CP
conservation and 
the only source of phases can be the VEVs
of fields which break $A_4\times Z_5$. 
In the model of interest, all the physically interesting CP phases will arise from
$Z_5$ breaking as in \cite{Antusch:2013wn}. 

For example, consider the $A_4$ singlet field $\xi$ 
which carries a $Z_5$ charge $\alpha^4$.
The VEV of this field arises from 
$Z_5$ invariant quintic terms in the superpotential \cite{Antusch:2013wn},
\be
 g P\left(\frac{\xi^5}{\Lambda^3}  -m^2\right) 
\label{Rflavon}
\ee
where, as in \cite{Antusch:2013wn}, 
$P$ denotes a singlet and the coupling $g$ and 
mass $m$ are real due to CP conservation.
The F-term condition from Eq.\ref{Rflavon} is,
\begin{equation}
 \left| \frac{\langle \xi \rangle^5}{\Lambda^3} - m^2\right|^2 
 =  0 .
\end{equation} 
This is satisfied, for example, by
$\langle \xi \rangle =  |(\Lambda^3m^2)^{1/5}|e^{-4i\pi/5}$,
where we arbitrarily select the phase to be $-4\pi /5$ from amongst a discrete set of five possible choices,
which are not distinguished by the F-term condition, as in \cite{King:2013xba}.
We emphasise that CP breaking is controlled by the Abelian $Z_5$ symmetry rather than the non-Abelian $A_4$ symmetry.

\subsection{Vacuum alignment}

Let us now consider the
$A_4$ triplet fields $\phi$ which also carry $Z_5$ charges.
In the full model there are four such triplet fields, or ``flavons'',
denoted as $\phi^u_1$, $\phi^u_2$, $\phi^d_1$, $\phi^d_2$.
The idea is that $\phi^u_i$ are responsible for up-type quark flavour,
while $\phi^d_i$ are responsible for down-type quark flavour. 
These VEVs are driven by the superpotential terms,
\beq
g_{21}P_{21}( \phi^u_2 \phi^d_1 \pm M_{21}^2)+g_{12}P_{12}( \phi^u_1 \phi^d_2 \pm M_{12}^2)
+P_{ii}\left( g^u_{ii}\frac{(\phi^{u}_i)^5}{\Lambda^3} + g^d_{ii}\frac{(\phi^{d}_i)^5}{\Lambda^3} \pm M_{ii}^{2}\right),
\label{triplet_driving}
\eeq
where $P_{ij}$ are linear combinations of singlets as in \cite{King:2013xba}. 
The coupling constants $g_{ij}$,
mass parameters $M_{ij}$ and cut-off scale $\Lambda$ are enforced to be real by CP
while the fields $\phi^{u}_i$ and $\phi^{d}_i$ will develop VEVs with quantised phases.
If we assume that $\phi^{u}_i$
both have the same phase, $e^{im\pi/5}$, 
then Eq.\ref{triplet_driving} implies that $\phi^{d}_i$ should have
phases $e^{in\pi/5}$ such that 
\beq
{\rm arg} (\phi^{u}_i) = \frac{m\pi}{5}, \ \ {\rm arg} (\phi^{d}_i) = \frac{n\pi}{5}, \ \ n+m=0  \ \ ({\rm mod }\ 5),
\label{phases}
\eeq
where $n,m$ are positive or negative integers. 

The structure of the Yukawa matrices depends on the so-called CSD4 vacuum alignments of these flavons
which were first derived in \cite{King:2013xba}, and we assume a similar set of alignments here,
although here the overall phases are quantised due to $Z_5$,
\be
\langle \phi^u_1 \rangle =
\frac{V^u_1}{\sqrt{2}}e^{im\pi/5} \begin{pmatrix}0 \\ 1 \\ 1\end{pmatrix}  \ , \qquad
\langle \phi^u_2 \rangle =
\frac{V^u_2}{\sqrt{21}} e^{im\pi/5}  \begin{pmatrix}1 \\ 4 \\ 2\end{pmatrix} \ , \label{phiu}
\ee
and
\be
\langle \phi^d_1 \rangle =V^d_1 e^{in\pi/5}
\begin{pmatrix} 1\\0\\0 \end{pmatrix}  \ , \ \qquad
\langle \phi^d_2 \rangle =V^d_2 e^{in\pi/5}
\begin{pmatrix} 0\\1\\0 \end{pmatrix} \ .
 \label{phid}
\ee
We note here that the vacuum alignments in Eq.\ref{phid} and the first alignment in Eq.\ref{phiu} are 
fairly ``standard'' alignments that are encountered in tri-bimaximal mixing models, while the second
alignment in Eq.\ref{phiu} is obtained using orthogonality arguments, as discussed in 
\cite{King:2013xba}, to which we refer the interested reader for more details. 

\subsection{Two light Higgs doublets}
The model will involve Higgs bi-doublets of two kinds, $h_u$ which lead to up-type quark and neutrino Yukawa couplings and $h_d$ which lead to down-type quark and charged lepton Yukawa couplings.
In addition a Higgs bidoublet $h_3$, which is also an $A_4$ triplet, is used to give the third family
Yukawa couplings. 

After the PS and $A_4$ breaking, most of these Higgs bi-doublets 
will get high scale masses and will not appear in the low energy spectrum. In fact only two
light Higgs doublets will survive down to the TeV scale, namely $H_u$ and $H_d$.
The precise mechanism responsible for this is quite intricate and is discussed in Appendix~\ref{Higgs}.
Analogous Higgs mixing mechanisms are implicitly assumed in many models, but are rarely discussed explicitly 
(however for an example within $SO(10)$ see \cite{Albright:1996hx}).

The basic idea is that the light Higgs doublet $H_u$ 
with hypercharge $Y=+1/2$, which couples to up-type quarks and neutrinos,
is a linear combination of components of the Higgs bi-doublets of the kind $h_u$ and $h_3$,
while the light Higgs doublet $H_d$ with hypercharge $Y=-1/2$, 
which couples to down-type quarks and charged leptons,
is a linear combination of components of Higgs bi-doublets of the kind $h_d$ and $h_3$,
\beq
h_u,h_3 \rightarrow H_u, \ \ \ \ 
h_d,h_3 \rightarrow H_d.
\label{H}
\eeq

\subsection{Yukawa operators}
The renormalisable Yukawa operators, which respect PS and $A_4$ symmetries, have the following form, 
leading to the third family Yukawa couplings shown, using Eqs.\ref{ql},\ref{H},
\beq
F.h_3F_3^c  
\rightarrow 
Q_3H_uu_3^c + Q_3H_dd_3^c + 
L_3H_u\nu_3^c+L_3H_de_3^c  ,\label{3rd} 
\eeq
where we have used Eqs.\ref{ql},\ref{H}.
The non-renormalisable operators, which respect PS and $A_4$ symmetries, have the following form, 
\bea
F.\phi^u_ih_uF_i^c  
&\rightarrow& 
Q.\vev{\phi^u_i}H_uu_i^c +
L.\vev{\phi^u_i}H_u\nu_i^c ,\label{up0} \\
F.\phi^d_ih_dF_i^c 
&\rightarrow &
Q.\vev{\phi^d_i}H_dd_i^c+
L.\vev{\phi^d_i}H_de_i^c,
\label{down0}
\eea
where $i=1$ gives the first column of each Yukawa matrix, while $i=2$ gives the second column
and we have used Eqs.\ref{ql},\ref{H}. Thus the third family masses are naturally larger since they correspond to renormalisable operators, while the hierarchy between first and second families arises from a hierarchy of 
flavon VEVs.

\subsection{Yukawa matrices}
Inserting the vacuum alignments in Eqs.\ref{phiu} and \ref{phid} into Eqs.\ref{up0} and \ref{down0},
together with the renormalisable third family couplings in Eq.\ref{3rd},
gives the Yukawa matrices of the form,
\begin{equation} \label{Y}
 Y^u = Y^{\nu} = \begin{pmatrix}  0 & b   & 0  \\ 
a & 4b & 0\\  a  & 2b
 & c\end{pmatrix}, \ \ \ \ 
 Y^d \sim Y^e \sim \begin{pmatrix}  y^0_d & 0  & 0  \\ 0 & y^0_s & 0\\ 0  & 0 & y^0_b\end{pmatrix}.
\end{equation}
The PS unification predicts the equality of Yukawa matrices $Y^u = Y^{\nu}$ and $Y^d \sim Y^e$,
while the $A_4$ vacuum alignment predicts the structure of each Yukawa matrix,
essentially identifying the first two columns with the vacuum alignments in Eqs.\ref{phiu} and \ref{phid}.
With a diagonal right-handed Majorana mass matrix, $Y^{\nu}$ leads to a successful
prediction of the PMNS mixing parameters \cite{King:2013xba}. Also
the Cabibbo angle is given by $\theta_C\approx 1/4$ \cite{King:2013hoa}.
Thus Eq.\ref{Y} is a good starting point for a theory of quark and lepton masses and
mixing, although the other 
quark mixing angles and the quark CP phase are approximately zero.
However above discussion ignores the effect of Clebsch factors which will alter the relationship
between elements of $Y^d$ and $Y^e$, which also include off-diagonal elements 
responsible for small quark mixing angles in the full model.

\section{The Model}
\label{model}

The most important fields appearing in the model are defined in Table~\ref{tab-fields}.
In addition to the fields introduced in the previous overview, 
the full model involves Higgs bi-doublets $h_{15}$ in the adjoint of $SU(4)_C$,
as well as messenger fields $X$ with masses given by the VEV of dynamical fields
$\Sigma$. The effective
non-renormalisable Yukawa operators therefore arise from a renormalisable high energy theory,
where heavy messengers $X$ with dynamical masses $\vev{\Sigma}$ are integrated out,
below the energy scale $\vev{\Sigma}$.

\begin{table}
{	\centering
\begin{tabular}{||c|c|c|c|c|c||}
\hline \hline
name &field & $SU(4)_C\times SU(2)_{L}\times SU(2)_{R}$ & $A_4$ & $Z_5$ & $R$ \\
\hline
\hline
Quarks&${F}$ &  $(4,2,1)$ & $3$ & $1$ & 1\\
and leptons&$F^{c}_{1,2,3}$   & $(\overline{4},1,2)$ & $1$ & $\alpha$,$\alpha^3$,$1$ & 1\\
\hline
PS Higgs &$\overline{H^c}$, $H^c$  & $(4,1,2)$, $(\overline{4},1,2)$ & $1$ & $1$ & 0\\
\hline
$A_4$ triplet &${\phi}^{u}_{1,2}$  & $(1,1,1)$ & $3$ & $\alpha^4,\alpha^2$ & 0\\
flavons &${\phi}^{d}_{1,2}$  & $(1,1,1)$ & $3$ & $\alpha^3,\alpha$ & 0\\
\hline
&$h_3$  & $(1,2,2)$  &$3$ & $1$ & 0\\
Higgs &$h_u$  & $(1,2,2)$  &$1''$ & $\alpha$ & 0\\
bidoublets&$h_d$,$h^d_{15}$  & $(1,2,2)$, $(15,2,2)$  &$1'$ & $\alpha^3$,$\alpha^4$ & 0\\
&$h^u_{15}$  & $(15,2,2)$  &$1$ & $\alpha$ & 0\\
\hline
Dynamical  &$\Sigma_u$  & $(1,1,1)$  &$ 1'' $ & $\alpha$ & 0\\
masses &$\Sigma_d$,$\Sigma^d_{15}$  & $(1,1,1)$, $(15,1,1)$  &$ 1' $ & $\alpha^3$,$\alpha^2$ & 0\\
\hline
Majoron &$\xi$ & $(1,1,1)$  &$ 1$ &  $\alpha^4$ & 0\\
\hline
&$X_{F''_{1,3}}$ & $(4,2,1)$  &$ 1''$&  $\alpha$,$\alpha^3$& 1 \\
Fermion &$X_{F'_{1,3}}$ & $(4,2,1)$  &$1'$&  $\alpha$,$\alpha^3$ & 1 \\
Messengers &$X_{\overline{F_i}}$ & $(\overline{4},2,1)$  & $1$ &  $\alpha^i$ & 1\\
 &$X_{\xi_{i}}$ & $(1,1,1)$  &$ 1$ &  $\alpha^i$ & 1\\
\hline
\hline
\end{tabular}
\caption{\label{tab-fields}  The basic Higgs, matter, flavon and messenger
content of the model, where $\alpha = e^{2\pi i /5}$ under $Z_5$.
$R$ is a supersymmetric R-symmetry.}
}
\end{table}

\subsection{Operators from Messengers}

Although the Yukawa operators in the up sector of the full model turn out to be the same as in Eq.\ref{up0},
the Yukawa operators in the down sector of the full model will involve Clebsch factors which 
will imply that $Y^d$ and $Y^e$
are not equal. In addition $Y^d$ and $Y^e$ will involve off-diagonal elements which 
however are ``very small'' in the sense that they will give rise to the small quark mixing angles
of order $V_{ub}$ and $V_{cb}$. The Cabibbo angle arises predominantly from the second column of 
$Y^u$,
with the prediction $V_{us}\sim 1/4$ being corrected by the very small off-diagonal elements of $Y^d$.

The allowed Yukawa operators arise from integrating out heavy fermion fields called ``messengers''
and will depend on the precise choice of fermion messengers.
In Table~\ref{tab-fields} we have allowed messengers of the form $X_{\overline F_i}$
for charges $\alpha^i$ ($i=1,\ldots, 4$), with a very restricted set of messengers
$X_{F'_1}$ ($X_{F''_1}$) and $X_{F'_3}$ ($X_{F''_3}$) with charges $\alpha$ and $\alpha^3$,
in the $1'$ ($1''$) representation of $A_4$.

The assumed messengers $X_{\overline F_i}$ have allowed couplings to $\phi F$
as follows,
\beq
X_{\overline F_1}\phi^u_1F 
+X_{\overline F_2}\phi^d_1F
+X_{\overline F_3}\phi^u_2F
+X_{\overline F_4}\phi^d_2F.
\label{XFbarphiF}
\eeq
The messengers $X_{F'}$ and $X_{F''}$ have allowed couplings to $hF^c_i$ as follows,
 \beq
X_{F'_1} h_uF^c_2
+X_{F'_3}h_uF^c_1
+X_{F''_1}h_dF^c_1
+X_{F''_1}h^d_{15}F^c_3
+X_{F''_3}h^d_{15}F^c_2.
\label{XFhF}
\eeq
The messengers couple to each other and become heavy via 
the dynamical mass fields 
$\Sigma$ which appear in Table~\ref{tab-fields},
\beq
X_{F'_1}\Sigma_u X_{\overline F_3}+X_{F'_3}\Sigma_u X_{\overline F_1}
+X_{F''_1}(\Sigma_d  X_{\overline F_1}+\Sigma^d_{15}  X_{\overline F_2})
+X_{F''_3}\Sigma_d  X_{\overline F_4}.
\label{XFSigmaXF}
\eeq

\begin{figure}[t]
\centering
\includegraphics[width=0.3\textwidth]{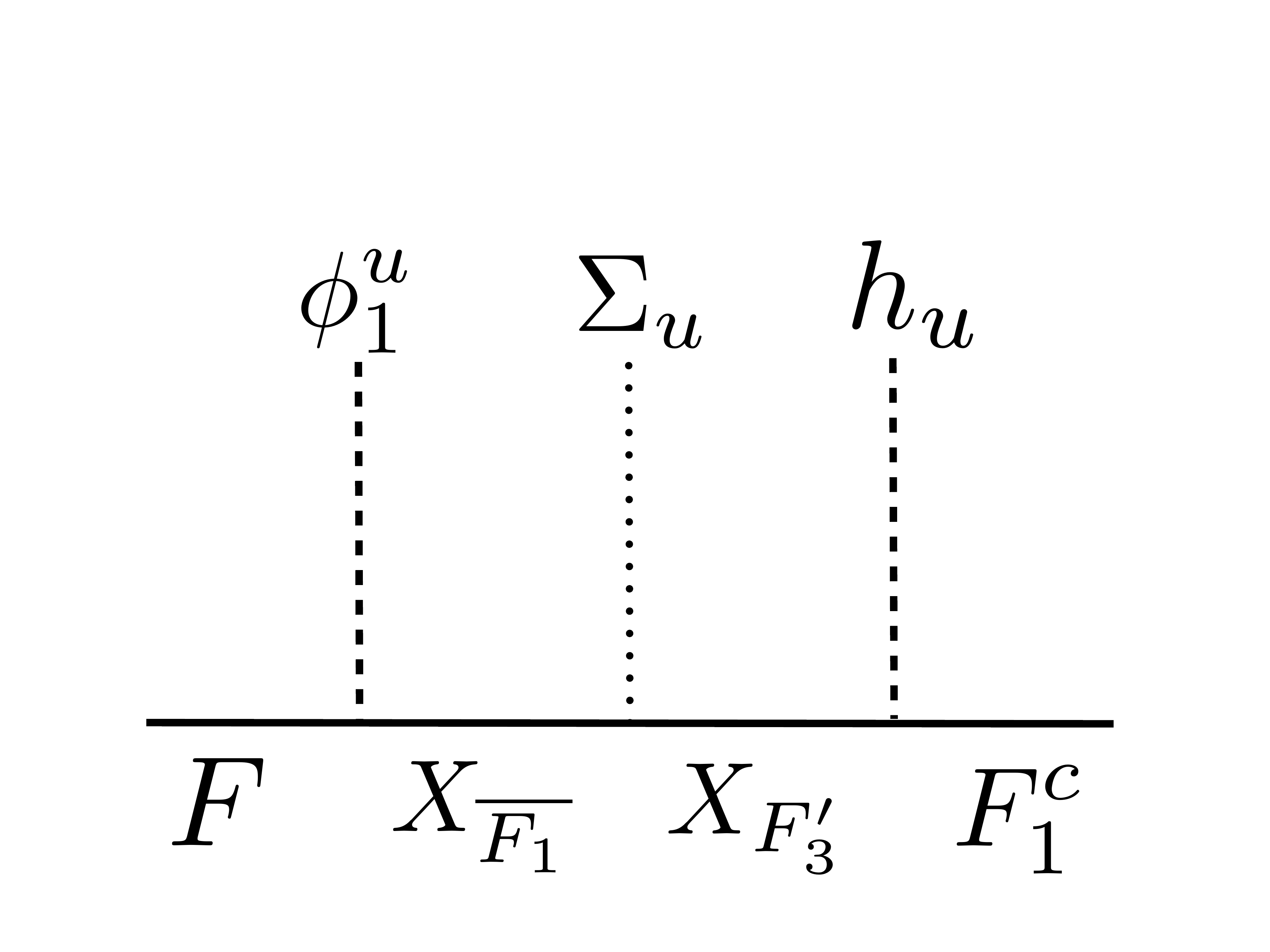} \hspace{3mm}
\includegraphics[width=0.3\textwidth]{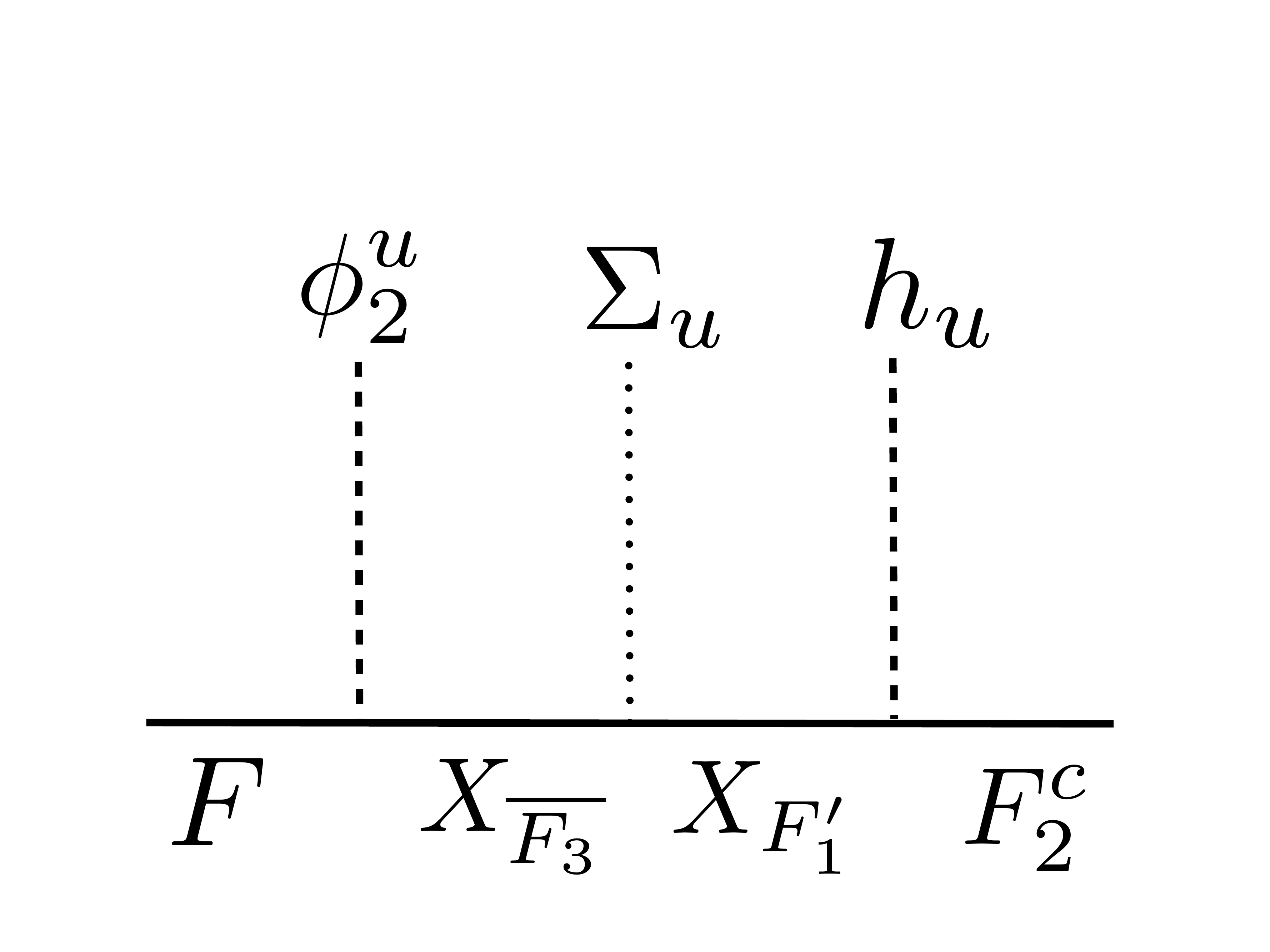}
\includegraphics[width=0.3\textwidth]{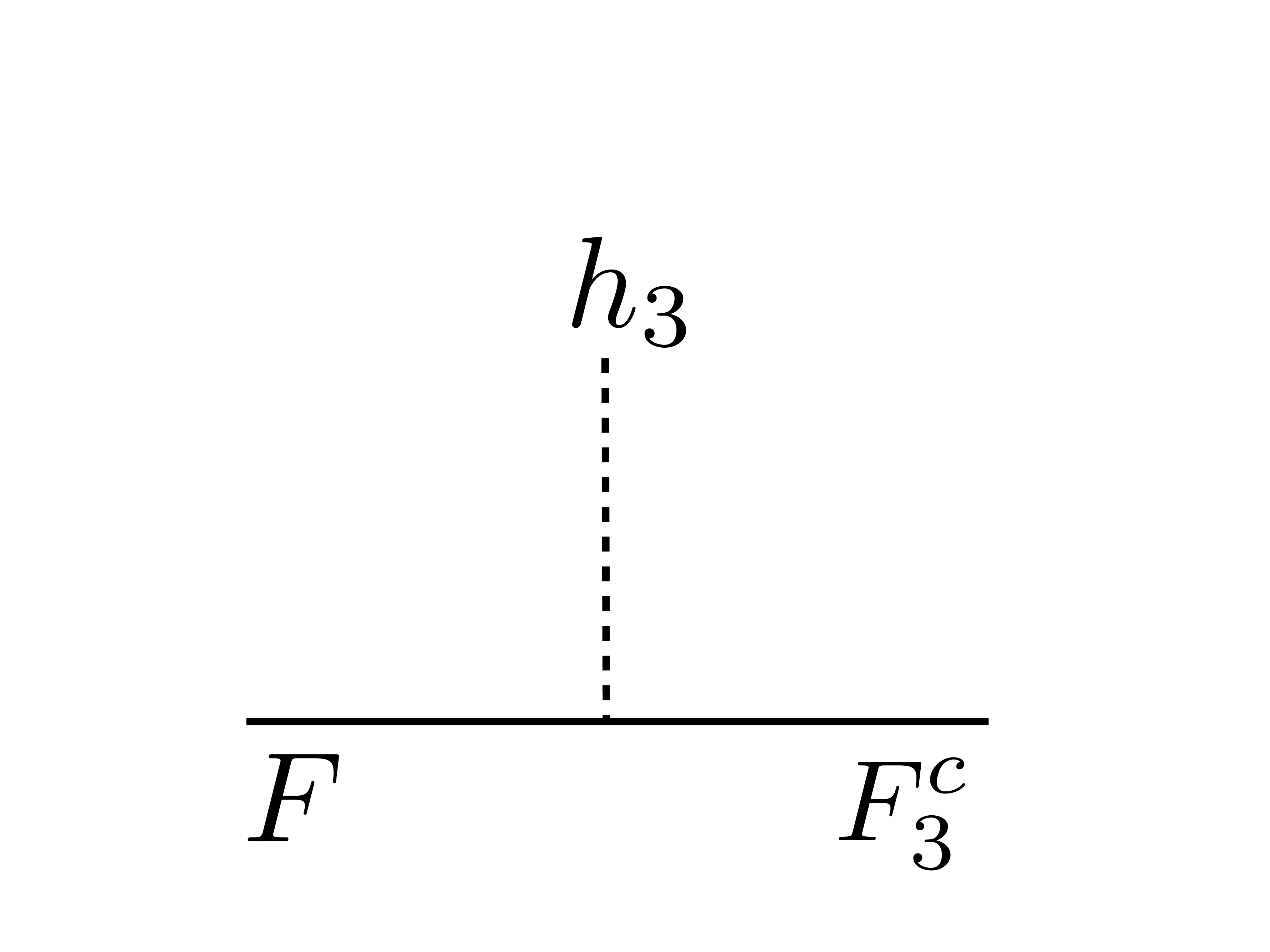}
\vspace*{-4mm}
    \caption{The fermion messenger diagrams responsible for the 
    operators leading to the
     up type quark and Dirac neutrino masses. The fermions depicted by the solid line have even R-parity.    } \label{mess2}
\vspace*{-2mm}
\end{figure}

\begin{figure}[t]
\centering
\includegraphics[width=0.3\textwidth]{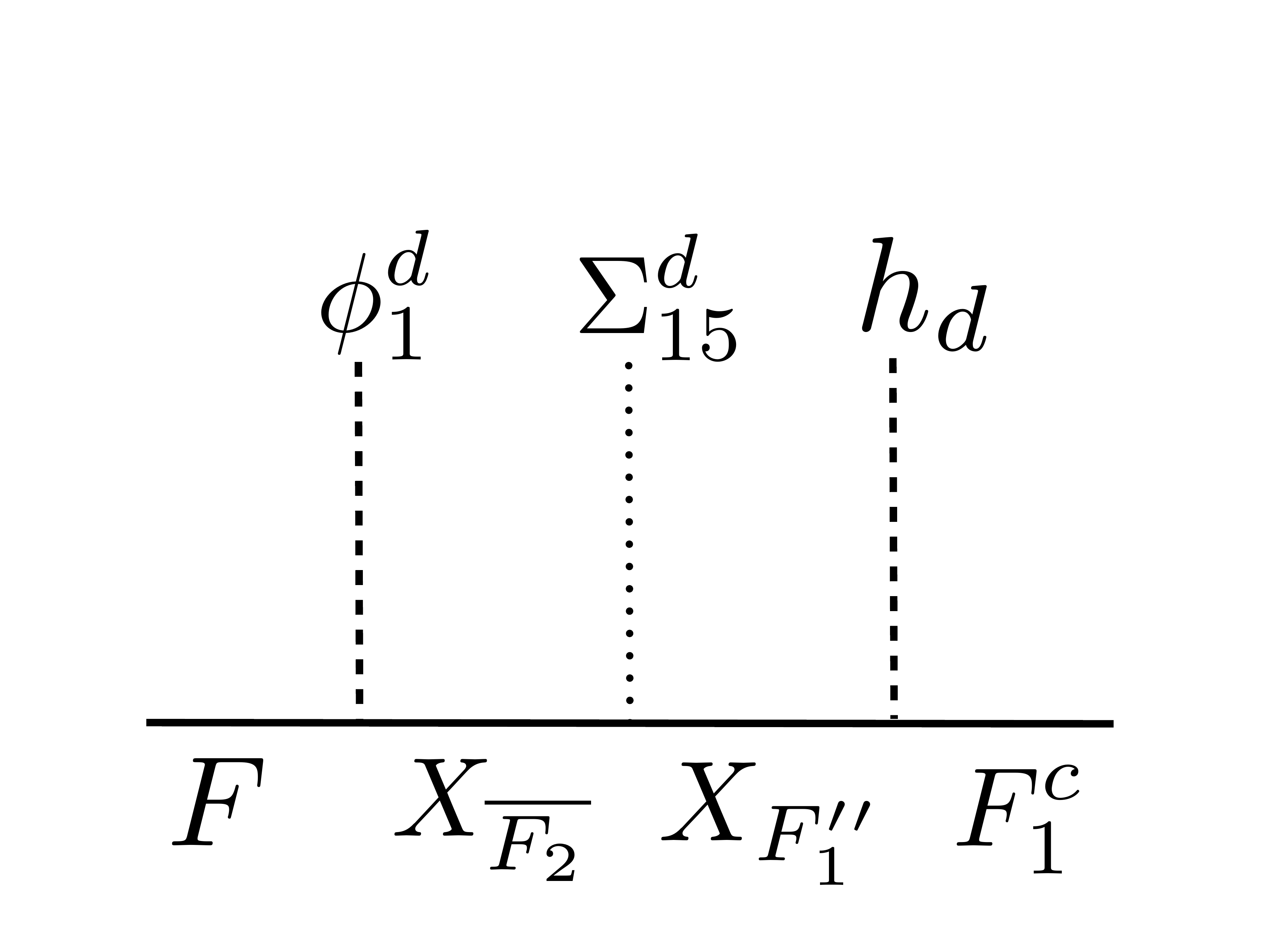}
\includegraphics[width=0.3\textwidth]{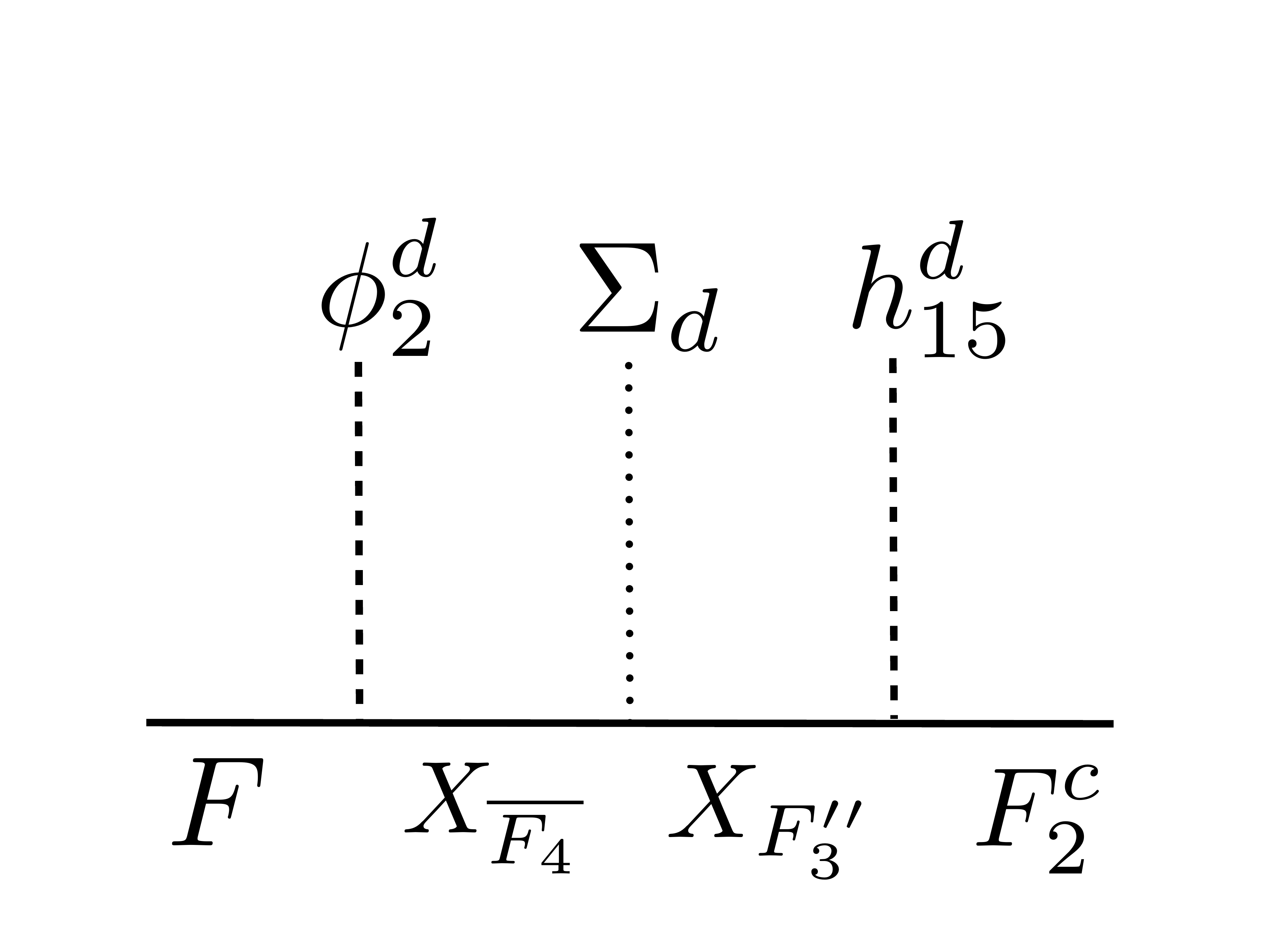}
\includegraphics[width=0.3\textwidth]{fig2c.pdf}
\vspace*{-4mm}
    \caption{The fermion messenger diagrams responsible for the 
    operators which lead to the diagonal
    charged lepton and down type quark masses. The fermions depicted by the solid line have even R-parity.     } \label{mess1}
\vspace*{-2mm}
\end{figure}

The leading order operators responsible for the Yukawa couplings involving the first and second families
to Higgs fields are obtained by integrating out
the heavy messengers, leading to effective operators.

The diagrams in Fig.\ref{mess2} yield the following operators which will be responsible
for the up-type quark and neutrino Yukawa couplings,
\beq
W^u_{Yuk}=
F.\frac{\phi^u_1}{\Sigma_u}h_uF^c_1
+F.\frac{\phi^u_2}{\Sigma_u}h_uF^c_2
+ F.h_3F^c_3.
\label{Yuku}
\eeq
The above operators are similar to those in Eq.\ref{up0} and will yield a Yukawa matrix
$Y^u=Y^{\nu}$ as in Eq.\ref{Y}.

The diagrams in Fig.\ref{mess1} yield the operators which will be responsible
for the diagonal down-type quark and charged lepton Yukawa couplings,
\bea
W^{d, {\rm diag}}_{Yuk}&=& 
 F.\frac{\phi^d_1}{\Sigma_{15}^d}h_dF^c_1
+F.\frac{\phi^d_2}{\Sigma_d}h_{15}^dF^c_2
+ F.h_3F^c_3.
\label{Yukd_diag}
\eea
These operators are similar to those in Eq.\ref{down0} and will yield Yukawa matrices
similar to those in Eq.\ref{Y} but with $Y^d\neq Y^{e}$ due to the Clebsch-Gordan coefficients
from the Higgs in the $15$ dimensional representation of $SU(4)_C$.
In addition, the above messenger sector generates further effective operators which 
give rise to off-diagonal down-type quark and charged lepton Yukawa couplings,
\bea
W^{d, {\rm off-diag}}_{Yuk}&=& 
F.\frac{\phi^d_1}{\Sigma_{15}^d}h_{15}^dF^c_3
+ F.\frac{\phi^u_1}{\Sigma_d}(h_dF^c_1+h_{15}^dF^c_3).
\label{Yukd_offdiag}
\eea

\begin{figure}[t]
\centering
\includegraphics[width=0.3\textwidth]{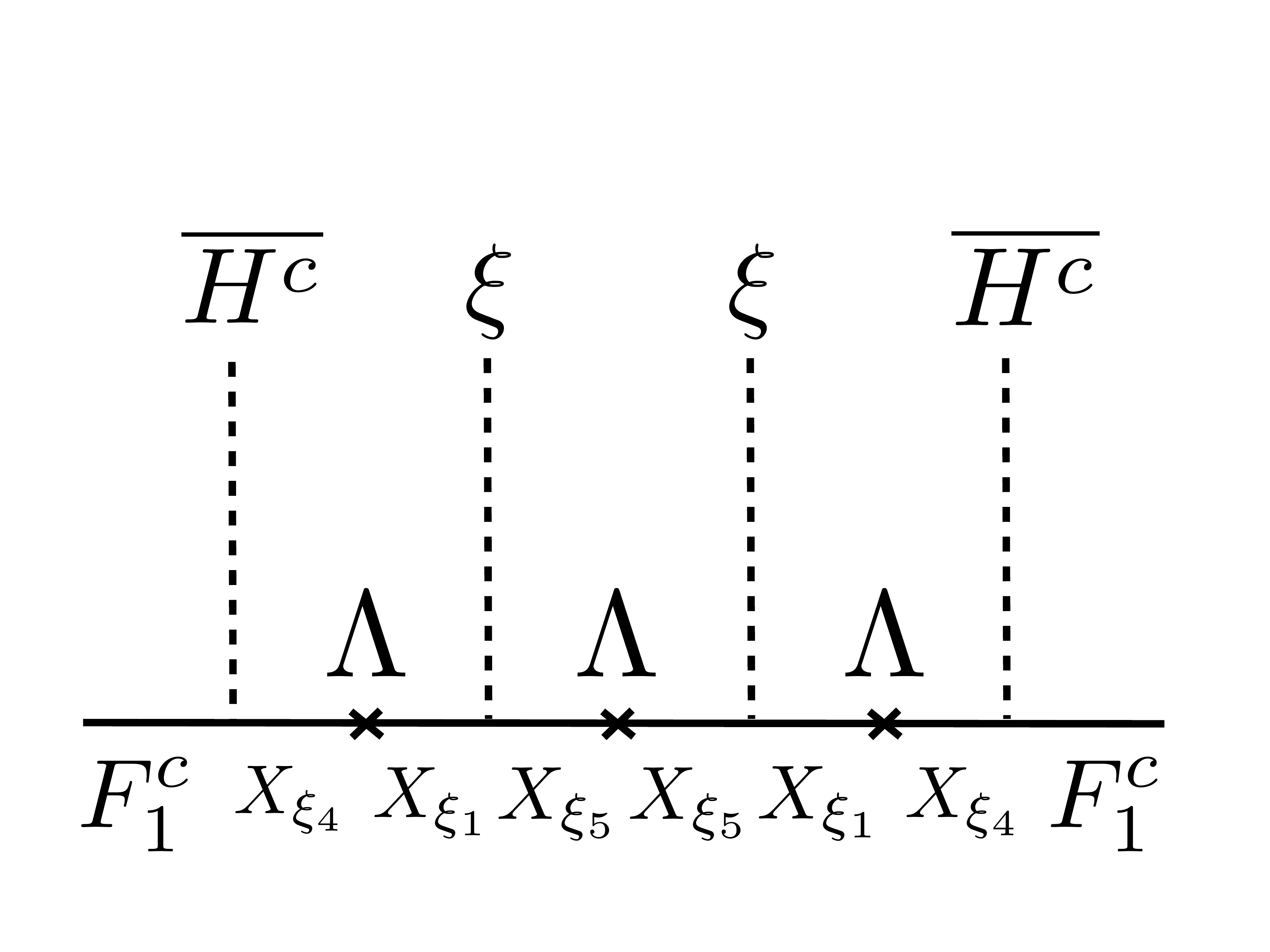} \hspace{3mm}
\includegraphics[width=0.3\textwidth]{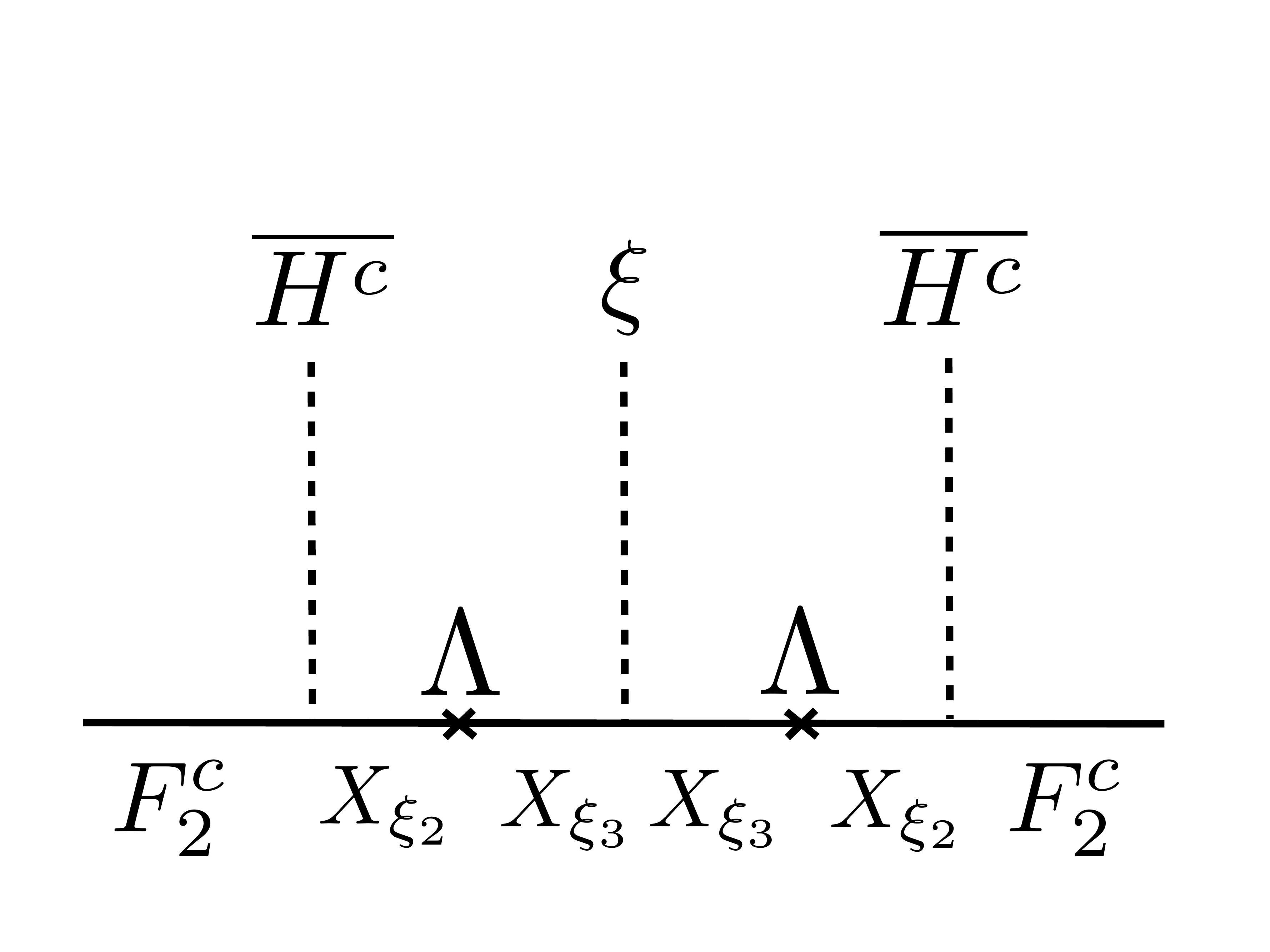}
\includegraphics[width=0.3\textwidth]{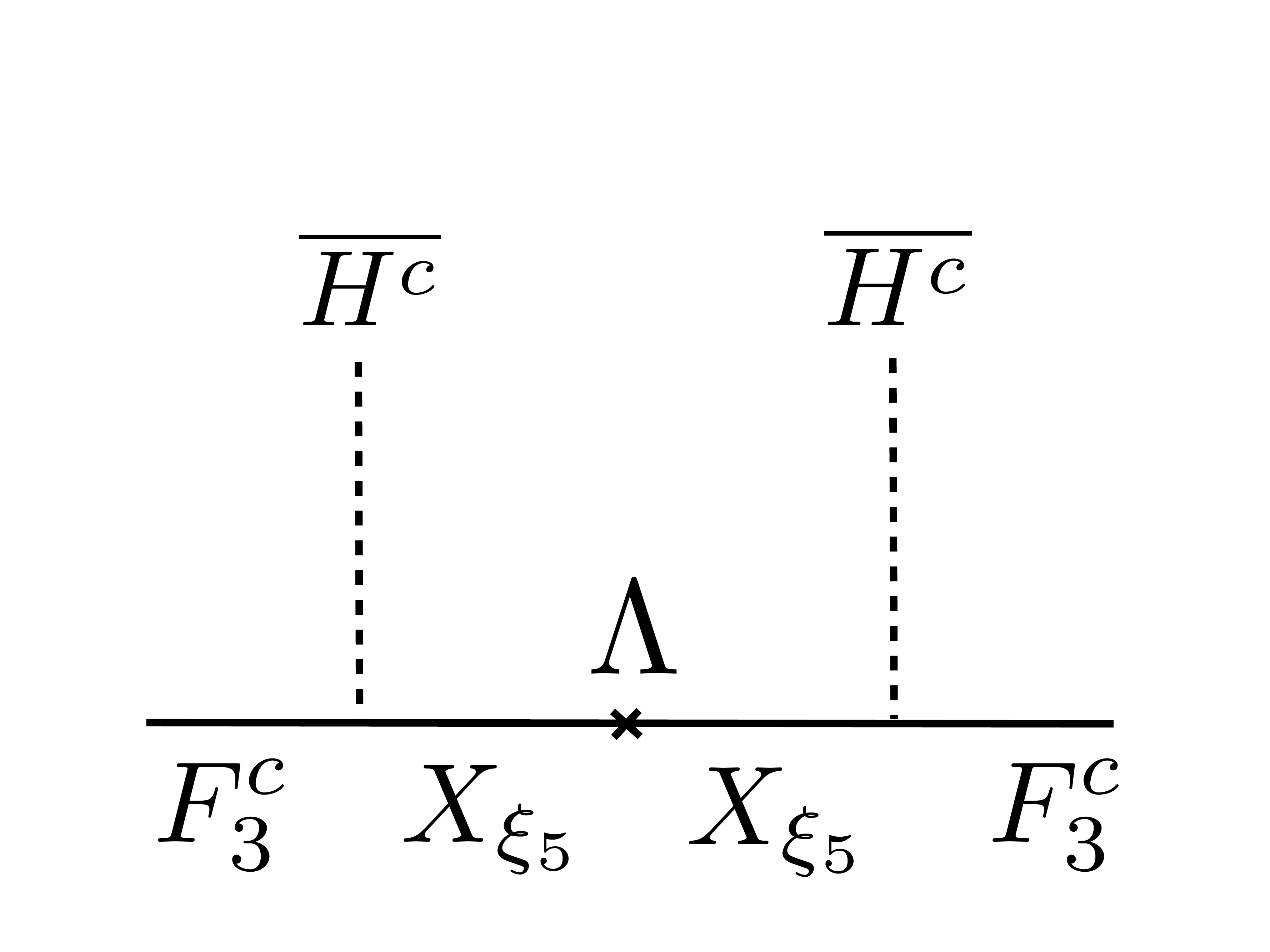}
\vspace*{-4mm}
    \caption{The fermion messenger diagrams responsible for the 
    effective operators in Eq.\ref{Maj} leading to the diagonal heavy (right-handed)
     Majorana neutrino masses. The fermions depicted by the solid line have even R-parity.    } 
     \label{mess3}
\vspace*{-2mm}
\end{figure}

The operators responsible for the heavy Majorana neutrino masses are given by,
\beq
W_{Maj}=  \frac{\xi^2}{\Lambda^2}  \frac{\overline{H^c}\overline{H^c}}{\Lambda}F^c_1F^c_1
+ \frac{\xi}{\Lambda}  \frac{\overline{H^c}\overline{H^c}}{\Lambda}F^c_2F^c_2
+ \frac{\xi}{\Lambda}  \frac{\overline{H^c}\overline{H^c}}{\Lambda}F^c_1F^c_3
+ \frac{\overline{H^c}\overline{H^c}}{\Lambda}F^c_3F^c_3,
\label{Maj}
\eeq
corresponding to the diagrams in Fig.\ref{mess3}.
These operators are mediated by the singlet messengers $X_{\xi_i}$
and involve the explicit messenger 
mass scale $\Lambda$ which may take values higher than the 
$A_4\times Z_5$ and Pati-Salam breaking scales.
The first three of these operators are controlled by the Majoron fields $\xi_i$
in Table~\ref{tab-fields}, which carries
a non-trivial phase due to the $Z_5$ symmetry, as discussed later.

Note that the  
dynamical mass $\Sigma$ fields
do not enter the Majorana sector since they 
transform under $A_4$ as $1',1''$ and hence 
do not couple to 
pairs of $X_{\xi_i}$.
Also note that the Majoron $\xi$ fields 
which transform under $A_4\times Z_5$ as $\xi\sim (1,\alpha^4)$
do not enter the charged fermion sector since they do not couple
$X_{\overline F_i}$ to the messengers $X_{F'}$ and $X_{F''}$
which transform under $A_4$ as $1'$ and $1''$.

\subsection{Yukawa and Majorana mass matrices}
According to the mechanism discussed in Appendix~\ref{Higgs}, 
the four Higgs multiplets in the fourth block of Table~\ref{tab-fields},
$h_3$, $h_u$, $h_d$, $h^d_{15}$,
result in two low energy light Higgs doublets 
$H_u$,$H_d$,
\beq
h_3 \rightarrow H_{u,d}, \ \ \ h_u\rightarrow \epsilon_uH_u, \ \ \ h_d\rightarrow \epsilon_d H_d, \ \ \ \ h^d_{15}\rightarrow H_d,
\label{admixtures}
\eeq
where $H_u$ is predominantly composed of the Higgs doublet from third component of $h_3$
with a small admixture $\epsilon_u$ of the Higgs doublet from $h_u$.
$H_d$ is predominantly composed of the Higgs doublet from $h^d_{15}$ 
plus the third component of $h_3$,
together with a small admixture $\epsilon_d$ of the Higgs doublet from $h_d$.
The particular admixtures assumed in Eq.\ref{admixtures} correspond to a 
particular choice of masses in Appendix~\ref{Higgs}.

With the vacuum alignments in Eq.~\ref{phiu},
the operators in Eq.~\ref{Yuku} then result in non-diagonal and equal up-type quark and neutrino Yukawa matrices,
\begin{equation} \label{Yunu2}
 Y^u = Y^{\nu} = \begin{pmatrix}  0 & b   & 0  \\ 
a & 4b & 0\\  a  & 2b
 & c\end{pmatrix},
\eeq
where,
\beq
a\sim \epsilon_u \frac{V^u_1}{\langle \Sigma_u  \rangle}, \ \ 
b\sim \epsilon_u \frac{V^u_2}{\langle \Sigma_u \rangle}, \ \
c\sim 1.
\label{yu0}
\eeq
Note that since $ Y^u = Y^{\nu}$, the up-type quark masses
are equal to the Dirac neutrino masses, 
\beq
m_u=m^D_{\nu1}, \ \ m_c=m^D_{\nu2}, \ \ 
m_t=m^D_{\nu3}.
\label{Dirac}
\eeq
From Eq.\ref{Yunu2} the up-type quark masses are given to excellent approximation by,
\beq
m_u= y_u v_u=a v_u / \sqrt{17}, \ \ m_c= y_c v_u=\sqrt{17} b v_u , \ \ 
m_t = c v_u.
\label{up}
\eeq
The Yukawa coupling eigenvalues for up-type quarks are given by,
\beq
y_u\sim\frac{a}{\sqrt{34}}\sim 4.10^{-6}, \ \ 
y_c\sim b \sqrt{ \frac{17}{21} }\sim 10^{-3}, \ \
y_t\sim c\sim 1,
\label{yu}
\eeq
where we have inserted some typical up-type quark Yukawa couplings, hence,
\beq
 a \sim 2.10^{-5}, \ \ b \sim 10^{-3} \longrightarrow \frac{a}{b}\sim \frac{V_1^u}{V_2^u}\sim  2.10^{-2},
\label{ratios2}
\eeq
where the ratio of up to charm masses is accounted for by the 2\% ratio of flavon VEVs.

Similarly, with the vacuum alignments in Eqs.~\ref{phiu},\ref{phid},
the operators in Eqs.~\ref{Yukd_diag},\ref{Yukd_offdiag} then result in
down-type quark and charged lepton Yukawa matrices
related by Clebsch factors,
\begin{equation} \label{Yed2}
Y^d = \begin{pmatrix} 
y_d^0 & 0  & Ay_d^0  \\ 
By_d^0 & y_s^0 & Cy_d^0 \\ 
By_d^0  & 0 & y_b^0 + Cy_d^0 \end{pmatrix},
\ \ \ \
Y^e =  \begin{pmatrix}  
-y_d^0/3 & 0  & Ay_d^0   \\ 
By_d^0 & -3y_s^0 & -3Cy_d^0 \\ 
By_d^0 & 0 & y_b^0-3Cy_d^0 \delta
\end{pmatrix},
\end{equation}
where the diagonal Yukawa couplings for down-type quarks are given by,
\beq
y_d^0\sim  \epsilon_d \frac{V^d_1}{\vev{\Sigma^d_{15}}} \sim 5.10^{-5}, \ \ 
y_s^0\sim  \frac{V^d_2}{\vev{\Sigma_d}} \sim 10^{-3}, \ \ 
y_b^0\sim 5.10^{-2},
\label{yd_diag}
\eeq
where $\epsilon_{u,d}$ were defined in Eq.\ref{admixtures} and for low $\tan \beta$ we have
inserted some typical down-type quark Yukawa couplings, assuming that the mixing angles are small.
The off-diagonal entries to the down-type quark and charged lepton Yukawa matrices are given by,
\beq
Ay_d^0\sim \frac{V^d_1}{\vev{\Sigma^d_{15}}}, \ \ 
By_d^0 \sim \epsilon_d\frac{V^u_1}{\vev{\Sigma_d}} , \ \ 
Cy_d^0 \sim  \frac{V^u_1}{\vev{\Sigma_d}},
\label{yd_offdiag}
\eeq
where,
\beq
A\sim \frac{C}{B} \sim \frac{1}{\epsilon_d}.
\label{ABC}
\eeq
From Eq.\ref{Yed2} the diagonal down-type quark and charged lepton Yukawa couplings are related by,
\beq
y_e^0=\frac{y_d^0}{3}, \ \ y_{\mu}^0=3y_s^0, \ \ y_{\tau}^0=y_b^0 .
\label{GJ}
\eeq
These are the well-known Georgi-Jarlskog (GJ) relations \cite{Georgi:1979df}, although the factor of 1/3 
which appears in the first relation above arises from a new mechanism, namely due to non-singlet fields which appear
in the denominator of effective operators
as discussed in detail in \cite{Antusch:2013rxa}.
The viablity of the GJ relations for mass eigenstates is discussed in 
\cite{Antusch:2013jca}. However here there are small off-diagonal entries in the Yukawa matrices which 
will provide corrections to the mass eigenstates, as well as other corrections to the GJ relations,
as discussed later.

Finally, from Eq.\ref{Maj}, we find the heavy Majorana mass matrix,
\begin{equation} \label{Maj0}
 M_R  =  \begin{pmatrix} M_1 & 0 &M_{13} \\ 0 &  
M_2 &0 \\  M_{13}& 0& M_3 \end{pmatrix}.
\eeq
The heavy Majorana neutrino masses from Eq.\ref{Maj} are in the ratios,
\beq
M_1:M_2:M_3\sim  \tilde{\xi}^2: \tilde{\xi} :1,
\label{ratios3}
\eeq
where, 
\beq
\tilde{\xi}=  \frac{\langle \xi \rangle }{\Lambda}.
\eeq
There is a competing correction to $M_1$ coming from 
the off-diagonal element, namely $M_{13}^2/M_3\sim \tilde{\xi}^2$ with the same phase,
which may be absorbed into the definition of the lightest right-handed neutrino mass.
Since we need to have a strong hierarchy of right-handed neutrino masses 
we shall require (see later),
\beq
M_1\sim 5.10^5 \ {\rm GeV}, \ \ 
M_2\sim 5.10^{10} \ {\rm GeV}, \ \
M_3\sim 5.10^{15} \ {\rm GeV}, 
\label{RHN}
\eeq
which may be achieved for example by,
\beq
\tilde{\xi} \sim 10^{-5}.
\label{eps2}
\eeq
Typically the heaviest right-handed neutrino mass is given by,
\beq
M_3 \sim  \frac{\vev{\overline{H^c}}^2}{\Lambda} \sim 5.10^{15} \ {\rm GeV},
\eeq
which is within an order of magnitude of the Pati-Salam breaking scale in Eq.\ref{PS}.
This implies that $\Lambda \sim 5.10^{16}$ GeV and hence, from Eq.\ref{eps2},
\beq
\langle \xi \rangle \sim 5.10^{11} {\rm GeV}.
\label{phaser_vev}
\eeq
The Majoron  fields $\xi$ act like a dynamical mass for $M_2$, with an effective coupling $\xi N^c_2N^c_2$ with a coupling constant
of about 0.1. In principle they could play a role in leptogenesis.
For example, the effect of Majorons on right-handed neutrino annihilations,
leading to possibly significantly enhanced efficiency factors, was recently discussed in \cite{Sierra:2014sta}.

\section{Quark Masses and Mixing}
\label{quarks}
\subsection{Convention}
We shall use the convention for the quark Yukawa matrices, 
\beq
\mathcal{L}=-v^uY^u_{ij}\overline u^i_{\mathrm{L}} u^j_{\mathrm{R}}
-v^dY^d_{ij}\overline d^i_{\mathrm{L}} d^j_{\mathrm{R}} + h.c.  
\label{convention}
\eeq
which are diagonalised by,
\begin{eqnarray}\label{DiagYu}
U_{u_\mathrm{L}} \, Y^u \,U^\dagger_{u_\mathrm{R}} =
\left(\begin{array}{ccc}
\!y_{u}&0&0\!\\
\!0&y_{c}&0\!\\
\!0&0&y_{t}\!
\end{array}
\right),\ \ \ \ 
U_{d_\mathrm{L}} \, Y^d \,U^\dagger_{d_\mathrm{R}} =
\left(\begin{array}{ccc}
\!y_{d}&0&0\!\\
\!0&y_{s}&0\!\\
\!0&0&y_{b}\!
\end{array}
\right)\! .
\end{eqnarray}  
The CKM matrix is then given by,
\begin{eqnarray}
U_{\mathrm{CKM}} = U_{u_\mathrm{L}} U^\dagger_{d_\mathrm{L}}. 
\end{eqnarray}
In the PDG parameterization \cite{PDG}, in the standard notation, 
$U_{\mathrm{CKM}} = R^q_{23} U^q_{13} R^q_{12}$
in terms of $s^q_{ij}=\sin (\theta^q_{ij})$ and 
$c^q_{ij}=\cos(\theta^q_{ij})$ and the CP violating phase $\delta^q$. 

\subsection{Analytic estimates for quark mixing}
In the above convention, the quark Yukawa matrices differ from those given in
Eqs.\ref{Yunu2},\ref{Yed2} by a complex conjugation,
\footnote{The complex conjugation of the Yukawa matrices arises from the fact that the Yukawa matrices 
given in Eqs.\ref{Yunu2},\ref{Yed2} correspond to the Lagrangian
$\mathcal{L}=-v^uY^u_{ij}u^i_{\mathrm{L}} u^c_j
-v^dY^d_{ij}d^i_{\mathrm{L}} d^c_j + h.c. $ involving the 
unbarred left-handed and CP conjugated right-handed fields.
Note that our LR convention for the quark Yukawa matrices in Eq.\ref{convention}
differs by an Hermitian conjugation compared to that used in the
Mixing Parameter Tools package \cite{Antusch:2005gp} due to the RL convention used there.} 
\begin{equation} \label{Yq}
 Y^u  = \begin{pmatrix}  0 & b   & 0  \\ 
a & 4b & 0\\  a  & 2b
 & c\end{pmatrix}, \ \ 
Y^d = \begin{pmatrix} 
y_d^0 & 0  & Ay_d^0  \\ 
By_d^0 & y_s^0 & Cy_d^0 \\ 
By_d^0  & 0 & y_b^0 + Cy_d^0 \end{pmatrix},
\end{equation}
where the parameters defined in Eqs.\ref{yu},\ref{yd_diag},\ref{yd_offdiag}
are given below, 
\beq
a\sim \epsilon_u e^{-im\pi/5} \frac{V^u_1}{\langle \Sigma_u  \rangle} \sim 2.10^{-5}, \ \ 
b\sim \epsilon_u e^{-im\pi/5}\frac{V^u_2}{\langle \Sigma_u \rangle} \sim 10^{-3}, \ \
c\sim 1,
\label{yu1}
\eeq
\beq
y_d^0\sim  \epsilon_d e^{-in\pi/5}\frac{V^d_1}{\vev{\Sigma^d_{15}}} \sim 5.10^{-5}, \ \ 
y_s^0\sim  e^{-in\pi/5}\frac{V^d_2}{\vev{\Sigma_d}} \sim 10^{-3}, \ \ 
y_b^0\sim 5.10^{-2},
\label{yd_diag1}
\eeq
\beq
Ay_d^0\sim e^{-in\pi/5}\frac{V^d_1}{\vev{\Sigma^d_{15}}}, \ \ 
By_d^0 \sim \epsilon_de^{-im\pi/5}\frac{V^u_1}{\vev{\Sigma_d}} , \ \ 
Cy_d^0 \sim e^{-im\pi/5} \frac{V^u_1}{\vev{\Sigma_d}},
\label{yd_offdiag1}
\eeq
where we have displayed the phases from Eqs.\ref{phiu},\ref{phid} explicitly in the new convention.

Cabibbo mixing clearly arises predominantly from the up-type quark Yukawa matrix $Y^u$,
which leads to a Cabibbo angle $\theta_C \approx 1/4$ 
or $\theta_C \approx 14^{\circ}$. 
The other quark mixing angles and CP violating phase arise from the off-diagonal elements 
of $Y^d$, which also serve to correct the Cabibbo angle to yield eventually $\theta_C \approx 13^{\circ}$.

Recall that any 3$\times$3 unitary matrix $U^{\dagger}$ can be written in
terms of three angles $\theta_{ij}$, three phases $\delta_{ij}$ (in all cases $i<j$) and three phases $\rho_{i}$
in the form \cite{King:2002nf},
\be
\label{eqparam1}
U^{\dagger}=
U_{23} U_{13} U_{12}\,{\rm diag}(e^{i\rho_1},e^{i\rho_2},e^{i \rho_3})\;,
\ee
where
\begin{eqnarray}\label{eqU12}
U_{12}= \left(\begin{array}{ccc}
  c_{12} & s_{12}e^{-i\delta_{12}} & 0\\
  -s_{12}e^{i\delta_{12}}&c_{12} & 0\\
  0&0&1\end{array}\right)
\end{eqnarray}
and similarly for $U_{13},U_{23}$, where $s_{ij}=\sin \theta_{ij}$ and 
$c_{ij}=\cos \theta_{ij}$ and the angles can be made
positive by a suitable choice of the $\delta_{ij}$ phases.
We use this
parameterisation for both $U^\dagger_{u_L}$ and $U^\dagger_{d_L}$,
where the phases $\rho_i$ can be absorbed into the quark mass
eigenstates, leaving
\be
\label{eqparam2}
U^{\dagger}_{u_L}=
U^{u_L}_{23} U^{u_L}_{13} U^{u_L}_{12}\;,\ \ \ \
U^{\dagger}_{d_L}=U^{d_L}_{23} U^{d_L}_{13} U^{d_L}_{12},
\ee
where $U^{\dagger}_{u_L}$ contains 
$\theta^u_{ij}$ and $\delta^u_{ij}$,
while $U^{\dagger}_{d_L}$ contains 
$\theta^d_{ij}$ and $\delta^d_{ij}$.
The CKM matrix before phase removal may be
written as
\be
\label{eqparam3}
U'_{\rm CKM} = {U^{u_L}_{12}}^\dagger
{U^{u_L}_{13}}^\dagger {U^{u_L}_{23}}^\dagger
U^{d_L}_{23}U^{d_L}_{13} U^{d_L}_{12}
 \;.
\ee
On the other hand, $U'_{\rm CKM}$ can be also parametrised as in Eq.~(\ref{eqparam1}),
\be
\label{eqparam4}
U'_{\rm CKM}={\rm diag}(e^{i\rho_1},e^{i\rho_2},e^{i \rho_3})\,
U_{23} U_{13} U_{12}\;.
\ee
The angles $\theta_{ij}$ are the standard PDG ones in $U_{CKM}$, and
five of the six phases of  $U'_{\rm CKM}$ in Eq.~(\ref{eqparam4}) may be removed leaving the 
standard PDG phase in $U_{\rm CKM}$ identified as \cite{King:2002nf}:
\be\label{eqdeltafromparam1}
\delta^q= \delta^q_{13}-\delta^q_{23}-\delta^q_{12}.
\ee
In the present case, given $Y^u$, it is clear that
$\theta_{13}^u \approx \theta_{23}^u \approx 0$.
Similary, given $Y^d$, we see that $\theta_{12}^d \approx 0$.
This implies that Eq.~(\ref{eqparam3}) simplifies to:
\be
\label{eqparam5}
U'_{\rm CKM} \approx 
{U^{u_L}_{12}}^\dagger
U^{d_L}_{23}U^{d_L}_{13} 
 \;.
\ee
Then, by equating the right-hand sides of Eqs.~(\ref{eqparam4}) and (\ref{eqparam5})
and expanding to leading order in the small mixing angles,
we obtain the following relations:
\bea \label{F1}
{\theta^q_{23}}e^{-i\delta^q_{23}}&\approx &
{\theta_{23}^{d}}e^{-i\delta_{23}^{d}}\;,
\\
\label{F2} {\theta^q_{13}}e^{-i\delta^q_{13}}&\approx &
{\theta_{13}^{d}}e^{-i\delta_{13}^{d}}-{\theta_{12}^{u}}e^{-i\delta_{12}^{u}}
{\theta_{23}^{d}}e^{-i\delta_{23}^{d}}
\;,\\
\label{F3} {\theta^q_{12}}e^{-i\delta^q_{12}}&\approx &
-{\theta_{12}^{u}}e^{-i\delta_{12}^{u}} \;,
\eea
from which we deduce,
\beq
\theta^q_{12} \approx  \theta^u_{12}, \ \ 
\theta^q_{23} \approx  \theta^d_{23}, \ \ 
|\theta^q_{13}-\theta^q_{12}\theta^q_{23}e^{i\delta^q}|\approx \theta^d_{13},
\label{1}
\eeq
where,
\beq
\theta^u_{12}\sim \frac{1}{4}, 
\ \ \theta^d_{23} \sim \left| \frac{Cy_d^0}{y_b^0}\right|, 
\ \ \theta^d_{13} \sim \left| \frac{Ay_d^0}{y_b^0}\right|.
\label{2}
\eeq
Notice from Eqs.\ref{1},\ref{2} that the magnitudes of the Yukawa matrix elements are all approximately fixed in terms of physical quark mixing parameters,
\beq
\theta^q_{23}\sim \left| \frac{Cy_d^0}{y_b^0}\right| \sim  0.040, 
\ \ |\theta^q_{13}-\theta^q_{12}\theta^q_{23}e^{i\delta^q}|\sim \left| \frac{Ay_d^0}{y_b^0}\right| \sim  0.009.
\label{3}
\eeq
Since $|{y_d^0}/{y_b^0}|\sim 0.001$, Eq.\ref{3} implies that,
\beq
A\sim 9, \ \ \ \ C\sim 40,
\ \ \ \ B\sim C/A\sim 4,
\label{ABC2}
\eeq
where the last relation uses Eq.\ref{ABC}.

Concerning the phases, from Eq.\ref{yd_offdiag1} we find, in the convention of Eq.\ref{eqU12},
\beq
\delta^u_{12}\sim 0, 
\ \ \delta^d_{23} \sim -\arg\left( \frac{Cy_d^0}{y_b^0}\right)\sim m\pi/5, 
\ \ \delta^d_{13} \sim -\arg\left( \frac{Ay_d^0}{y_b^0}\right)\sim n\pi/5,
\eeq
where, from Eq.\ref{phases}, $n+m$ is a multiple of $5$.
Hence, from Eqs.\ref{F1},\ref{F2},\ref{F3},
\beq
\delta^q_{12}\sim0, \ \ 
\delta^q_{23}\sim  \delta^d_{23} \sim m\pi/5, \ \ 
\delta^q_{13}\sim -\arg (0.009e^{-in\pi/5}-\frac{1}{4}0.04e^{-im\pi/5}),
\eeq
so the physical CP phase is given by the very approximate expression,
\beq
\delta^q= \delta^q_{13}-\delta^q_{23}-\delta^q_{12}\sim
-\arg (0.009e^{-in\pi/5}-0.010e^{-im\pi/5})
-\frac{m\pi}{5}.
\eeq
Clearly CP violation requires $n\neq m$, indeed $\delta^q$ only depends on the 
difference $n-m$ with a positive value of $\delta^q\sim \frac{7\pi}{18}$ in the first quadrant requiring $n<m$. 
Since $n+m$ must be a multiple of $5$,
then the only possibility is $n=2,m=3$ which corresponds to one of the discrete choices of phases
in Eq.\ref{phases}.

\subsection{Numerical results for quark mixing}
With the phases fixed by the choice of discrete choice of phases $n=2,m=3$, 
as discussed in the previous subsection,
the only free parameters are 
$a,b,c$ in the up sector, and $A,B,C$ and $y_d^0,y_s^0,y_b^0$ in the down sector matrices,
where we have explicitly removed the phases from these parameters, in order to 
make them real,
\beq
Y^u  = \begin{pmatrix}  
0 & be^{-i3\pi/5}   & \epsilon c  \\ 
ae^{-i3\pi/5} & 4be^{-i3\pi/5} & 0\\  
ae^{-i3\pi/5}  & 2be^{-i3\pi/5} & c\end{pmatrix}.
\label{Yu}
\eeq
\beq
Y^d = \begin{pmatrix} 
y_d^0e^{-i2\pi/5} & 0  & Ay_d^0e^{-i2\pi/5}  \\ 
By_d^0e^{-i3\pi/5} & y_s^0e^{-i2\pi/5} & Cy_d^0e^{-i3\pi/5} \\ 
By_d^0e^{-i3\pi/5}  & 0 & y_b^0 + Cy_d^0e^{-i3\pi/5} \end{pmatrix}
\label{Yd}
\eeq
Note that we have introduced a small correction term $\epsilon$ in the $(1,3)$ entry of $Y^u$
which will mainly affect $\theta^q_{13}$.
Physically this corresponds to a small admixture of the first component of the Higgs triplet 
$h_3$ contributing to the physical light Higgs state $H_u$, as discussed in Appendix~\ref{Higgs}.
The previous analytic results were for $\epsilon=0$, but we find numerically that the best fit to CKM
parameters requires a non-zero value of $\epsilon$.

\begin{figure}[h]
\centering
\includegraphics[width=0.45\textwidth]{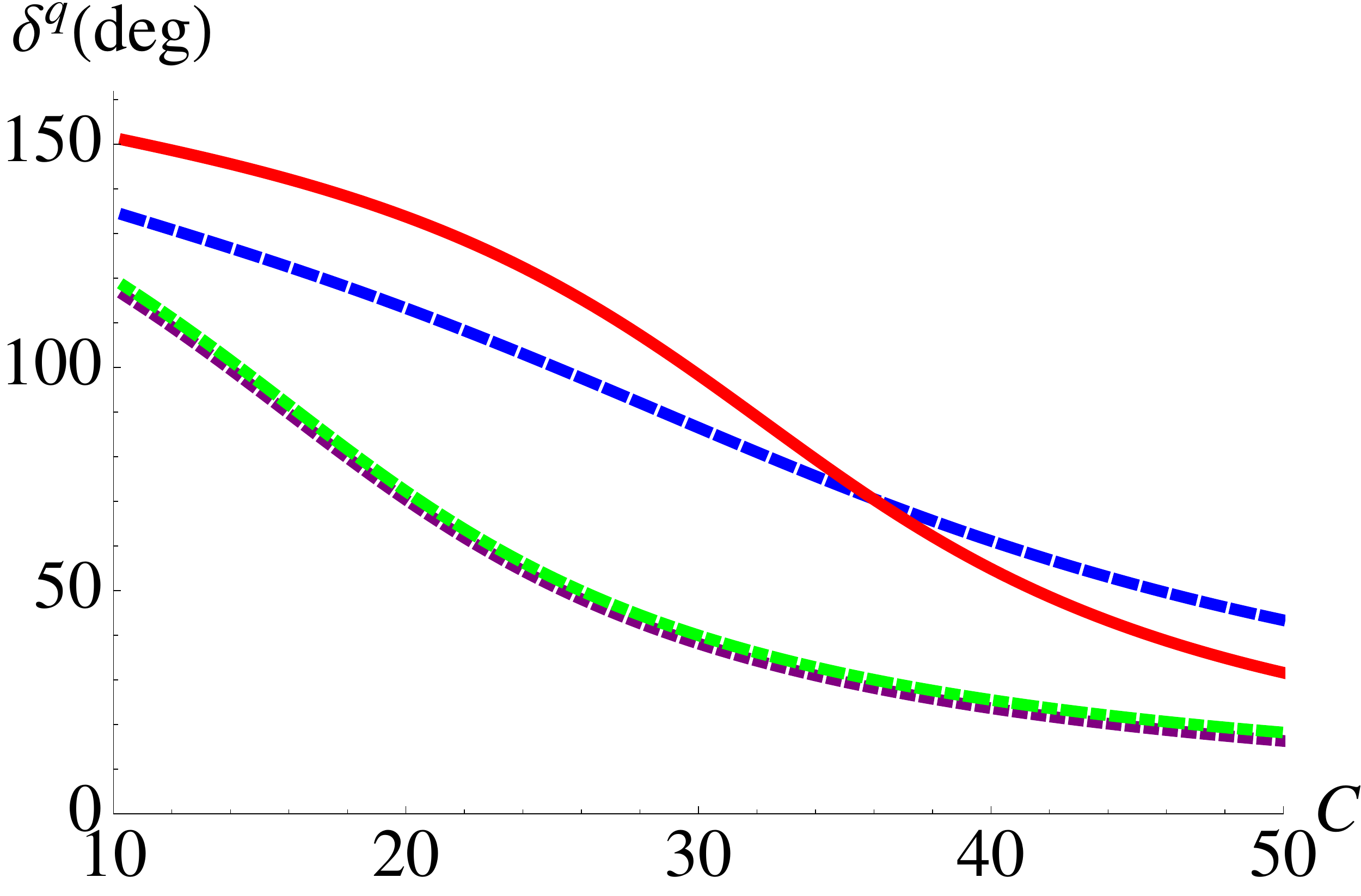} 
\ \ \ \ 
\includegraphics[width=0.45\textwidth]{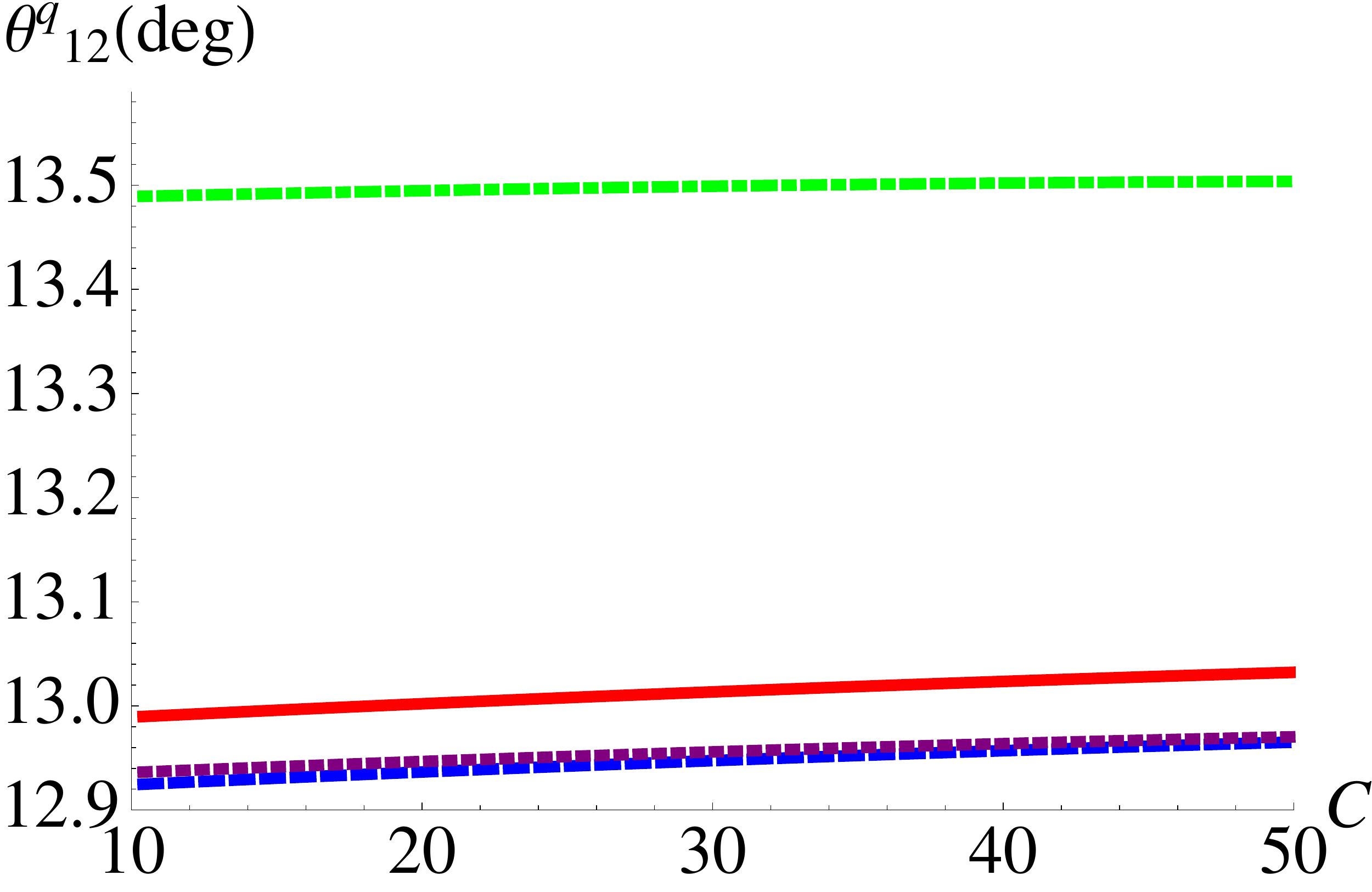} 
\ \ \ \ 
\includegraphics[width=0.45\textwidth]{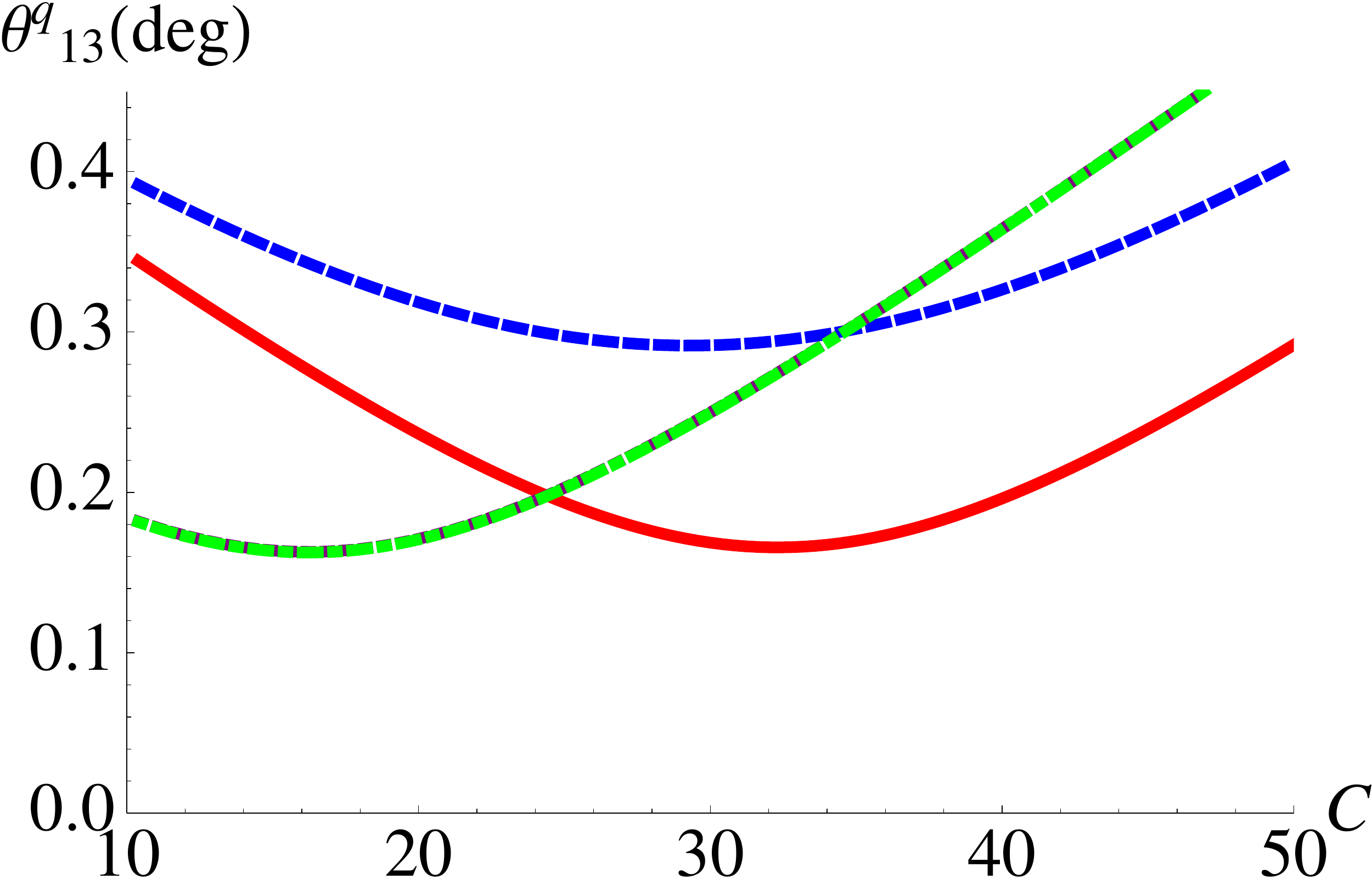} 
\ \ \ \ 
\includegraphics[width=0.45\textwidth]{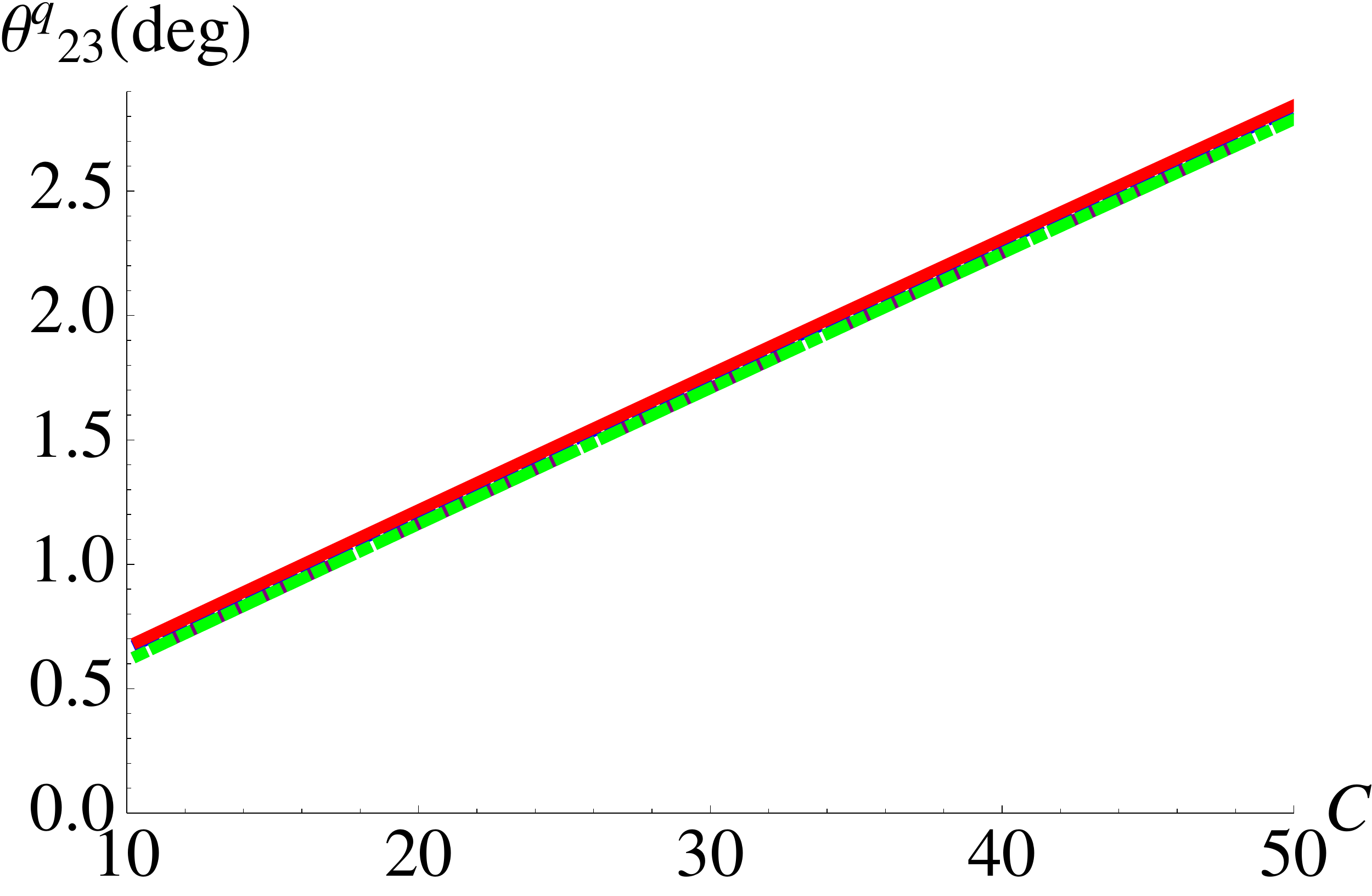} 
    \caption{CKM parameters resulting from Eqs.\ref{Yu},\ref{Yd}, all plotted in degrees as a function of $C$,
    using the parameters in Eqs.\ref{yu2},\ref{yd_diag2}.
    Green dot dashed lines are for $A=5,B=3$ with  $\epsilon=0$.
    Purple dotted lines are for $A=5,B=7$ with  $\epsilon=0$.
    Blue dashed lines are for $A=9,B=7$ with  $\epsilon=0$. 
    Red solid lines are for $A=9,B=7$ with  $\epsilon=-2.4\times 10^{-3}$. } 
     \label{quark}
\vspace*{-2mm}
\end{figure}

For the following results, we shall fix the parameters which approximately determine the six quark masses 
at the high scale to be,
\beq
a= 1.6.10^{-5}, \ \ 
b= 0.8.10^{-3}, \ \
c=0.75,
\label{yu2}
\eeq
\beq
y_d^0=0.9.10^{-5}, \ \ 
y_s^0=  1.4.10^{-4}, \ \ 
y_b^0= -0.9.10^{-2},
\label{yd_diag2}
\eeq
Although the quark results are insensitive to the sign of $y_b^0$, the lepton sector results 
lead to a better fit with the negative sign of $y_b^0$ as discussed later.
Using the Mixing Parameter Tools (MPT) package \cite{Antusch:2005gp},
in Fig.\ref{quark} we show the CKM parameters for different choices of 
$A,B$ as a function of $C$.
$\theta^q_{23}$ is really only sensitive to $C$ only, while 
$\theta^q_{12}$ is mainly sensitive to $B$.
$\theta^q_{13}$ and $\delta^q$ are both sensitive $A$.
The effect of the correction $\epsilon$ is to shift the blue dashed curve to the red solid curve,
lowering $\theta^q_{13}$ while leaving 
$\theta^q_{23}$ almost unchanged,
allowing the best fit of the CKM parameters for $C=36$.

To take a concrete example, for the red solid at the value $C=36$, with the above input parameters
$A=9,B=7$ (c.f. Eq.\ref{ABC2}) and 
$\epsilon=-2.4\times 10^{-3}$, we find
the quark Yukawa eigenvalues at the high scale,
\bea
\label{example_yq}
&&y_u = 3.9.10^{-6}, \
y_c=3.3.10^{-3}, \ 
y_t=0.75, \\
&&y_d=0.81.10^{-5}, \
y_s=1.5.10^{-4}, \
y_b=0.91.10^{-2}
\eea
and the CKM parameters at the high scale,
\beq
\label{example_CKM}
\theta^q_{12}=13.02^\circ, \ 
\theta^q_{13}=0.17^\circ, \ 
\theta^q_{23}=2.09^\circ, \ 
\delta^q=70.4^\circ .
\eeq
These parameters are consistent with those given, for example, in \cite{Antusch:2013jca},
after including RG corrections, in particular due to the large top Yukawa coupling.
Notice that there are as many input parameters as there are physical observables in the quark sector,
so no prediction is claimed.
However we emphasise two interesting features, firstly that the Cabibbo angle
is understood to arise from $Y^u$ leading to $\theta_C\approx  1/4$ or
$\theta_C\approx 14^0$, with a small (one degree) correction mainly controlled by $B$.
Secondly the phases which appear are quantised according to $Z_5$,
which also controls the leptonic phases as discussed in the following subsection. 
Indeed, with $Y^{\nu}=Y^u$ fixed by the quark sector, the entire neutrino sector 
only depends on three additional right-handed neutrino masses, 
which determine the three physical neutrino masses, with the entire neutrino mixing matrix
then being fully determined, with only very small charged lepton mixing corrections appearing
in the PMNS mixing matrix.

\section{Lepton masses and mixing}
\label{leptons}
In this section we discuss the leading order predictions for PMNS mixing which arise from 
the neutrino Yukawa and Majorana matrices in Eq.\ref{Yunu2} which result in a very simple
form of effective neutrino mass matrix, after the see-saw mechanism has been applied.

\subsection{Convention}
The neutrino Yukawa matrix $Y^{\nu}$ is defined in a LR convention by
\footnote{This LR convention for the Yukawa matrix
differs by an Hermitian conjugation compared to that used in the
Mixing Parameter Tools package \cite{Antusch:2005gp} due to the RL convention used there.} 
$$\mathcal{L}=-v^uY^{\nu}_{\alpha i }\overline \nu^{\alpha}_{\mathrm{L}}
\nu^{i}_{\mathrm{R}} + h.c.  $$
where $\alpha = e, \mu , \tau$ labels the three left-handed neutrinos and $i=1,2,3$ labels the three
right-handed neutrinos. 

The physical effective neutrino Majorana mass matrix $m^{\nu}$ is determined
from the columns of $Y^{\nu}$ via the see-saw mechanism,
\begin{eqnarray}
m^{\nu} = - v_u^2\, Y^{\nu} M^{-1}_\mathrm{R} Y^{\nu T} \; ,
\label{seesaw}
\end{eqnarray} 
where the light Majorana neutrino mass
matrix $m^\nu$ is defined by 
\footnote{Note that this convention for the 
light effective Majorana neutrino mass matrix $m^{\nu}$
differs by an overall complex conjugation compared to that used in the
Mixing Parameter Tools package \cite{Antusch:2005gp}.} 
$\mathcal{L}_\nu=-\tfrac{1}{2} m^\nu \overline \nu_{\mathrm{L}} 
\nu^{c}_{\mathrm{L}}$ + h.c., while the heavy right-handed Majorana neutrino
mass matrix $M_R$ is defined by
$\mathcal{L}^R_\nu=-\tfrac{1}{2} M_R \overline \nu^c_{\mathrm{R}} 
\nu_{\mathrm{R}}$ + h.c. and $m^\nu$ is diagonalised by
\begin{eqnarray}\label{eqDiagMnu}
U_{\nu_\mathrm{L}} \,m^\nu\,U^T_{\nu_\mathrm{L}} =
\left(\begin{array}{ccc}
\!m_1&0&0\!\\
\!0&m_2&0\!\\
\!0&0&m_3\!
\end{array}
\right)\! .
\end{eqnarray}  
The PMNS matrix is then given by
\begin{eqnarray}
U_{\mathrm{PMNS}} = U_{e_\mathrm{L}} U^\dagger_{\nu_\mathrm{L}}\; .
\label{PMNSmatrix}
\end{eqnarray}
We use a standard parameterization 
$
U_{\mathrm{PMNS}} = R^l_{23} U^l_{13} R^l_{12} P^l
$ 
in terms of $s^l_{ij}=\sin (\theta^l_{ij})$,
$c^l_{ij}=\cos(\theta^l_{ij})$, the Dirac CP violating phase $\delta^l$ and
further Majorana phases contained in $P^l={\rm diag}(e^{i\frac{\beta^l_1}{2}},e^{i\frac{\beta^l_2}{2}},1)$.
The standard PDG parameterization  \cite{PDG} differs slightly due to the definition of Majorana phases which are by given by $P^l_{\rm PDG}={\rm diag}(1,e^{i\frac{\alpha_{21}}{2}},e^{i\frac{\alpha_{31}}{2}})$.
Evidently the PDG Majorana 
phases are related to those in our convention by $\alpha_{21}=\beta_2^l-\beta_1^l$ and $\alpha_{31}=-\beta_1^l$,
after an overall unphysical phase is absorbed by $U_{e_\mathrm{L}}$.

\subsection{See-saw mechanism}
The neutrino Yukawa and Majorana matrices are as in Eq.\ref{Yunu2},
with $Y^{\nu}=Y^u$ in Eq.\ref{Yu},
\beq
Y^{\nu}  = \begin{pmatrix}  
0 & be^{-i3\pi/5}   & 0 \\ 
ae^{-i3\pi/5} & 4be^{-i3\pi/5} & 0\\  
ae^{-i3\pi/5}  & 2be^{-i3\pi/5} & c\end{pmatrix},
\ \ 
 M_R  \approx   \begin{pmatrix} M_1e^{8i\pi/5} & 0 & 0 \\ 0 &  
M_2e^{4i\pi/5} &0 \\  0 & 0& M_3 \end{pmatrix},
\label{Ynu}
\eeq
where we have ignored the small off-diagonal Majorana mass $M_{13}$
which gives a tiny mixing correction of order $10^{-5}$ from Eq.\ref{eps2},
and dropped the correction $\epsilon$ which is completely negligible in the lepton sector
due to sequential dominance (see below).
We have also assumed a phase in the Majoron VEV
$\langle \xi \rangle \sim e^{4i\pi/5}$ in the operators in Eq.\ref{Maj} responsible for the 
right-handed neutrino masses, as discussed below.

Using Eq.\ref{Ynu}, the see-saw formula in Eq.\ref{seesaw} leads to 
the neutrino mass matrix $m^{\nu}$, 
\beq
m^{\nu} 
= m_a \begin{pmatrix} 0 & 0 & 0 \\ 0 & 1 & 1 \\ 0 & 1 & 1 \end{pmatrix} 
+ m_be^{2i\eta}\begin{pmatrix} 1 & 4 & 2 \\ 4 & 16 & 8 \\ 2 & 8 & 4  \end{pmatrix}
+m_ce^{2i\eta} \begin{pmatrix} 0 & 0 & 0 \\ 0 & 0 & 0 \\ 0 & 0 & 1 \end{pmatrix}, 
\label{seesaw3}
\eeq
where,
\beq
m_a=\frac{a^2v_u^2}{ {M}_1}, \ m_b=\frac{b^2v_u^2}{ {M}_2}, \ m_c=\frac{c^2v_u^2}{ {M}_3},
\label{abc}
\eeq
are three real parameter combinations which determine the three physical neutrino masses 
$m_1,m_2,m_3$, respectively. 
According to sequential dominance $m_c$ will determine the lightest neutrino mass $m_1$ where 
we will have $m_1\ll m_2 < m_3$, so that the third term 
arising from the heaviest right-handed neutrino of mass $M_3$
is approximately decoupled from the see-saw mechanism.
(This is why the correction $\epsilon$ is completely negligible in the lepton sector.)

In order to understand 
the origin of the relative phases $\eta =2\pi /5$
which enter the neutrino mass matrix $m^{\nu}$, it is worth recalling that 
the see-saw operators responsible for the dominant first two terms of the 
neutrino mass matrix in Eq.\ref{seesaw3} have the form
\beq
\label{seesaw2}
m^{\nu}\sim  \frac{\vev{\phi_{\rm atm}}\vev{\phi_{\rm atm}}^T}{\langle \xi\rangle^2}
+  \frac{\vev{\phi_{\rm sol}}\vev{\phi_{\rm sol}}^T}{\langle \xi \rangle},
\eeq
where we have written $\phi_{\rm atm}=\phi_1^u$, $\phi_{\rm sol}=\phi_2^u$
to highlight the fact that the first term gives the dominant contribution to the atmospheric neutrino mass $m_3$,
while the second term controls the solar neutrino mass $m_2$.
The mild neutrino hierarchy between $m_3$ and $m_2$ emerges due to the
choice of Majoron VEV $\langle \xi \rangle$ in Eq.\ref{eps2} which partly cancels the hierarchy in the 
square of the flavon
VEVs in Eq.\ref{ratios2}.
The lightest neutrino mass $m_1$ arises from smaller terms (not shown),
leading to a normal neutrino mass hierarchy, where 
the heaviest atmospheric neutrino mass $m_3$ is associated with the lightest right-handed neutrino mass $M_1$
as in light sequential dominance \cite{King:1998jw}.

Since $\vev{\phi_{\rm atm}}$ and $\vev{\phi_{\rm sol}}$ have the same phase, $e^{-i3\pi/5}$,
and $\langle \xi \rangle $ has a phase 
\footnote{This phase is the complex conjugate of the phase given in the previous convention
in Eq.\ref{Maj}.}
$e^{4i\pi/5}$,
Eq.\ref{seesaw2} shows that the atmospheric term
has a phase $(e^{-i3\pi/5})^2/(e^{4i\pi/5})^2=e^{-14i\pi/5}$,
while the solar term is real.
After multiplying $m^{\nu}$ by an overall phase $e^{4i\pi/5}$, which we are allowed to do since overall phases are irrelevant, the atmospheric term becomes real, while the other two terms pick up phases
of $e^{4i\pi/5}$.
This is equivalent to having a phase $\eta  = 2\pi /5$ in Eq.~\ref{seesaw3}.
Different choices of phase for $\eta$ are theoretically possible, but the phenomenologically
successful choice for the relative phase of the atmospheric and solar terms 
(the first and second terms in Eq.\ref{seesaw3}) is
$\eta  = 2\pi /5$, whereas for example $\eta  = -2\pi /5$ leaves the mixing angles unchanged
but reverses the sign of the CP phases
\cite{King:2013iva,King:2013xba,King:2013hoa}. 
The dependence on see-saw phases was fully discussed in 
\cite{King:2013iva}. Here we only note that in this model the see-saw phases are restricted to
a discrete choice corresponding to the fifth roots of unity due to the $Z_5$ symmetry.
The fact that the decoupled
third term proportional to $m_c$ (responsible  
for the lightest neutrino mass $m_1$) has the same phase as the second term proportional to $m_b$
(responsible for the solar neutrino mass)
is a new prediction of the current model and will affect the $m_1$ dependence of the results.

From Eqs.\ref{Dirac},\ref{up}, the Dirac neutrino masses are equal to the up-type quark masses 
which are related to $a,b,y_t$ and hence
Eq.\ref{abc} becomes,
\beq
m_a= 17\frac{m_u^2}{M_1}, \ m_b = \frac{m_c^2}{17M_2}, \ m_c=\frac{m_t^2}{M_3}.
\label{M_i}
\eeq
Using Eq.\ref{M_i}, the three right-handed neutrino masses $M_1$, $M_2$, $M_3$ may be determined
for particular values of $m_a$, $m_b$, $m_c$, and the known quark masses $m_u,m_c,m_t$
(evaluated at high scales).

The neutrino mass matrix in Eq.\ref{seesaw3} may be diagonalised numerically to determine the physical neutrino masses and the PMNS mixing matrix as in 
Eq.\ref{eqDiagMnu}.
We emphasise that, at leading order, with the phase $\eta =2\pi /5$ fixed by the previous argument,
the neutrino mass matrix involves just 
3 real input parameters $m_a$, $m_b$, $m_c$ from which 
12 physical parameters in the lepton sector are predicted, comprising 9
lepton parameters from diagonalising the neutrino mass matrix $m^{\nu}$
in Eq.\ref{seesaw3}
(the 3 angles $\theta^l_{ij}$, 3 phases $\delta^l, \beta^l_1,\beta^l_2$ and the 3 light neutrino masses $m_i$) 
together with the 3 heavy right-handed
neutrino masses $M_i$ from Eq.\ref{M_i}. The model is clearly highly predictive, 
involving 12 predictions in the lepton sector from only 3 input parameters.

\subsection{A first numerical example}
To take a numerical example, diagonalising the neutrino mass matrix in Eq.\ref{seesaw3}, with the 
three input parameters
\beq
\label{example}
m_a = 0.035 \ {\rm eV}, \
m_b = 0.002  \ {\rm eV},\
m_c=0.002 \ {\rm eV}, 
\eeq
the Mixing Parameter Tools package \cite{Antusch:2005gp} gives the physical neutrino masses,
\beq
\label{example_m_i}
m_1 = 3.29.10^{-4} \ {\rm eV}, \
m_2=8.62.10^{-3} \ {\rm eV}, \
m_3=4.93.10^{-2} \ {\rm eV},
\eeq
corresponding to the mass squared differences,
\beq
\label{example_m_i2}
\Delta m^2_{21} = 7.42.10^{-5} \ {\rm eV^2}, \
\Delta m^2_{31} = 2.43.10^{-3} \ {\rm eV^2}, \
\Delta m^2_{32} = 2.36.10^{-3} \ {\rm eV^2}, 
\eeq
and the lepton mixing parameters,
\beq
\label{example_PMNS}
\theta^l_{12}=32.2^\circ, \ 
\theta^l_{13}=9.3^\circ, \ 
\theta^l_{23}=41.6^\circ, \ 
\delta^l=248^\circ, \ 
\beta^l_1=114^\circ, \
\beta^l_2=90^\circ.
\eeq
The PDG Majorana phases \cite{PDG} are given by 
$\alpha_{21}=\beta^l_2-\beta^l_1$ and $\alpha_{31}=-\beta^l_1$.
For the choice of input parameters in Eq.\ref{example} and the high scale quark masses, 
\beq
m_u= 1 \ {\rm MeV}, \
m_c= 400 \ {\rm MeV}, \
m_t= 100 \ {\rm GeV},
\eeq
Eq.\ref{M_i} then determines the three right-handed neutrino masses to be,
\beq
\label{example_M_i}
M_1 = 5\times 10^5 \ {\rm GeV}, \
M_2= 5\times 10^9\ {\rm GeV}, \
M_3= 5\times 10^{15}\ {\rm GeV}.
\eeq
Eq.\ref{example} shows the 3 input parameters, while Eqs.\ref{example_m_i}, \ref{example_PMNS},
\ref{example_M_i} shows the 12 output predictions. One may regard the 3 input parameters in Eq.\ref{example} 
as fixing the 3 light physical neutrino masses in Eq.\ref{example_m_i}, with all the 6 PMNS matrix parameters
in Eq.\ref{example_PMNS} as being independent predictions, along with the 3 right-handed neutrino masses
in Eq.\ref{example_M_i}.

So far we have ignored charged lepton corrections which are expected in the model to be small.
However the corrections are not entirely negligible as the following example shows.
The charged lepton Yukawa matrix is given from Eq.\ref{Yed2},
\beq
Y^e = \begin{pmatrix} 
-(y_d^0/3)e^{-i2\pi/5} & 0  & Ay_d^0e^{-i2\pi/5}  \\ 
By_d^0e^{-i3\pi/5} & -3y_s^0e^{-i2\pi/5} & -3Cy_d^0e^{-i3\pi/5} \\ 
By_d^0e^{-i3\pi/5}  & 0 & y_b^0 - 3Cy_d^0e^{-i3\pi/5} \end{pmatrix}.
\label{Ye}
\eeq
which should be compared to the down quark Yukawa matrix in Eq.\ref{Yd}.
The off-diagonal elements of $Y^e$ are small, similar to those of $Y^d$
which are responsible for the small quark mixing angles and a correction to the Cabibbo angle
of one degree.
The quark mixing angles fix the three real parameters 
to be for example $A=9,B=7,C=36$ and the down quark couplings in Eq.\ref{yd_diag2}.
Including the charged lepton Yukawa matrix with these parameters 
and the same neutrino mass parameters as in Eq.\ref{example},
the MPT package gives the lepton mixing parameters,
\beq
\label{example_PMNS2}
\theta^l_{12}=32.15^\circ , \ 
\theta^l_{13}=8.9^\circ , \ 
\theta^l_{23}=45.2^\circ , \ 
\delta^l=259^\circ , \ 
\beta^l_1=92^\circ , \
\beta^l_2=70^\circ .
\eeq
Comparing the results in Eq.\ref{example_PMNS2} to those in Eq.\ref{example_PMNS}, we see that 
the atmospheric angle has increased by about $3^\circ$ to become maximal
due to the $(2,3)$ element in the charged lepton Yukawa
matrix, which is enhanced by a Clebsch factor of 3 relative to the same element in the down Yukawa matrix.
The reactor angle has decreased slightly, and the CP oscillation phase has increased.
With $y_b^0$ taken to be positive instead of negative, 
and all the other parameters unchanged, we find the results below,
\beq
\label{example_PMNS3}
\theta^l_{12}=32.27^\circ , \ 
\theta^l_{13}=9.65^\circ , \ 
\theta^l_{23}=37.3^\circ , \ 
\delta^l=240^\circ , \ 
\beta^l_1=132^\circ , \
\beta^l_2=106^\circ .
\eeq
The main effect of the sign of $y_b^0$ is on the atmospheric and reactor angles.

\subsection{Modified Georgi-Jarlskog relations}
Since the charged lepton masses are known with much higher precision than the down type quark masses,
the down Yukawa couplings in practice will be predicted from inputting the charged lepton masses
in order to accurately fix $y_d^0$, $y_s^0$, $y_b^0$.
Comparing $Y^e$ in Eq.\ref{Ye} to $Y^d$ in Eq.\ref{Yd}, we find that we do not get exactly the GJ relations
in Eq.\ref{GJ}
due to the off-diagonal elements which also involve Clebsch factors. Numerically we find that,
for $y_b^0$ negative and the other parameters as above, the 
Yukawa eigenvalues at the GUT scale are approximately related as,
\beq
y_e=\frac{y_d}{2.6}, \ \ y_{\mu}=2.8y_s, \ \ y_{\tau}=0.97y_b ,
\label{GJ2}
\eeq
while for $y_b^0$ positive we find,
\beq
y_e=\frac{y_d}{3.0}, \ \ y_{\mu}=2.7y_s, \ \ y_{\tau}=1.05y_b .
\label{GJ3}
\eeq
These may be compared to the phenomenological relation \cite{Antusch:2013jca},
\beq
\left|  \frac{y_{\mu}}{y_s} \frac{y_d}{y_e} \right| =10.7^{+1.8}_{-0.8}.
\label{ratio1}
\eeq
For example for $y_b^0$ negative we find the RHS to be 7.3 which differs by more than 4 sigma.
In order to bring this relation into better agreement with experiment 
we would need to increase this ratio, for example by increasing the muon Yukawa eignenvalue compared to the strange quark Yukawa eigenvalue. One way to do this is to introduce a flavon $\phi^{d15}_2$
with the same charges as $\phi^{d}_2$ but in the adjoint $15$ of $SU(4)_C$. 
The middle diagram in Fig.\ref{mess1} involving $\phi^{d15}_2$ 
involves a Clebsch factor of +9 as compared to  the factor of -3 with $\phi^{d}_2$ \cite{Antusch:2013rxa}.
Below the PS the colour singlet component of $\phi^{d15}_2$ mixes with $\phi^{d}_2$, to yield a light flavon combination,
\beq
 \phi^{d'}_2 =\phi^{d15}_2 \cos \gamma +  \sin \gamma  \phi^{d}_2.
\eeq
Hence middle diagram in Fig.\ref{mess1} involving $\phi^{d'}_2$ implies the relation,
\beq
\frac{y^0_{\mu}}{y^0_s} = 9\cos \gamma -3  \sin \gamma  .
\eeq
For example by suitable choice of the mixing angle $\gamma$ we can arrange $y^0_{\mu}=4.5y^0_s$,
\beq
Y^e = \begin{pmatrix} 
-(y_d^0/3)e^{-i2\pi/5} & 0  & Ay_d^0e^{-i2\pi/5}  \\ 
By_d^0e^{-i3\pi/5} & -4.5y_s^0e^{-i2\pi/5} & -3Cy_d^0e^{-i3\pi/5} \\ 
By_d^0e^{-i3\pi/5}  & 0 & y_b^0 - 3Cy_d^0e^{-i3\pi/5} \end{pmatrix}.
\label{Ye2}
\eeq
By comparing $Y^e$ in Eq.\ref{Ye2} to $Y^d$ in Eq.\ref{Yd}, 
we find the modified GJ relations,
\beq
y_e=\frac{y_d}{2.6}, \ \ y_{\mu}=4.1y_s, \ \ y_{\tau}=0.97y_b ,
\label{GJ4}
\eeq
and hence,
\beq
\left|  \frac{y_{\mu}}{y_s} \frac{y_d}{y_e} \right| =   10.7   ,
\eeq
which reproduces the central value in Eq.\ref{ratio1}.
In the above estimate we have assumed 
$A=9,B=7,C=36$ and the other couplings in Eq.\ref{yd_diag2}.
Using the same neutrino mass parameters as in Eq.\ref{example},
the MPT package gives the same lepton mixing parameters as for the GJ form
in Eq.\ref{example_PMNS2}, to very good accuracy. 

\subsection{Numerical results for neutrino masses and lepton mixing}

In our numerical results 
we shall use the charged lepton Yukawa matrix in Eq.\ref{Ye2}, together with the neutrino mass matrix in Eq.\ref{seesaw3}, as summarised below,
\beq
m^{\nu} 
= m_a \begin{pmatrix} 0 & 0 & 0 \\ 0 & 1 & 1 \\ 0 & 1 & 1 \end{pmatrix} 
+ m_be^{i4\pi/5}\begin{pmatrix} 1 & 4 & 2 \\ 4 & 16 & 8 \\ 2 & 8 & 4  \end{pmatrix}
+m_ce^{i4\pi/5} \begin{pmatrix} 0 & 0 & 0 \\ 0 & 0 & 0 \\ 0 & 0 & 1 \end{pmatrix}, 
\label{seesaw4}
\eeq
\beq
Y^e = \begin{pmatrix} 
-(y_d^0/3)e^{-i2\pi/5} & 0  & Ay_d^0e^{-i2\pi/5}  \\ 
By_d^0e^{-i3\pi/5} & -4.5y_s^0e^{-i2\pi/5} & -3Cy_d^0e^{-i3\pi/5} \\ 
By_d^0e^{-i3\pi/5}  & 0 & y_b^0 - 3Cy_d^0e^{-i3\pi/5} \end{pmatrix}.
\label{Ye3}
\eeq
As discussed previously, the lepton mixing depends on 
predominantly on $m^{\nu}$ which involves the three real mass parameters $m_a$, $m_b$, $m_c$,
which are effectively fixed by the neutrino masses.
However there are small corrections coming from $Y^e$,
which involves the real parameters $A,B,C$ which determine the quark mixing angles
and the real Yukawa couplings
$y_d^0$, $y_s^0$, $y_b^0$ which were previously determined from the down-type quark masses.

\begin{figure}[h]
\centering
\includegraphics[width=0.48\textwidth]{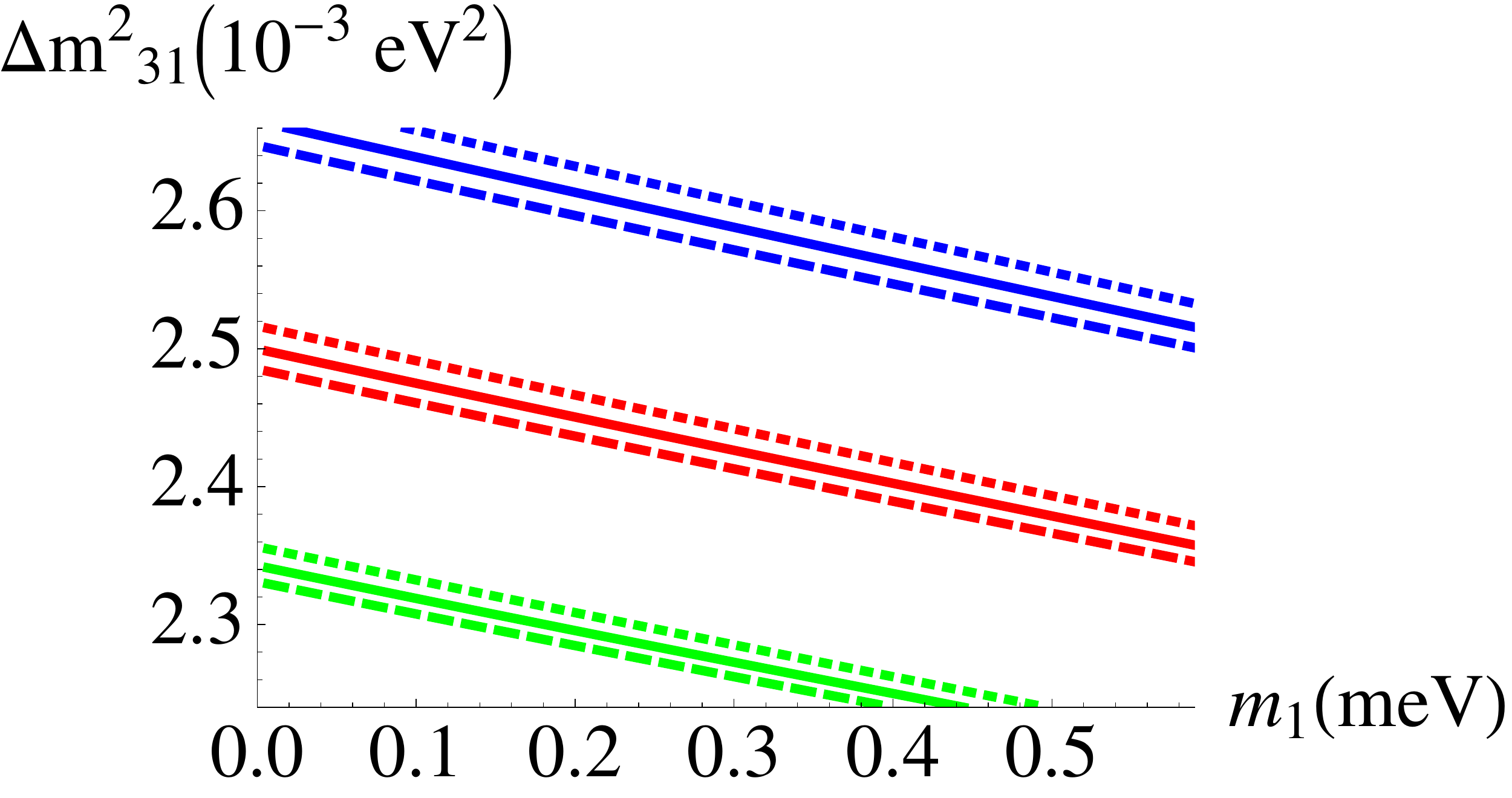} 
\ \ \ \ 
\includegraphics[width=0.48\textwidth]{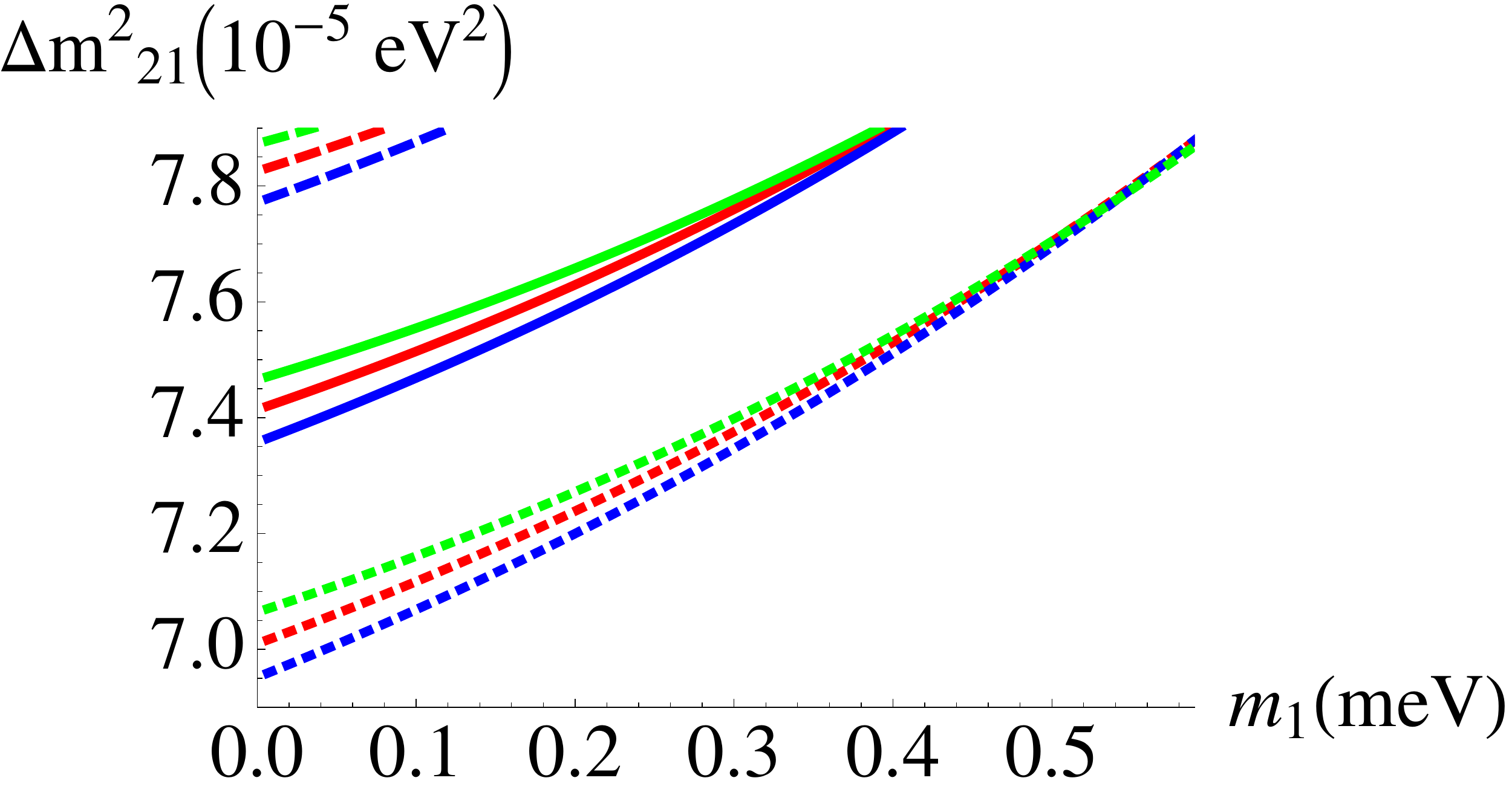} 
    \caption{The neutrino mass squared parameters $\Delta m_{31}^2$ and $\Delta m_{21}^2$
      resulting from Eq.\ref{seesaw4}, plotted as a function of the lightest neutrino mass $m_1$.
           Each line corresponds to a fixed $m_a$ and $m_b$ with varied $m_c$.
          The (Blue, Red, Green) coloured lines correspond to $m_a=(0.036,0.035,0.034)$ eV, respectively,
    and give (High, Central, Low) values of $\Delta m_{31}^2$. 
    The (Dashed, Solid, Dotted) styles correspond to $m_b=(0.00210,0.00205,0.00200)$ eV, respectively,
    and yield (High, Central, Low) values of $\Delta m_{21}^2$.
    The parameter $m_c$ is varied from $0-0.004$ eV corresponding to $m_1=0-0.006$ eV.
    } 
     \label{neutrinomasssquareds}
\vspace*{-2mm}
\end{figure}

\begin{figure}[h]
\centering
\includegraphics[width=0.48\textwidth]{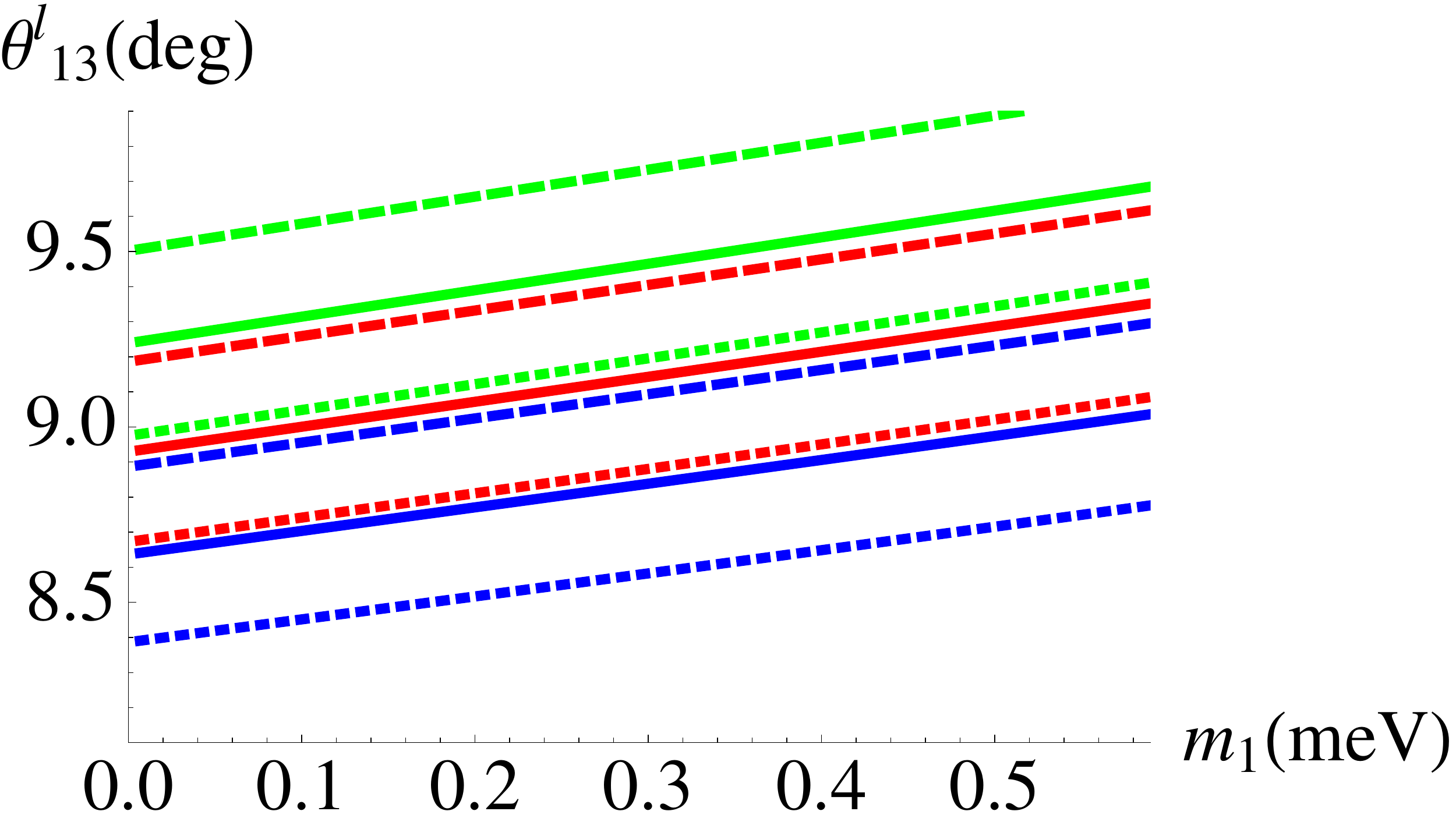} 
\ \ \ \ 
\includegraphics[width=0.48\textwidth]{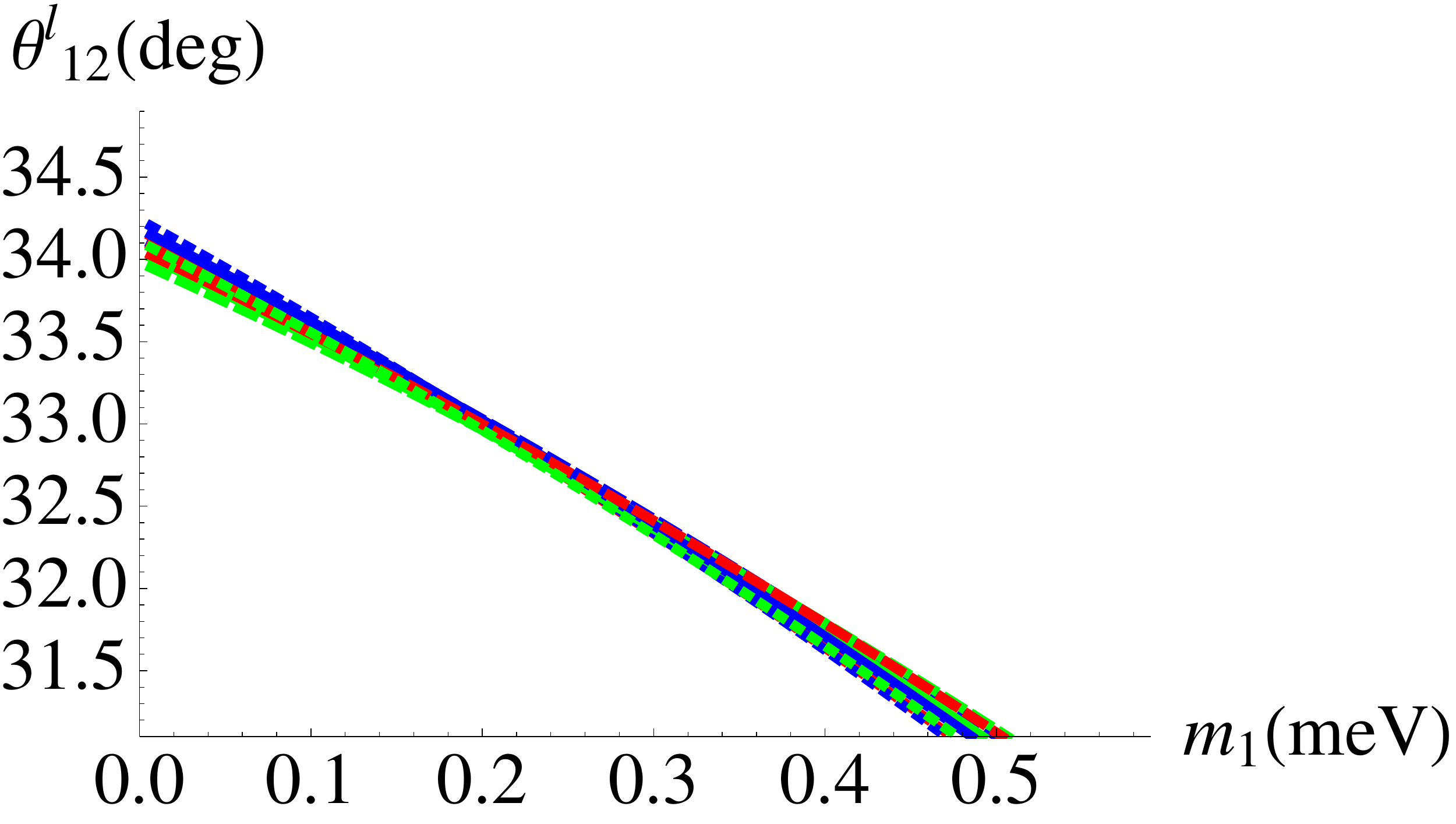} 
\ \ \ \ 
\includegraphics[width=0.48\textwidth]{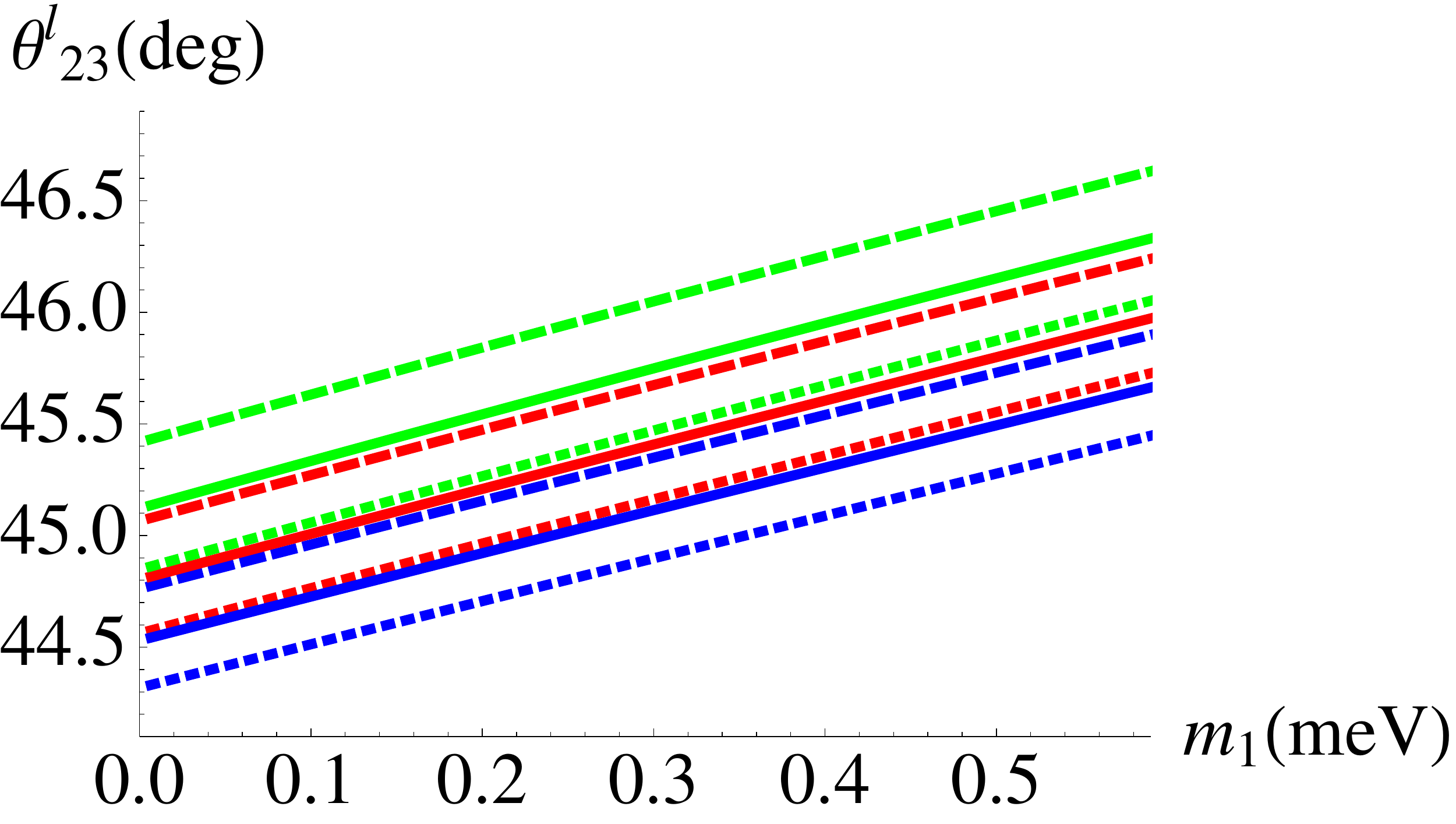} 
\ \ \ \ 
\includegraphics[width=0.48\textwidth]{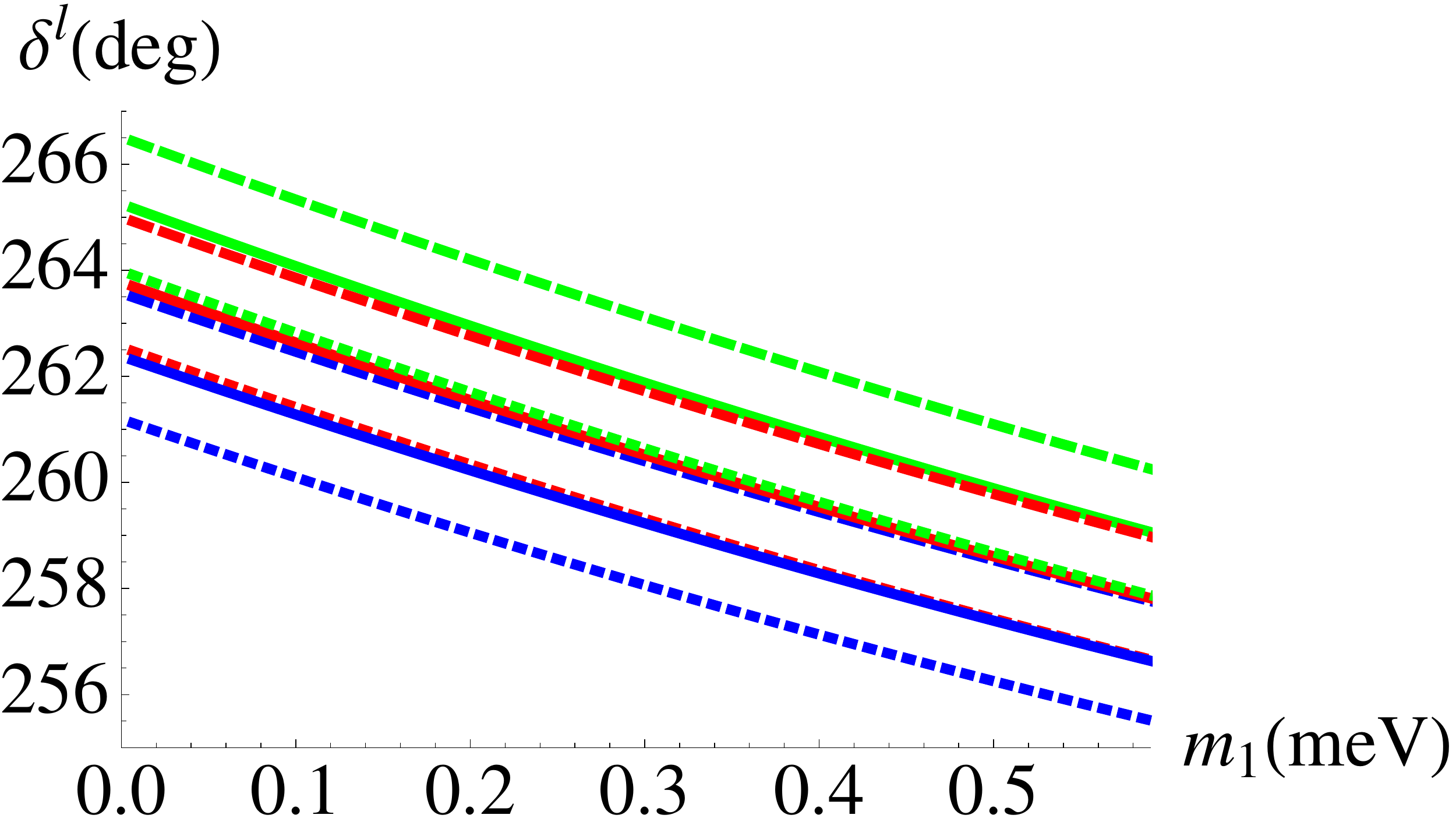} 
    \caption{PMNS predictions of the model, resulting from Eqs.\ref{seesaw4},\ref{Ye3}, plotted as a function of the lightest neutrino mass $m_1$ for charged lepton parameters given by
    $A=9,B=7,C=36$ and the down quark couplings in Eq.\ref{yd_diag2}.
    Each line corresponds to a fixed $m_a$ and $m_b$ with varied $m_c$, using the same values
    as in Fig.\ref{neutrinomasssquareds}, with the colour coding and line styles as before.} 
     \label{lepton}
\vspace*{-2mm}
\end{figure}

As discussed previously (c.f. Eqs.\ref{example_PMNS}, 
\ref{example_PMNS2}, \ref{example_PMNS3}) the effect on 
lepton mixing depends on the sign of $y_b^0$ where the negative sign pushes up the
atmospheric angle towards maximal, while also decreasing the reactor angle, while the 
positive sign has the opposite effect. Here we shall show results for the negative sign of
$y_b^0$, as in Eq.\ref{yd_diag2}.
We shall also use the same real parameters $A=9,B=7,C=36$
which gave a good fit to the quark mixing angles and CP phase in Eq.\ref{example_CKM}.
Since lepton mixing depends mainly on the three real mass parameters $m_a$, $m_b$ and $m_c$
which also determine the neutrino masses, we shall show results as a function of the
neutrino mass parameters.
Here we shall restrict ourselves to showing results where we keep the parameters appearing in $Y^e$ fixed at the above ``benchmark'' values, and vary only $m_a$, $m_b$ and $m_c$.
The parameter $m_a$ is mainly responsible for the atmospheric neutrino mass and hence 
$\Delta m_{31}^2$, while $m_b$ is mainly responsible for the solar neutrino mass and hence 
$\Delta m_{21}^2$, with $m_c$ being mainly responsible for the lightest neutrino mass $m_1$,
which is zero for $m_c=0$. Once the parameters $m_a$ and $m_b$ are chosen to 
fix $\Delta m_{31}^2$ and $\Delta m_{21}^2$ for $m_c=0$ , then all neutrino parameters 
are predicted as a function of $m_c$ and hence $m_1$, as described below.

Using the Mixing Parameter Tools package \cite{Antusch:2005gp},
in Fig.\ref{neutrinomasssquareds} we show the neutrino mass squared differences
as a function of the lightest physical neutrino mass $m_1$, corresponding to varying
$m_c$ for various fixed values of $m_a,m_b$ as given in the figure caption.
Note that $\Delta m_{21}^2$ actually increases with $m_1$.
This is because, with fixed $m_a$ and $m_b$, switching on $m_c$ also increases $m_2$.
Since $m_2^2$ increases linearly with $m_c$, after expanding,
this has a more significant effect on $\Delta m_{21}^2$ than the quadratic 
increase of $m_1^2$, in the region of small $m_c$.
In Fig.\ref{lepton} we show the resulting 
model predictions for the lepton mixing angles and CP oscillation
phase. In all the plots (blue, red, green) coloured lines correspond to (high, central, low) values of 
$\Delta m_{31}^2$, while the (dashed, solid, dotted) styles correspond to (high, central, low) values of 
$\Delta m_{21}^2$.
Note that the presently $3\sigma$ allowed
range of mass squared parameters are \cite{Capozzi:2013csa,GonzalezGarcia:2012sz,update}:
$\Delta m_{31}^2=(2.25-2.65).10^{-3}$ eV$^2$, $\Delta m_{21}^2=(7.0-8.0).10^{-5}$ eV$^2$,
and our choice of parameters covers most of these ranges.
Thus the red solid curve corresponds to central values of both
$\Delta m_{31}^2$ and $\Delta m_{21}^2$ for low values of $m_1$, while the other curves
reflect the uncertainty in the PMNS predictions due to the present precision
in the neutrino mass squared differences. 

Using the Mixing Parameter Tools package \cite{Antusch:2005gp},
in Fig.\ref{lepton} we show the PMNS predictions of the model, resulting from Eqs.\ref{seesaw4},\ref{Ye3}, plotted as a function of the lightest neutrino mass $m_1$. 
From Fig.\ref{lepton}, the PMNS parameters are predicted to be in the following ranges:
\beq
\theta^l_{12}=34^\circ-31^\circ, \ \ \theta^l_{13}=8.4^\circ-9.7^\circ, \ \ \theta^l_{23}=44.4^\circ-46.4^\circ,\ \ \delta^l=266^\circ-256^\circ.
\eeq
These predictions should be compared to
the presently $3\sigma$ allowed
ranges \cite{update}:
\beq
\theta^l_{12}=31^\circ-36^\circ, \ \ \theta^l_{13}=5.5^\circ-10^\circ, \ \ \theta^l_{23}=37^\circ-55^\circ,\ \ 
\delta^l=0^\circ-360^\circ,
\eeq
and the best fit values for a normal hierarchy with $1\sigma$ errors \cite{Forero:2014bxa}:
\beq
\theta^l_{12}={34.63^\circ}^{+1.02^\circ}_{-0.98^\circ}, \ \ \theta^l_{13}= {8.80^\circ}^{+0.37^\circ}_{-0.39^\circ}, \ \ 
\theta^l_{23}={48.9^\circ}^{+1.6^\circ}_{-7.4^\circ}, \ \ 
\delta^l={241^\circ}^{+115^\circ}_{-68^\circ}.
\eeq
The solar angle prediction is 
$34^\circ \gsim \theta^l_{12}\gsim 31^\circ$, for the lightest neutrino mass in the range 
$0 \lsim m_1 \lsim 0.5$ meV, corresponding to a normal neutrino mass hierarchy.
Since the solar angle is very insensitive to $\Delta m_{31}^2$ and $\Delta m_{21}^2$ values,
and decreases as $m_1$ increases, an accurate determination of the solar angle
will accurately determine $m_1$ in this model.
The model also predicts a reactor angle $\theta^l_{13}=9^\circ\pm 0.5^\circ$, close to its best fit value, 
with a significant dependence on $\Delta m_{31}^2$ and $\Delta m_{21}^2$.
A striking prediction of the model 
is the atmospheric angle which is
predicted to be close to maximal to within about one degree for 
nearly all allowed $\Delta m_{31}^2$ and $\Delta m_{21}^2$.
The bulk of the parameter space for low $m_1$
predicts in fact $\theta^l_{23}=45^\circ\pm 0.5^\circ$.
It is worth noting that the most recent 
fit \cite{Forero:2014bxa} is quite compatible with maximal atmospheric mixing to within $1\sigma$
for the case of a normal mass squared ordering, when 
the latest T2K disappearance data is included.
The model also predicts accurately the CP phase with the bulk of the parameter space around
$\delta^l=260^\circ \pm 5^\circ$, compatible with the best fit value,
although the latter has a much larger error.

In general one can expect corrections coming from
renormalisation group (RG) running \cite{King:2000hk,Boudjemaa:2008jf}
as well as canonical normalisation corrections \cite{Antusch:2007ib}.
For a SUSY GUT with light sequential dominance, as in the present model, 
the RG corrections for high $\tan \beta\sim 50$ have been shown to be \cite{Boudjemaa:2008jf}: 
$\Delta \theta^l_{23}\sim +1^\circ$,
$\Delta \theta^l_{12}\sim +0.4^\circ$, $\Delta \theta^l_{13}\sim - 0.1^\circ$,
where the positive sign means that the value increases in running from the GUT scale to low energy,
while for low $\tan \beta \lsim 10$ the RG corrections are negligible compared to the range of the predictions. In particular the effect of right-handed neutrino thresholds \cite{King:2000hk} is expected to be negligible in this model since the heaviest right-handed neutrino mass is close to the GUT scale,
while the lighter right-handed neutrinos have very small Yukawa couplings given by 
$a\sim 2.10^{-5}$ and $b\sim 10^{-3}$ from Eq.\ref{ratios2}.

We emphasise that, since the parameters in $Y^e$ in Eq.\ref{Ye3} are fixed from the quark sector,
and the light neutrino masses are determined by three real parameters 
$m_a$, $m_b$, $m_c$ in Eq.\ref{seesaw4}, the entire PMNS matrix containing
3 mixing angles and 3 CP phases emerges as a prediction of the model,
although 2 of these CP phases will be difficult to measure for a normal neutrino mass
hierarchy, so we have not plotted their predictions.
The model may be tested most readily by its prediction of maximal atmospheric
mixing and a normal neutrino mass hierarchy. 
It would be interesting to perform a $\chi^2$ analysis of the quark and lepton masses
and mixing angles predicted by the model, but that is beyond the scope of the present paper.

\subsection{Majorana phases, Neutrinoless double beta decay and Sum of Neutrino Masses relevant for Cosmology}
The Majorana phases $\alpha_{21},\alpha_{31}$ (in PDG convention defined below Eq.\ref{PMNSmatrix})
predicted by the model are displayed in Fig.\ref{Majorana},
using the same parameter sets and colour coding as for the other plots.
Note that $\alpha_{31}\approx -90^{\circ}$, similar to the oscillation phase $\delta^l$.

\begin{figure}[h]
\centering
\includegraphics[width=0.48\textwidth]{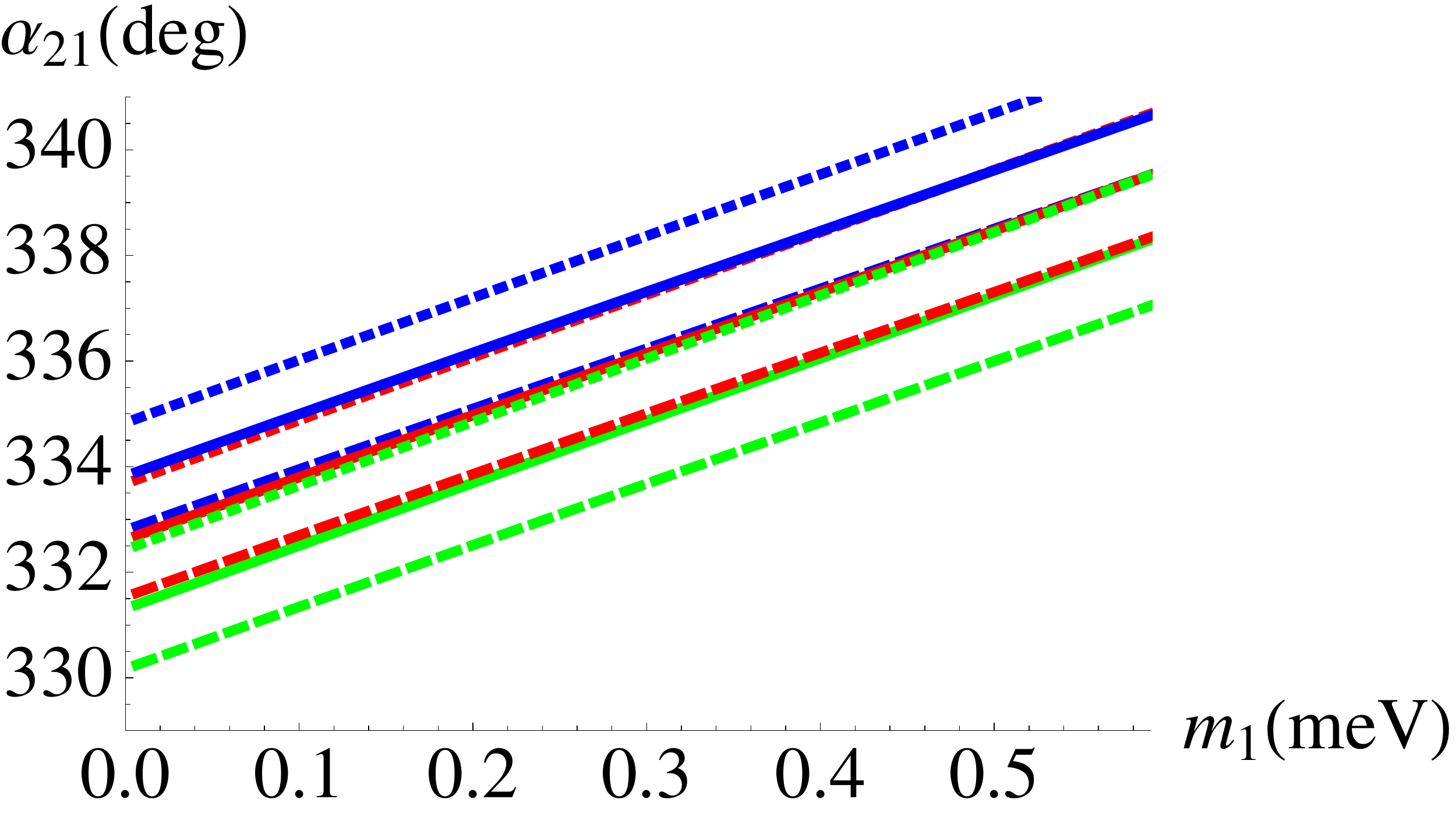} 
\ \ \ \ 
\includegraphics[width=0.48\textwidth]{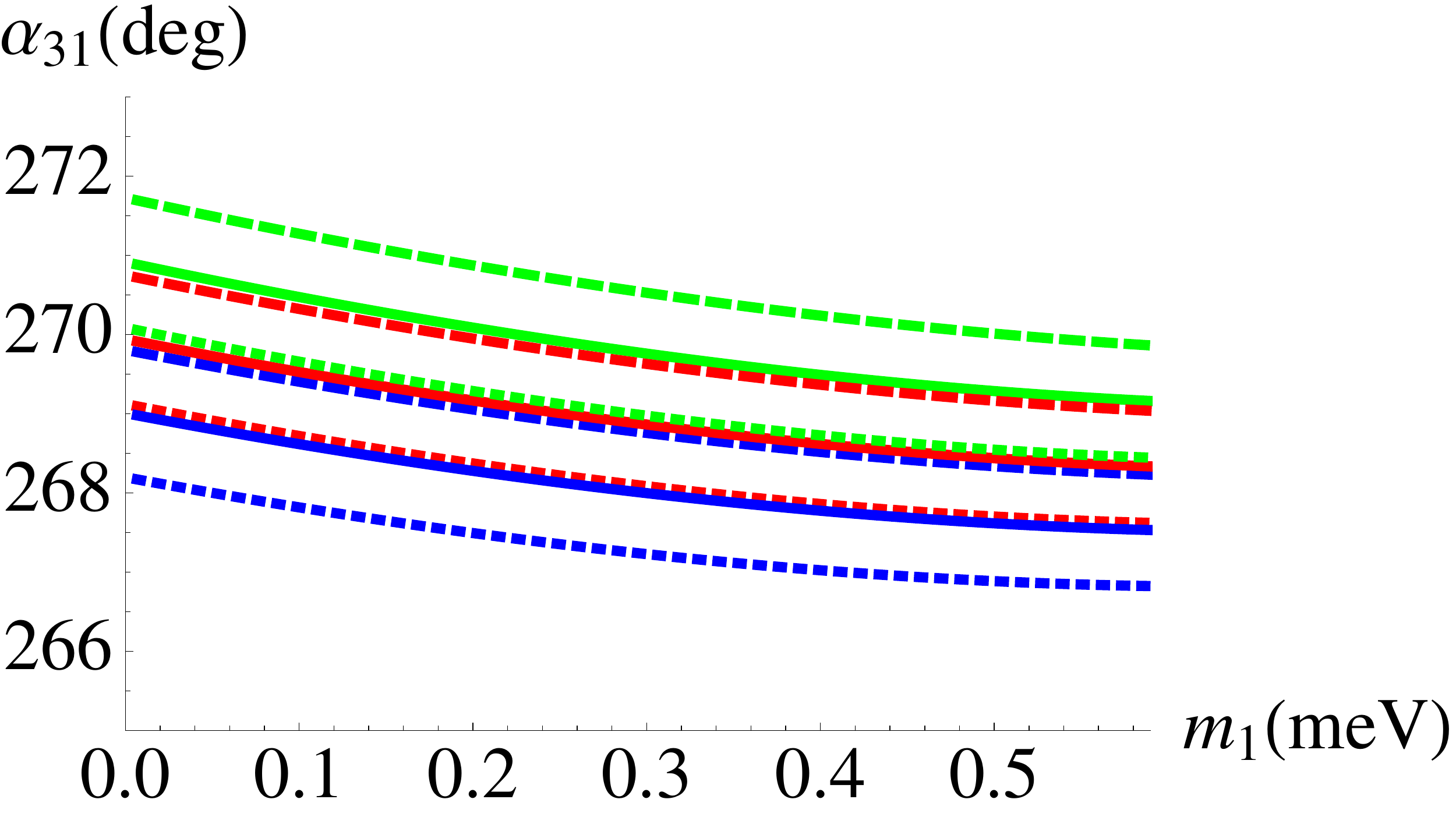} 
    \caption{Majorana phases (in PDG convention defined below Eq.\ref{PMNSmatrix})
    as predicted by the model, resulting from Eqs.\ref{seesaw4},\ref{Ye3}, plotted as a function of the lightest neutrino mass $m_1$ for charged lepton parameters given by
    $A=9,B=7,C=36$ and the down quark couplings in Eq.\ref{yd_diag2}.
    Each line corresponds to a fixed $m_a$ and $m_b$ with varied $m_c$, using the same values
    as in Fig.\ref{neutrinomasssquareds}, with the colour coding and line styles as before.} 
     \label{Majorana}
\vspace*{-2mm}
\end{figure}

The Majorana phases $\alpha_{21},\alpha_{31}$ enter the
effective mass $|m_{ee}|$ observable in neutrinoless double beta decay parameter given by,
\begin{equation}
 |m_{ee}| = |m_1 c_{12}^2 c_{13}^2 + m_2 s_{12}^2 c_{13}^2 e^{i \alpha_{21}} + m_3 s_{13}^2 e^{i (\alpha_{31} - 2\delta)}|.
 \label{eqmee_2}
\end{equation}
In the present model $|m_{ee}|$ is predicted to be always very small and unobservable in the foreseeable
future. For example, for the parameters in Eq.\ref{example}, \ref{example_m_i} and \ref{example_PMNS2},
we find,
\begin{equation}
 |m_{ee}| \approx |0.2 + 2.4 e^{-i 0.12\pi} + 1.2 e^{i0.61\pi}|\ {\rm meV} \approx 2.1\ {\rm meV}.
 \label{eqmee_3}
\end{equation}

The sum of neutrino masses is relevant for cosmology, since it contributes to hot dark matter,
leading to a constraint on its value and eventually a measurement.
This is defined by,
\beq
\Sigma m_i\equiv \Sigma_{i=1}^3 m_i = m_1 + m_2 +m_3 .
\label{sum_mi}
\eeq
Due to the rather strong normal hierarchy, this value is dominated by the value of $m_3$,
which is controlled by the parameter $m_a$ in the neutrino mass matrix in Eq.\ref{seesaw4}.

\begin{figure}[h]
\centering
\includegraphics[width=0.48\textwidth]{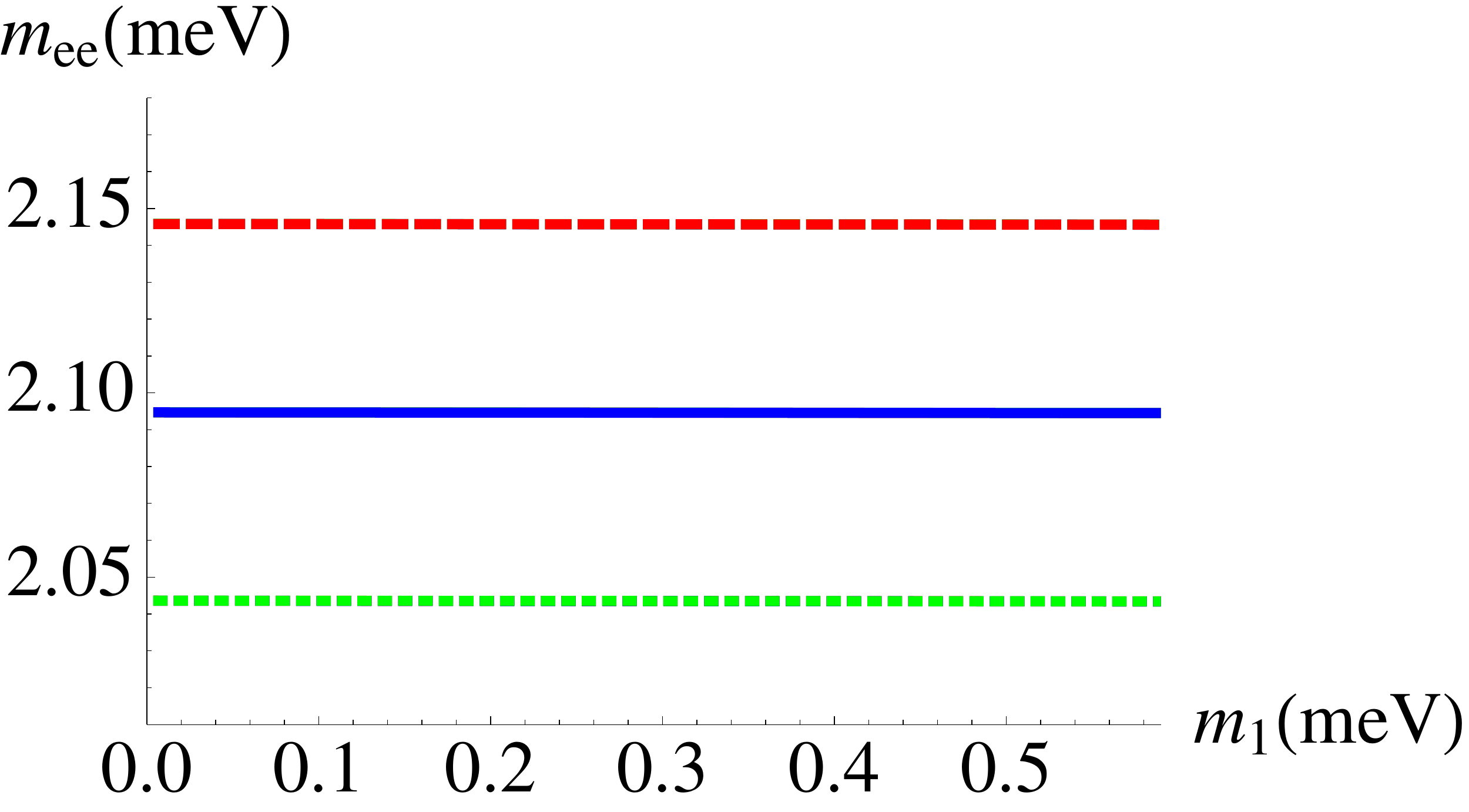} 
\ \ \ \ 
\includegraphics[width=0.48\textwidth]{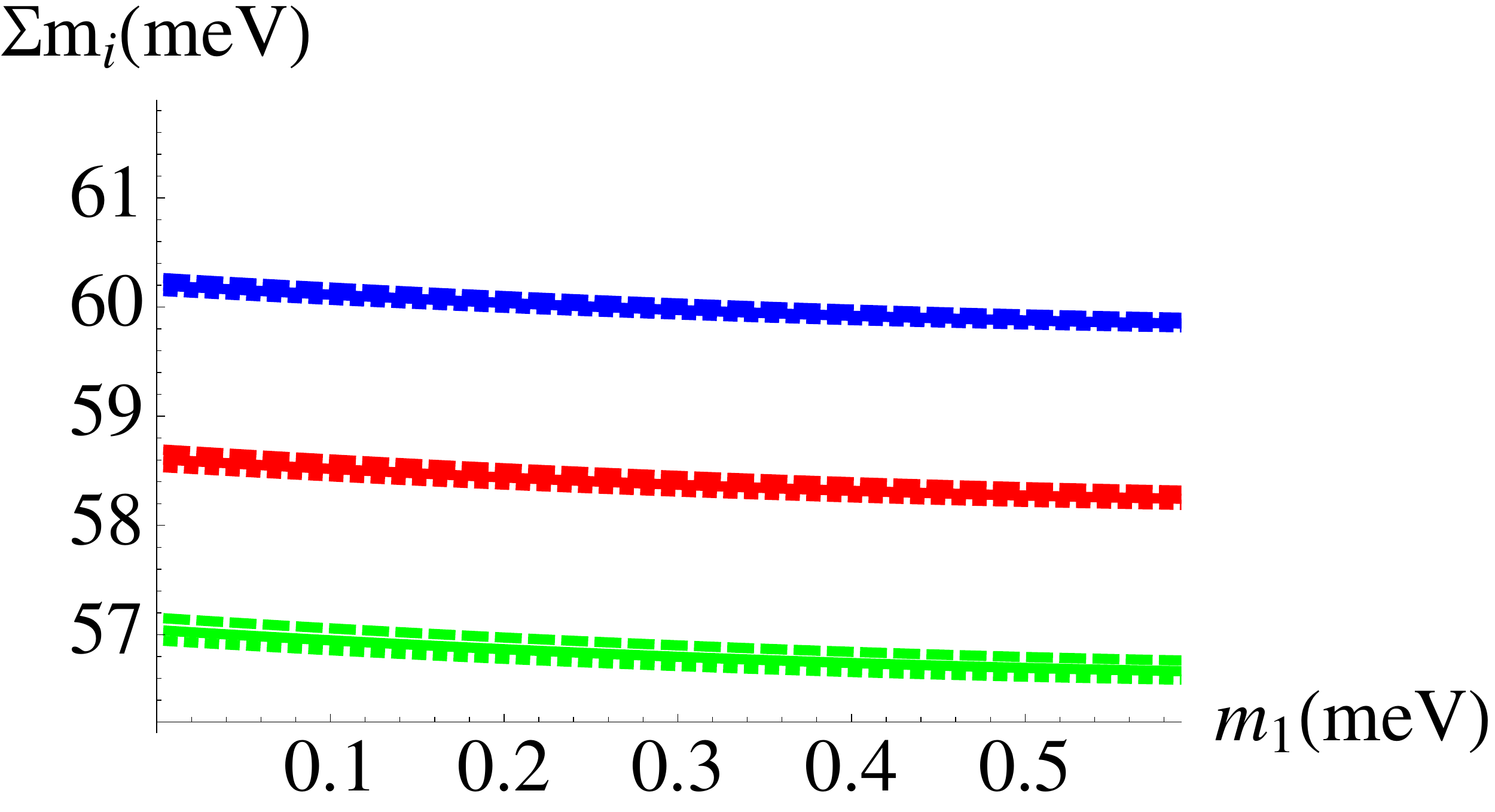} 
    \caption{The neutrinoless double beta decay parameter $|m_{ee}|$ (left panel) and the sum of neutrino
    masses $\Sigma m_i$ (right panel)
    as predicted by the model, resulting from Eqs.\ref{seesaw4},\ref{Ye3}, plotted as a function of the lightest neutrino mass $m_1$ for charged lepton parameters given by
    $A=9,B=7,C=36$ and the down quark couplings in Eq.\ref{yd_diag2}.
    Each line corresponds to a fixed $m_a$ and $m_b$ with varied $m_c$, using the same values
    as in Fig.\ref{neutrinomasssquareds}, with the colour coding and line styles as before.} 
     \label{m}
\vspace*{-2mm}
\end{figure}

In Fig.\ref{m} we show the neutrinoless double beta decay parameter $|m_{ee}|$ (left panel) and the sum of neutrino masses $\Sigma m_i$ (right panel) as predicted by the model,
using the same parameter sets and colour coding as for the other plots.
Note that for $|m_{ee}|$ (left panel) the three colours corresponding to different values of 
$m_a$ lie accurately on top of each other.
The three dashed curves predict
$|m_{ee}|\approx 2.15$ meV, the 
three solid  curves predict
$|m_{ee}|\approx 2.10$ meV and the
three dotted  curves predict
$|m_{ee}|\approx 2.05$ meV, 
corresponding to the three different values of $m_b=2.15,2.10,2.05$.
This can be understood from the neutrino mass matrix in Eq.\ref{seesaw4},
since $|m_{ee}|= |m^{\nu}_{11}|= m_b$, with the charged lepton matrix in Eq.\ref{Ye2}
providing only very small corrections to this result.
The fact that Eq.\ref{eqmee_2} was used to calculate the results and agrees very accurately
with the expectation $|m_{ee}|= |m^{\nu}_{11}|= m_b$ provides a highly non-trivial check on our calculation
of PMNS parameters and neutrino masses, and gives confidence to all our results.
Note that $|m_{ee}|$, being equal to $m_b$, is approximately fixed by $\Delta m^2_{21}$
in Fig.\ref{neutrinomasssquareds}.
Since $|m_{ee}|$ is predicted to be too small to measure in the foreseeable future,
an observation of neutrinoless double beta decay could exclude the model.
Similar comments apply to a cosmological observation of $\Sigma m_i$.

\section{Higher Order Corrections }
\label{higher}

\subsection{HO corrections to vacuum alignment}
The triplet vacuum alignments are achieved by renormalisable superpotentials, 
as discussed in \cite{King:2013xba}. Since the messenger scale
associated with any non-renormalisable corrections to vacuum alignment is unconstrained by the model, 
it is possible that any such terms may be highly suppressed.
In the present analysis we shall therefore ignore any HO 
corrections to the vacuum alignments
in Eqs.\ref{phiu},\ref{phid}.

\subsection{HO corrections to Yukawa operators}
Let us now consider HO corrections to the operators in Eqs.\ref{Yuku},\ref{Yukd_diag},
\ref{Yukd_offdiag}, consisting of extra insertions of $\phi$,
leading to effective operators of the type,
\beq
\Delta W_{Yuk}=F.\left(  \frac{\phi}{\Sigma} \right)^nhF^c,
\label{HO}
\eeq
for $n>1$.
For example, $\frac{ \phi^d_1}{\Sigma_d}$ and $\frac{ \phi^d_2}{\Sigma_u}$ are both 
singlets of $Z_5$, so either of these ratios may in principle be inserted into any of the LO operators
in Eqs.\ref{Yuku},\ref{Yukd_diag}
\ref{Yukd_offdiag}. However in practice, which HO insertions are allowed will depend on the details
of the messenger sector. 
In order for an effective operator to be allowed, it is necessary that that the  
messenger diagram responsible for it can be drawn, and whether this is possible or not will depend
on the choice of charges of the messenger fields $X_F$ and $X_{\overline F}$ under all the symmetries.

In order to allow such HO operators as in Eq.\ref{HO}, 
for $n>1$, at least one of the messenger fields $X_F$ and $X_{\overline F}$ would have to be a triplet of $A_4$ in order to permit the coupling $X_F\phi X_{\overline F}$ where $\phi$ is a triplet,
as is clear from Fig.\ref{mess_HO} (left panel). Such triplet messenger fields $X_F$ and $X_{\overline F}$ are not required in order to construct the LO operators and must be introduced for the sole purpose of allowing the HO operators of this
kind. 

Moreover, such triplet messenger fields would be dangerous since they may allow operators
of the kind in Eq.\ref{HO} for $n=1$ involving the Higgs triplet $h_3$
which could contribute to up and charm quark masses for example.

For these reasons we have chosen not to introduce any messenger fields $X_F$ and $X_{\overline F}$
which are triplets of $A_4$, thereby forbidding HO operators of the type shown in Eq.\ref{HO}
for $n\ge 2$ involving any Higgs fields or involving the $A_4$ triplet Higgs $h_3$ for $n\ge 1$.

The couplings in Eqs.\ref{XFbarphiF},\ref{XFhF},\ref{XFSigmaXF}
can also lead to HO operators of the generic kind, after integrating out the messengers,
as shown in Fig.\ref{mess_HO} (right panel).
\beq
\Delta W_{Yuk}=F.\left(  \frac{\phi}{\Sigma} \right)\left(  \frac{\Sigma}{\Sigma} \right)^nhF^c,
\label{H2O}
\eeq
where $n\geq 1$. At the order $n=1$, only a single operator of this kind is generated,
\beq
\Delta W_{Yuk}=F.\left(  \frac{\phi^d_1}{\Sigma^d_{15}} \right)\left(  \frac{\Sigma_d}{\Sigma_u} \right)h_uF_1^c,
\label{HOup}
\eeq
which gives a correction in the (1,1) entry of $Y^u$ and hence a contribution to the up quark Yukawa coupling,
\beq
\Delta y_u \sim  \epsilon_u \frac{V^d_1}{\vev{\Sigma^d_{15}}} \frac{\vev{\Sigma_d}}{\vev{\Sigma_u}} 
\sim \frac{ \epsilon_u}{ \epsilon_d}\frac{\vev{\Sigma_d}}{\vev{\Sigma_u}}y_d^0 ,
\eeq
where we have used $y_d^0$ given in Eq.\ref{yd_diag}.
The correction is small if $\epsilon_u\vev{\Sigma_d} \ll  \epsilon_d\vev{\Sigma_u}$.

\begin{figure}
\centering
\includegraphics[width=0.4\textwidth]{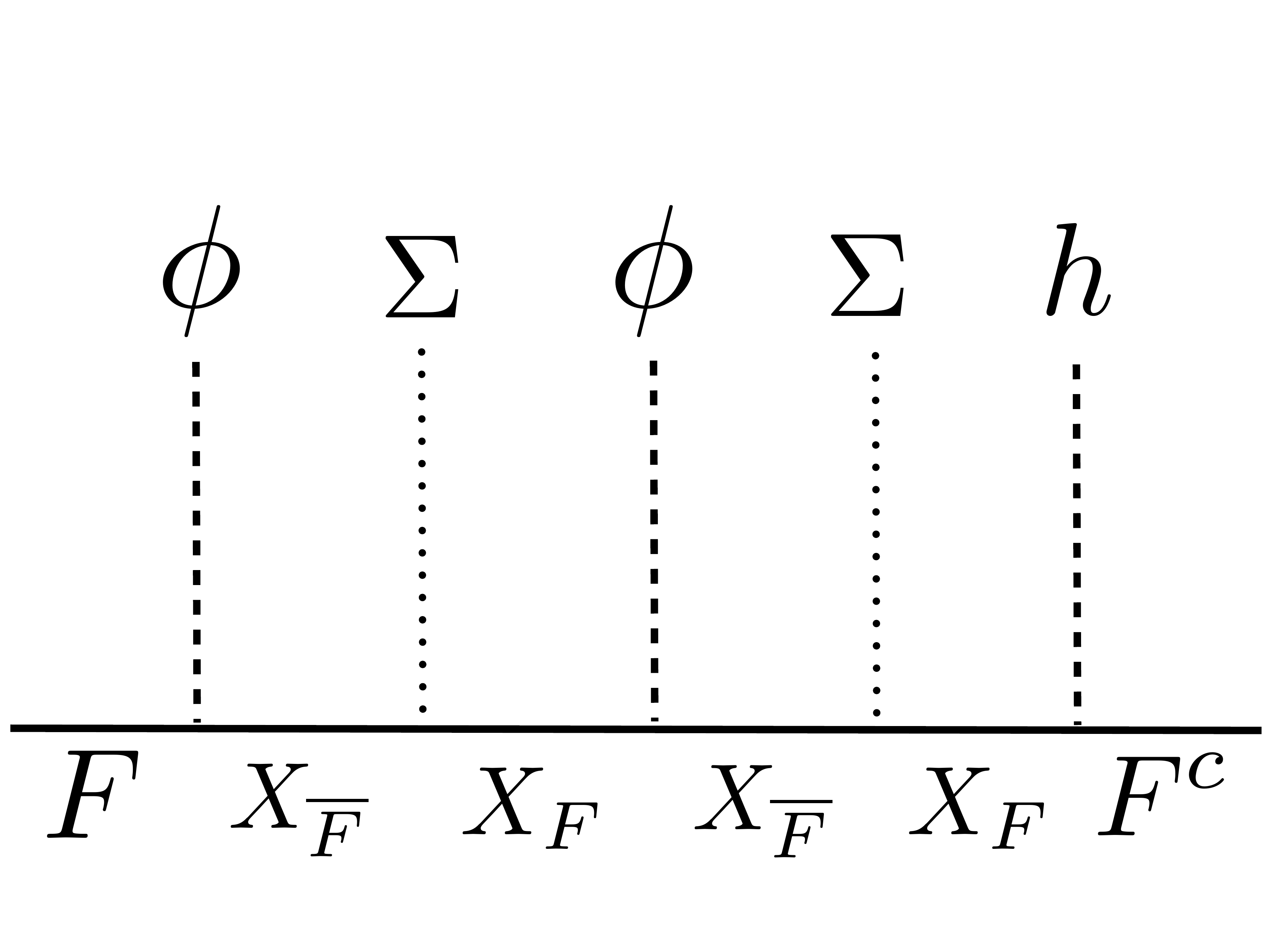} \hspace{5mm}
\includegraphics[width=0.4\textwidth]{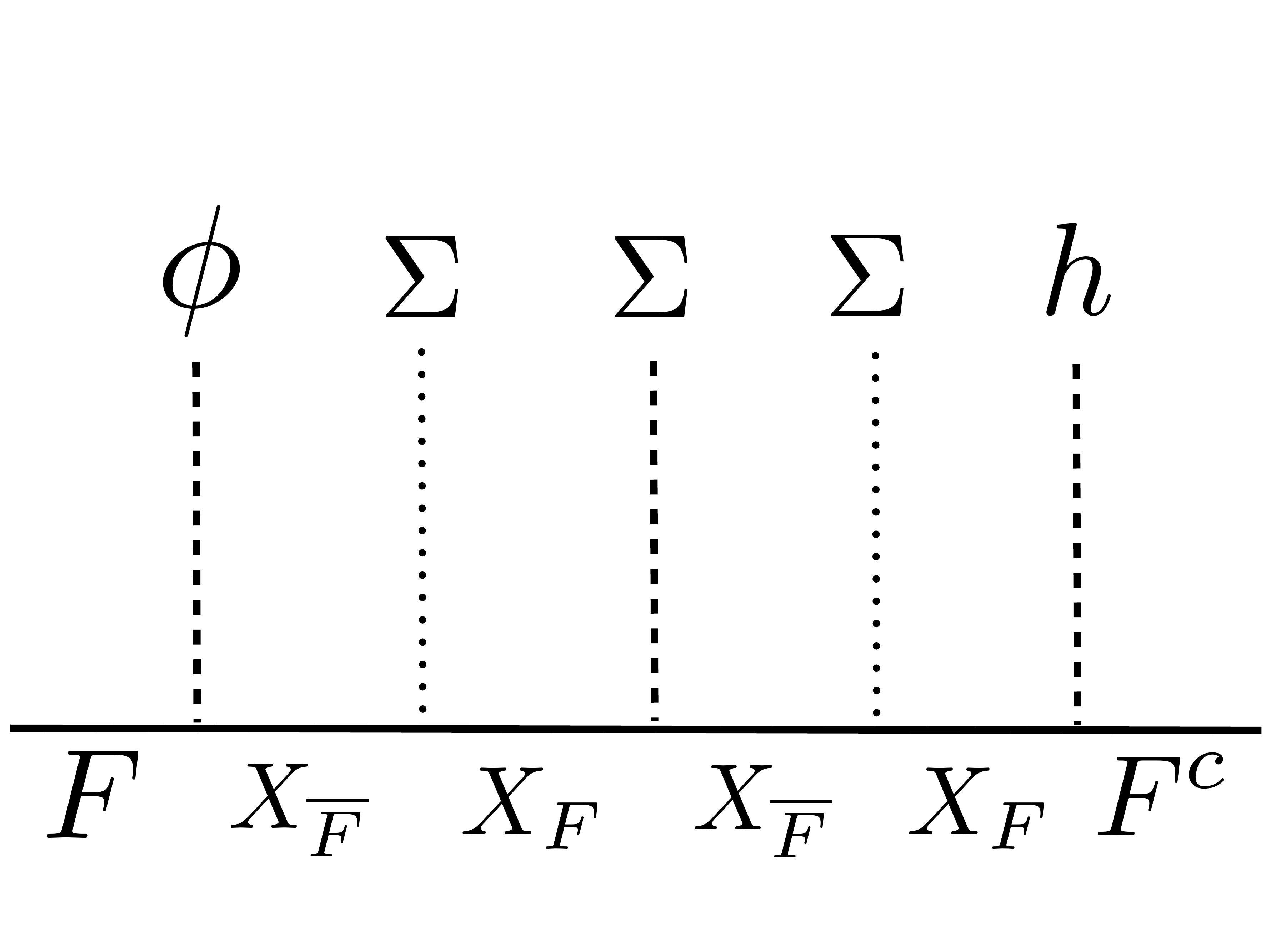} 
\vspace*{-4mm}
    \caption{Some possible higher order diagrams.
   The left panel shows a generic diagram involving triplet fermion messengers, which if present, would lead to effective higher order operators as in Eq.\ref{HO}. In our model we assume such triplet messengers
    to be absent which prevents diagrams with more than one $\phi$ field. 
    The right panel shows a generic diagram responsible for the 
    effective higher order operators as in Eq.\ref{H2O}.} 
     \label{mess_HO}
\vspace*{-2mm}
\end{figure}

\subsection{HO corrections to Majorana operators}

The relevant bilinear charges in the Majorana sector are
\be
F_1^cF_1^c\sim \alpha^2, \ \ 
F_1^cF_2^c\sim \alpha^4, \ \
F_1^cF_3^c\sim \alpha, \ \
F_2^cF_2^c\sim \alpha, \ \
F_2^cF_3^c\sim \alpha^3, \ \
F_3^cF_3^c\sim 1.
\eeq
The messengers which transform under $A_4\times Z_5$ as
$X_{\xi_i}\sim (1,\alpha^i)$ can couple to the Majoron field $\xi\sim (1,\alpha^4)$ 
leading to the LO operators in Eq.\ref{Maj}
(dropping $\overline{H^c}$ and $\Lambda$),
\be
F_1^cF_1^c\xi^2, \ \ 
F_1^cF_3^c\xi, \ \
F_2^cF_2^c\xi, \ \
F_3^cF_3^c\sim 1.
\eeq
Since each insertion of $\xi$ carries a suppression factor of $\vev{\xi}/\Lambda \sim 10^{-5}$, HO operators
involving more powers of $\xi$, such as $F_1^cF_2^c\xi^4$, are negligible.

\section{Conclusions}
\label{conclusions}
 
In this paper we have proposed a rather elegant theory of flavour
based on the Pati-Salam gauge group combined with $A_4\times Z_5$ family symmetry
which provides an 
excellent description of quark and lepton masses, mixing and CP violation.
Pati-Salam unification relates quark and lepton Yukawa matrices and in particular 
predicts $Y^u=Y^{\nu}$, leading to Dirac neutrino masses being equal to up, charm and top masses.
The see-saw mechanism involves very hierarchical right-handed
Majorana neutrino masses with sequential dominance.
The $A_4$ family symmetry
determines the structure of Yukawa matrices via CSD4 vacuum alignment,
with the three columns of $Y^u=Y^{\nu}$ being proportional to 
$(0,1,1)^T$, $(1,4,2)^T$ and $(0,0,1)^T$, respectively, where
each column has a multiplicative phase
determined by $Z_5$ breaking,
which controls CP violation in both the quark and lepton sectors.
The other Yukawa matrices $Y^d\sim Y^e$ 
are both approximately diagonal,
with charged lepton masses related to down quark masses by modified GJ relations,
and containing small off-diagonal elements
responsible for the small quark mixing angles $\theta^q_{13}$ and $\theta^q_{23}$.
The model hence predicts the Cabibbo angle $\theta_C\approx 1/4$, up to such small
angle corrections.

The main limitation of the model is that it 
does not predict the charged fermion masses. However the third family masses are naturally larger since they arise at renormalisable order, while the hierarchy between first and second family masses 
can be understood to originate from
hierarchies between flavon VEVs. Although the model does not predict the small quark mixing angles, it does offer a
qualitative understanding of both CP violation and the Cabibbo angle $\theta_C\approx 1/4$, 
which, as discussed above, is closely related to the lepton mixing angles via the CSD4 vacuum alignment.
Moreover, the model contains 6 fewer parameters in the flavour sector than the 22 parameters of the SM,
and hence predicts the entire PMNS matrix, as is clear from Eqs.\ref{seesaw4},\ref{Ye3}
where all the parameters which appear there are fixed by fermion (including neutrino) masses 
and small quark mixing angles.
Hence the model predicts the entire PMNS lepton mixing matrix with no free parameters,
including the three lepton mixing angles and 
the three leptonic CP phases with negligible theoretical error from HO corrections.
The resulting PMNS matrix turns out to have an approximate TBC form 
as regards maximal atmospheric mixing and the reactor angle $ \theta^l_{13}\approx 9^\circ$,
although the solar angle deviates somewhat from its tri-maximal value, corresponding to a negative deviation
parameter $s\sim -0.03$ to $-0.1$, where $\sin \theta^l_{12}=(1+s)/\sqrt{3}$
\cite{King:2007pr}.

The predictions of a normal neutrino mass hierarchy and maximal atmospheric angle
will both be either confirmed or excluded over the next few years by current or near future neutrino experiments
such as SuperKamiokande, T2K, NO$\nu$A and PINGU \cite{Winter:2013ema}. 
The Daya Bay II reactor upgrade, including the 
short baseline experiment JUNO \cite{Zhan:2013cxa}, will also test the normal 
neutrino mass hierarchy and 
measure the reactor and solar angles to higher accuracy,
enabling precision tests of the predictions $ \theta^l_{13}=9^\circ\pm 0.5^\circ$
and $34^\circ \gsim \theta^l_{12}\gsim 31^\circ$, for the lightest neutrino mass in the range 
$0 \lsim m_1 \lsim 0.5$ meV. With such a mass range, neutrinoless double beta decay 
will not be observable in the foreseeable future.
In the longer term, the superbeam proposals
\cite{Ballett:2013wya} would measure 
the atmospheric mixing angle to high accuracy, confronting the prediction
$\theta^l_{23}=45^\circ\pm 0.5^\circ$,
and ultimately testing the
prediction of the leptonic CP violating oscillation phase $\delta^l=260^\circ \pm 5^\circ$.

\section*{Acknowledgements}
SFK would like to thank Pasquale Di Bari, Simon King and Christoph Luhn for discussions
SFK also acknowledges partial support 
from the EU ITN grant INVISIBLES 289442 .

\appendix

\section{$A_4$}
\label{A4}
$A_4$ has four irreducible representations, three singlets
$1,~1^\prime$ and $1^{\prime \prime}$ and one triplet $3$. 
The products of singlets are:
\begin{equation}\begin{array}{llll}
1\otimes1=1&1^\prime\otimes1^{\prime\prime}=1&1^{\prime}\otimes1^{\prime}=1^{\prime\prime}
&1^{\prime\prime}\otimes1^{\prime\prime}=1^\prime.
\end{array}
\end{equation}
The generators of the $A_4$ group,
can be written as $S$ and $T$ with $S^2=T^3=(ST)^3=\mathcal{I}$.
We work in the Ma-Rajasakaran basis \cite{Ma:2001dn} where the triplet generators are, 
\begin{equation}\label{eqST}
S=\left(
\begin{array}{ccc}
1&0&0\\
0&-1&0\\
0&0&-1\\
\end{array}
\right), \ \ \ \ 
T=\left(
\begin{array}{ccc}
0&1&0\\
0&0&1\\
1&0&0\\
\end{array}
\right)\,.
\end{equation}
In this basis one has the following Clebsch rules for the multiplication of two triplets, 
\begin{equation}\label{pr}
\begin{array}{lll}
(ab)_1&=&a_1b_1+a_2b_2+a_3b_3\,;\\
(ab)_{1'}&=&a_1b_1+\omega a_2b_2+\omega^2a_3b_3\,;\\
(ab)_{1''}&=&a_1b_1+\omega^2 a_2b_2+\omega a_3b_3\,;\\
(ab)_{3_1}&=&(a_2b_3,a_3b_1,a_1b_2)\,;\\
(ab)_{3_2}&=&(a_3b_2,a_1b_3,a_2b_1)\,,
\end{array}
\end{equation}
where $\omega^3=1$, $a=(a_1,a_2,a_3)$ and $b=(b_1,b_2,b_3)$. 

Under a CP transformation in this basis we require \cite{Ding:2013bpa},
\beq
a\rightarrow (a^*_1,a^*_3,a^*_2), \ \ 
b\rightarrow (b^*_1,b^*_3,b^*_2),
\eeq
so that 
\begin{equation}\label{pr}
\begin{array}{lll}
(ab)_{1'}&\rightarrow &a^*_1b^*_1+\omega a^*_3b^*_3+\omega^2a^*_2b^*_2 = (a^*b^*)_{1''} \,\\
(ab)_{1''}&\rightarrow &a^*_1b^*_1+\omega^2 a^*_3b^*_3+\omega a^*_2b^*_2 = (a^*b^*)_{1'} \,.
\end{array}
\end{equation}

\section{Two light Higgs doublets $H_u$ and $H_d$}
\label{Higgs}

We have introduced five Higgs bi-doublet multiplets $h_3$, $h_u$, $h_d$, $h^d_{15}$, $h^u_{15}$, distinguished by $A_4$ and $Z_5$ charges.
Ignoring $SU(4)_C$ and $A_4$ quantum numbers, a generic Higgs bi-doublet 
under $SU(2)_L\times SU(2)_R$ may be written as
\begin{equation}
h=(2,2)=
\left(\begin{array}{cc}
{h_1}^0 & {h_2}^+ \\   
{h_1}^- & {h_2}^0      
\end{array} \right) \label{h}
\end{equation}
where $h_{1}$ and $h_{2}$ form two $SU(2)_L$ doublets with $U(1)_{T_{3R}}$ charges of $-1/2$ and $1/2$.
Henceforth it is convenient to use a slightly different notation as follows.
We label each of the Higgs bi-doublets 
as $h_a(2,2)$ and, below the $SU(2)_R$ breaking scale,
each of them will split into two Higgs doublets,
denoted as $h_a^{\pm}(2,\pm 1/2)$
labelled by their $U(1)_{T_{3R}}$ charges of $\pm 1/2$,
rather than their electric charges as shown in Eq.\ref{h}.
Thus the five bi-doublets above will yield eight Higgs doublets from 
$h_u^{\pm}$, $h_d^{\pm}$ and the colour singlet parts of  
$h^{d\pm}_{15}$, $h^{u\pm}_{15}$, plus additional colour triplet and octet Higgs doublets from $h^{d\pm}_{15}$, $h^{u\pm}_{15}$, together with the six Higgs doublets from $h^{\pm}_3$.
We shall arrange for nearly all of these Higgs doublets to have superheavy masses
near the GUT scale, leaving only the two light Higgs doublets $H_u$ and $H_d$, as follows.

The $h_3$ multiplet, which will be mainly responsible for the third family Yukawa couplings,
is a triplet of $A_4$. We introduce a triplet $\phi_3 \sim 3$ which is a PS and $Z_5$ singlet and couples as
$\phi_3 h_3 h_3$. If $\phi_3$ develops a VEV in the third direction
\footnote{Vacuum alignment may be achieved by a superpotential term $\zeta\phi_3\phi_3$
where $\zeta$ is an $A_4$ triplet driving field, leading to $\vev{\phi_3}\sim (0,0,V_3)$.
In general, small corrections to this vacuum alignment can lead to $\vev{\phi_3}\sim (\epsilon,0,1)V_3$
corresponding to a small admixture $\epsilon$ of the first component of the Higgs triplet 
$h_3$ contributing to the physical light Higgs state $H_u$,
and hence a small correction to $Y^u$ in Eq.\ref{Yu}.
Similar corrections to $Y^d$ may be absorbed into the existing parameters in Eq.\ref{Yd}.}, 
$\vev{\phi_3}\sim (0,0,V_3)$,
then, using the Clebsch rules in Eq.\ref{pr},
this gives a large 
mass to the first two $A_4$ components of $h_3$ while leaving the third component massless.
Introducing a TeV scale mass term $\mu h_3 h_3$
will give a light mass to the third component of $h_3$.
The Higgs bi-doublets in the third $A_4$ component of $h_3$ will mix with other Higgs bi-doublets 
as discussed below and two linear combinations of the mixed states, $H_u$ and $H_d$,
will remain light, allowing the renormalisable third family Yukawa couplings.

The operators involving the Higgs fields $h_u$, $h_d$, $h^d_{15}$, $h^u_{15}$,
collectively denoted as $h_a$,
have the general form,
\beq
\frac{(h_aH^c)(\overline{H^c}h_b)}{S_{ab}}\rightarrow \frac{\langle{H^c}\rangle \langle{\overline{H^c}}\rangle}
{\langle{S_{ab}}\rangle}h_a^+h_b^- \equiv M_{ab}h_a^+h_b^-
\label{hmix0}
\eeq
where $S_{ab}$ are Pati-Salam
singlet fields which develop VEVs somewhat higher than the Pati-Salam breaking scale. 
When $H^c$ gets a VEV in its right-handed neutrino component, it will project out the $T_{3R}=+1/2$
component of $h_a$, which we write as $h_a^+$. Similarly when $\overline{H^c}$ gets a VEV in its right-handed neutrino component, it will project out the $T_{3R}=-1/2$ component of $h_b$, which we write as $h_b^-$.

The diagrams responsible for generating the operators of the form in Eq.\ref{hmix0}
are shown in Fig.\ref{hmixfig}. These diagrams should be considered as Higgsino doublet mixing diagrams. The Higgsino messenger fields which couple to $(h_aH^c)$ are denoted as $X_{H_a}$
and those which couple to $(\overline{H^c}h_b)$ are denoted as $X_{\overline{H_b}}$,
where the messenger masses are generated by the couplings $X_{H_a}S_{ab}X_{\overline{H_b}}$
when $S_{ab}$ develops its VEV, leading to the effective operators in Eq.\ref{hmix0}.

\begin{figure}
\centering
\includegraphics[width=0.4\textwidth]{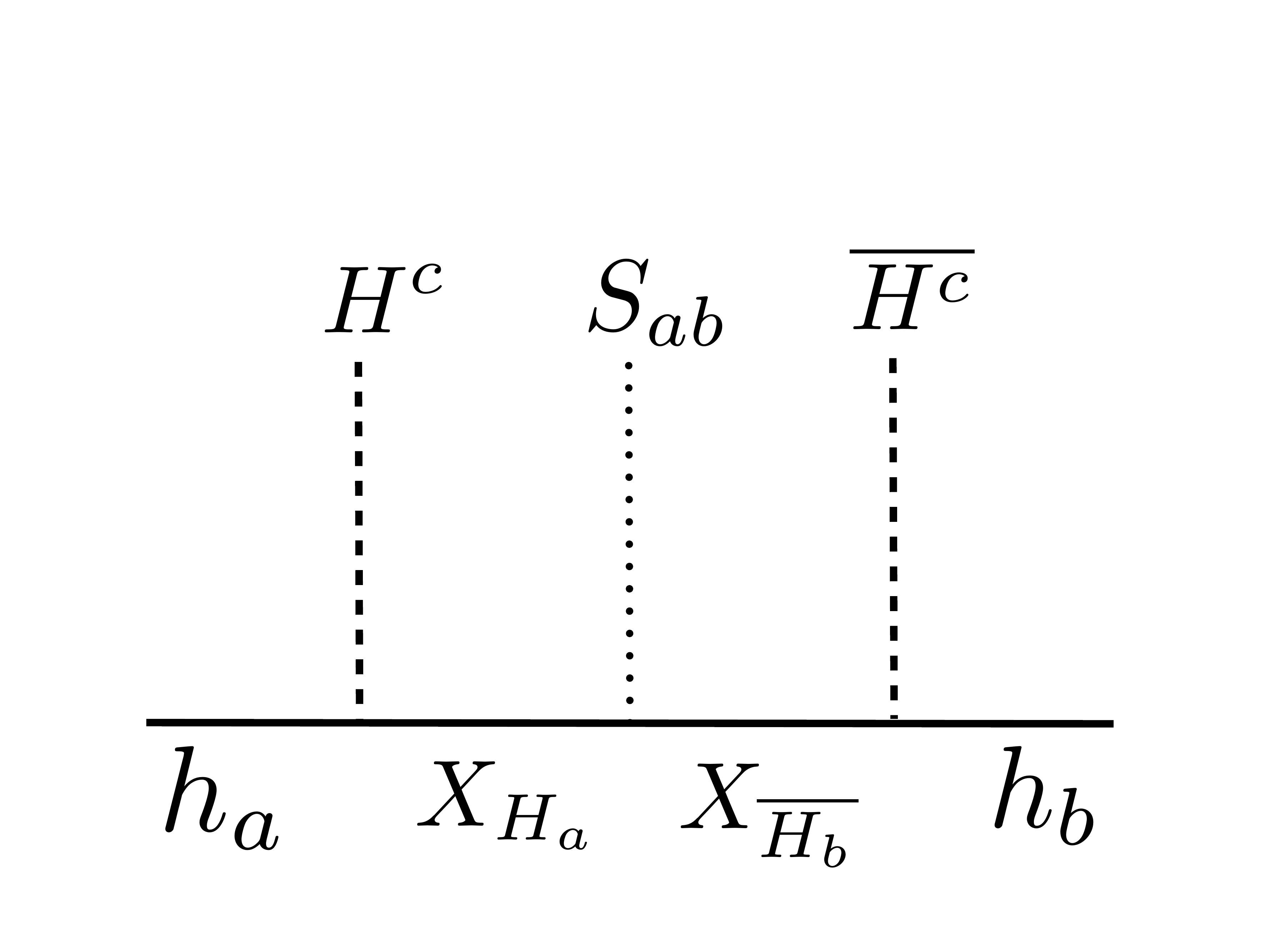} 
\vspace*{-4mm}
    \caption{The diagram shows the higgsino messenger diagrams responsible for the 
    effective operators in Eqs.\ref{hmix0},\ref{hmix} leading to GUT scale higgsino doublet masses. 
    The higgsinos depicted by the solid line have odd R-parity.    } \label{hmixfig}
\vspace*{-2mm}
\end{figure}

The choice of singlets $S_{11}$, $S_{33}$, $S_{24}$, $S_{34}$ with appropriate $Z_5$ 
and $A_4$ charges, lead to the following particular operators of the general form of Eq.\ref{hmix0}:
\bea
&& \frac{(h_uH^c)(\overline{H^c}h_u)}{S_{11}}+\frac{(h^d_{15}H^c)(\overline{H^c}h^d_{15})}{S_{33}} \\
&+& \frac{(h_dH^c)(\overline{H^c}h^u_{15})}{S_{24}}+\frac{(h^u_{15}H^c)(\overline{H^c}h_d)}{S_{24}}\nonumber\\
&+& \frac{(h^d_{15}H^c)(\overline{H^c}h^u_{15})}{S_{34}}+\frac{(h^u_{15}H^c)(\overline{H^c}h^d_{15})}{S_{34}}.
\label{hmix}
\eea
Note that $S_{ab}$ has the same $A_4\times Z_5$ charges as $S_{ba}$.

In addition we require the following three operators, involving 
the third component of $h_3$, given by $h_3.\phi_3$,
\beq
\frac{(\phi_3.h_3H^c)(\overline{H^c}h_u)}{\Lambda_3 S_{01}}+
\frac{(h_dH^c)(\overline{H^c}h_3.\phi_3)}{\Lambda_3 S_{30}}
+ \frac{(h^u_{15}H^c)(\overline{H^c}h_3.\phi_3)}{\Lambda_3 S_{40}}.
\label{h3mix}
\eeq
Since the matrix of charges is symmetric (since $S_{ab}$ has the same $A_4\times Z_5$ charges as $S_{ba}$) the operators above must be given by a particular messenger sector
which forbids similar operators with $H^c$ and $\overline{H^c}$ interchanged.

The operators in Eqs.\ref{hmix},\ref{h3mix} and the term $\mu h_3 h_3$ lead to the following
Higgsino mass matrix,
in the basis where the rows correspond to 
$h_3^{+},h_u^+,h_d^+,h^{d+}_{15},h^{u+}_{15}$ and the columns
correspond to $h_3^{-},h_u^-,h_d^-,h^{d-}_{15},h^{u-}_{15}$,
\begin{equation} \label{hmix2}
\begin{pmatrix}  
\mu & M_{01}  & 0  & 0 & 0 \\ 
0 & M_{11}  & 0  & 0 & 0 \\ 
0& 0 & 0 & 0 & M_{24}\\ 
M_{30} &0 & 0 & M_{33}  & M_{34}\\
M_{40} & 0& M_{42} & M_{43} & 0.
\end{pmatrix}.
\end{equation}
The Higgsino masses from Eq.\ref{hmix2} can be written explicitly as,
\bea
&& \mu h_3^+ h_3^- + (M_{01}h_3^++M_{11}h_u^+)h_{u}^-+h_d^+M_{24}h^{u-}_{15}\\
&& +h^{d+}_{15}(M_{30}h_3^-+M_{33}h^{d-}_{15}+M_{34}h^{u-}_{15})\\
&&+h^{u+}_{15}(M_{40}h_3^-+M_{42}h_{d}^-+M_{43}h^{d-}_{15}).
\eea
By studying these mass terms it is apparent that,
only one linear combination
of the Higgs doublet $h_u^+$ and the third component of the Higgs doublet in $h_3^{u+}$ has a large mass,
namely $M_{01}h_3^++M_{11}h_u^+$,
while the orthogonal linear combination will remain light. It is also clear that only two linear combinations
of the Higgs doublet $h_d^-$ and the colour singlet Higgs doublet in 
$h^{d-}_{15}$ and the third component of the Higgs doublet in $h_3^{-}$
has a large mass, while the orthogonal
linear combination will remain light. 
By contrast, the Higgs doublets in $h_u^-$, $h^{u-}_{15}$,
$h_d^+$, $h^{d+}_{15}$ and
$h^{u+}_{15}$ all appear in three different terms and will all become very heavy.
In particular the colour triplet and octet components of $h^{d\pm}_{15}$ will combine with those of 
$h^{u\mp}_{15}$ so that all coloured Higgs doublets become very massive.

In summary, most of the Higgs doublets will gain large masses near the GUT scale, leaving only two
light Higgs doublets, $H_u$ and $H_d$.
The light Higgs doublet which couples to up-type quarks and neutrinos, $H_u$, will be identified as 
a linear combination of the third component of the Higgs doublet in $h_3^{u+}$
and $h_u^+$.
The light Higgs doublet, $H_d$,
which couples to down-type quarks and charged leptons will be identified
as a linear combination of the Higgs doublet $h_d^-$,
the third component of the Higgs doublet in $h_3^{-}$
 and the colour singlet Higgs doublet from 
$h^{d-}_{15}$. 
The light mass term $\mu h_3^+ h_3^-$ will lead to the term $\mu H_uH_d$ term as in the MSSM.
This term may alternatively be induced by a singlet $S$ term
$S h_3^+ h_3^-$ which will lead to the term $S H_uH_d$ term as in the NMSSM,
generating a light Higgsino mass from the TeV scale singlet VEV.

\end{document}